%% file: 288P_inactive.tex
\documentclass{aa}  

\usepackage{graphicx}
\usepackage{color}
\usepackage{txfonts}
\usepackage{newtxtext}
\usepackage[varvw]{newtxmath}
\usepackage{float}
\usepackage{dblfloatfix}    
\begin{document}

\title{Component properties and mutual orbit of binary main-belt comet 288P/(300163) 2006 VW$_{139}$}

   \subtitle{}

   \author{J. Agarwal\inst{1,}\inst{2}
     \and
     Y. Kim\inst{1}
     \and
     D. Jewitt\inst{3,}\inst{4}
     \and 
     M. Mutchler\inst{5}
     \and
     H. Weaver\inst{6}
     \and
     S. Larson\inst{7}
   }

   \institute{Max-Planck-Institut f\"ur Sonnensystemforschung, Justus-von-Liebig-Weg 3, 37077 G\"ottingen, Germany
     \and Institut f\"ur Geophysik und extraterrestrische Physik, Technische Universit\"at Braunschweig, Mendelssohnstr. 3, 38106 Braunschweig, Germany\\
     \email j.agarwal@tu-braunschweig.de
     \and
     Dept. Earth, Planetary and Space Sciences, UCLA, 595 Charles Young Drive East, Box 951567 Los Angeles, CA 90095-1567, USA
     \and
     Dept. Physics and Astronomy, UCLA, 430 Portola Plaza, Los Angeles, CA 90095-1547, USA
     \and
     Space Telescope Science Institute, 3700 San Martin Drive, Baltimore, MD 21218, USA
     \and
     The Johns Hopkins University Applied Physics Laboratory, 11100 Johns Hopkins Road, Laurel, Maryland 20723, USA
     \and
     Lunar and Planetary Laboratory, University of Arizona, 1629 E. University Blvd., Tucson AZ 85721-0092, USA
   }

   \date{}

 
  \abstract
   {The binary asteroid 288P/(300163) is unusual both for its combination of wide-separation and high mass ratio and for its comet-like activity. It is not currently known whether there is a causal connection between the activity and the unusual orbit or if instead the activity helped to overcome a strong detection bias against such sub-arcsecond systems.}
   {We aim to find observational constraints discriminating between possible formation scenarios and to characterise the physical properties of the system components.}
   {We measured the component separation and brightness using point spread function fitting to high-resolution Hubble Space Telescope/Wide Field Camera 3 images from 25 epochs between 2011 and 2020. We constrained component sizes and shapes from the photometry, and we fitted a Keplerian orbit to the separation as a function of time.}
   {Approximating the components A and B as prolate spheroids with semi-axis lengths $a<b$ and assuming a geometric albedo of 0.07, we find $a_\mathrm{A}\leq$0.6\,km, $b_\mathrm{A}\geq$1.4\,km, $a_\mathrm{B}\leq$0.5\,km, and $b_\mathrm{B}\geq$0.8\,km. We find indications that the dust production may have concentrated around B and that the mutual orbital period may have changed by 1--2 days during the 2016 perihelion passage. Orbit solutions have semi-major axes in the range of (105 -- 109)\,km, eccentricities between 0.41 and 0.51, and periods of (117.3 -- 117.5)\,days pre-perihelion and (118.5 -- 119.5)\,days post-perihelion, corresponding to system masses in the range of (6.67 -- 7.23)$\times$10$^{12}$\,kg. The mutual and heliocentric orbit planes are roughly aligned.}
   {Based on the orbit alignment, we infer that spin-up of the precursor by the Yarkovsky-O’Keefe-Radzievskii-Paddack (YORP) effect led to the formation of the binary system. We disfavour (but cannot exclude) a scenario of very recent formation where activity was directly triggered by the break-up, because our data support a scenario with a single active component.}

   \maketitle
%

\section{Introduction}
The main-belt asteroid 288P (asteroidal designation 300163) combines comet-like activity with being a binary system having unusual properties. 
288P was discovered as an asteroid by the Spacewatch astronomical survey\footnote{https://www.minorplanetcenter.net/db\_search/show\_object?object\_id =300163} at Kitt Peak National Observatory on 15 November 2006 and given the preliminary designation 2006 VW$_{139}$. Activity in this object was first reported on 28 November 2011 \citep{CBET_288P_discovery} based on data from the Panoramic Survey Telescope And Rapid Response System (Pan-STARRS) 1 telescope. 288P has been emitting dust for periods of several months \citep{hsieh_288P-1, licandro_288P} during at least three perihelion passages \citep{CBET_288P_reactivation,hsieh-ishiguro2018}, suggesting that the activity is likely driven by a temperature-dependent process, such as the sublimation of ice.

The orbit of 288P is located in the outer asteroid belt (semi-major axis, $a$ = 3.047 AU, eccentricity, $e$ = 0.201, and inclination, $i$ = 3.2$\degr$) and characterised by a Tisserand parameter with respect to Jupiter of $T_\mathrm{J}$= 3.204, which is typical for asteroids \citep{kresak1982}. Numerical simulations have shown that 288P's position in orbital space is almost impossible to reach from an initial position in the Kuiper Belt \citep{hsieh-haghighipour2016}, such that 288P is most likely native to the asteroid belt.

The combination of comet-like activity and an asteroidal orbit makes 288P one of about 30 currently known active asteroids \citep{jewitt2012, jewitt_AIV} and a member of the sub-group of main-belt comets \citep[MBCs,][]{hsieh-jewitt2006} that is characterised by recurrent activity near perihelion and currently comprises seven known objects.
288P is also one of 11 known members of an asteroid family that formed by break-up of an 11\,km-diameter precursor asteroid about 7.5 million years ago \citep{novakovic-hsieh2012}. The nucleus has been classified as C-type by \citet{licandro_288P}. 
Hubble Space Telescope (HST) images show that 288P is a binary system with similarly-sized ($r \sim$1\,km) components and a wide separation (mutual semi-major axis $a \sim$100\,km) \citep{agarwal-jewitt2016,agarwal-jewitt2017}, while known binary asteroids typically have either similar sizes or wide separations, but not both together.

Most small ($<$20\,km) binary asteroids are thought to have formed by rotational fission \citep{pravec-harris2007, walsh-richardson2008, walsh_AIV} after acceleration by the Yarkovsky-O’Keefe-Radzievskii-Paddack (YORP) effect \citep{rubincam2000}. In 288P this formation hypothesis is supported by the observed alignment of the binary and the heliocentric orbital planes \citep{agarwal-jewitt2017} which is a likely consequence of the YORP effect \citep{vokrouhlicky-nesvorny2003, hanus-durech2011}. However, models indicate that direct formation by rotational fission limits the component separation to $a < 34 r_\mathrm{p}$, with $r_\mathrm{p}$ being the radius of the larger component \citep{jacobson-scheeres2011}. Hence, an additional process such as the binary YORP (BYORP) effect \citep{jacobson-scheeres2014} must have driven the evolution to the current wide separation.

The integrated rotational lightcurve indicates a 16\,h rotation period of at least one component \citep{waniak_288P_dps}. But the sub-escape speed velocity component of dust perpendicular to the orbital plane suggests that fast rotation (with a $\sim$3\,h period) may have augmented the action of gas drag in lifting the dust \citep{agarwal-jewitt2016}. 

The processes behind the formation and subsequent evolution of the 288P binary system and their possible interrelation with the activity are currently not well understood. \citet{agarwal-jewitt2017} conclude that the most likely scenario includes rotational splitting following YORP spin-up, leading to the formation of the binary system and exposing fresh ice, which in turn can have driven the orbital evolution to a wide binary through the recoil force from sublimation. However, they do not rule out other scenarios, including those in which binary formation (either by rotational splitting or by a collision such as the family-forming event 7.5 million years ago) and activation (for example by fast rotation of or impact onto one of the components) are unrelated processes. 
Key aspects that would make it possible to discriminate between the different scenarios include (1) whether one or both components show activity, (2) if the binary orbit has been subject to measurable non-gravitational forces during the last perihelion passage, and (3) the rotation period of the presumably smaller and/or less elongated component. 

We here present new HST/Wide Field Camera 3 (WFC3) observations obtained between August 2017 and May 2020 while 288P was inactive and moving out from perihelion. The observations are described in Section~\ref{sec:obs}. We have used least-squares fitting of the Point Spread Functions (PSFs) of the two components to measure their angular separation and brightness. With the same technique, we have re-analysed the earlier HST data sets from 2011 and 2016-17 \citep{agarwal-jewitt2016, agarwal-jewitt2017} (Section~\ref{sec:data_analysis}).
This technique allows us to reliably measure separations down to the linear pixel scale of the WFC3 (0.04\arcsec). The largest observed separation is 0.09\arcsec\ and the faintest apparent magnitude of an individual component is V=23.5\,mag. Such measurements are well beyond the capabilities of ground-based facilities and challenge even the superb resolution and PSF stability of the HST.
The photometry yields information on the component sizes and elongations, while the relative component positions constrain the orientation of the binary orbit plane and their mutual orbits. In Section~\ref{sec:orbit_fitting} we fit Keplerian orbits to the measured separations using the approach from \citet{agarwal-jewitt2017}. In Section~\ref{sec:discussion} we interpret our findings and discuss their implications for understanding the evolution of the 288P system.

\section{Observations}
\label{sec:obs}
We observed 288P during 12 epochs in HST Cycle 24 
\citep[2016 August -- 2017 February,][visits 1-12 in Table~\ref{tab:observations}, GO 14790, 14864, and 14884]{agarwal-jewitt2017}, during 5 epochs in Cycles 25 (GO 15328) and 26 (GO 15481), respectively (2017 August -- 2019 May, visits 13-23), and during one epoch of 6 HST orbits spread over 29 hours during Cycle 27 (GO 16073, 2020 May, visit 24). The heliocentric constellation of these observations is shown in Fig.~\ref{fig:plane_view}.
The Cycle 24 observations covered the active phase around the perihelion passage, while no activity was detected during Cycles 25 -- 27. We also included in our analysis two epochs of HST observations obtained during the previous active phase in 2011 December \citep[][visits A+B, GO 12597]{agarwal-jewitt2016}. All observations were made with the UVIS channel of the WFC3 (angular pixel scale of 0.04\arcsec) and the broad F606W filter \citep[central wavelength $\lambda_\mathrm{c}$=595.6\,nm, and FWHM=234.0\,nm,][]{baggett-boucarut2007}.
The Cycle 24 -- 26 observations were carried out using a 2$\times$2 dither pattern with sub-pixel offsets, to increase the effective resolution. At each dither station, two 230s exposures were obtained. We read out only the 1k$\times$1k C1K1C subarray, corresponding to a field of view (FOV) of 40\arcsec\ $\times$ 40\arcsec. The Cycle 27 observations were done in three pairs of consecutive orbits. We used a 2-point dither pattern with 8 exposures of 280s per station, and placed the shift between dither stations between the consecutive orbits. To minimise readout time, we used the C512C sub-array, corresponding to an FOV of 512x512 pixels (20\arcsec\ $\times$ 20\arcsec).
In 2011, three exposures of 350s at each of two dither stations were obtained with the full 4k$\times$4k (162\arcsec\ $\times$ 162\arcsec) FOV. The observational circumstances are listed in Table~\ref{tab:observations}.

\begin{table*}[h]
  \caption{Parameters of the HST observations. }
\vspace{2mm}
\centering
\label{tab:observations}
\begin{tabular}{llrrrrrrrrrr}
\hline\noalign{\smallskip}
N & Date & DOY16 & $r_h$ & $\Delta$ & $\alpha$ & $PA_{-\odot}$ & $PA_{-v}$ & $\epsilon$ & long & lat & $\nu$ \\
\hline\hline\noalign{\smallskip}
A & 2011-Dec-07 & --1484.75 & 2.53 & 1.76 & 16.47 & 66.44 & 247.41 & 0.26 & 28.02 & --2.94 & 39.25\\
B & 2011-Dec-15 & --1476.15 & 2.54 & 1.85 & 18.44 & 67.33 & 247.36 & 0.01 & 28.15 & --2.67 & 41.51\\
1 & 2016-Aug-22 & 235.16 & 2.47 & 1.50 &  8.95 & 259.53 & 246.55 &  2.00 & 351.19 &  5.23 & 337.94\\
2 & 2016-Sep-01 & 245.67 & 2.46 & 1.46 &  4.54 & 275.56 & 246.88 &  2.17 & 349.36 & --5.41 & 340.72\\
3 & 2016-Sep-09 & 253.50 & 2.45 & 1.45 &  2.25 & 330.82 & 247.19 &  2.24 & 347.81 & --5.46 & 342.96 \\
4 & 2016-Sep-20 & 264.16 & 2.45 & 1.46 &  5.33 & 42.83 & 247.64 &  2.22 & 345.69 & --5.43 & 346.05\\
5 & 2016-Sep-29 & 273.33 & 2.44 & 1.49 &  9.23 & 54.45 & 248.00 &  2.12 & 344.13 & --5.32 & 348.59\\
6 & 2016-Oct-26 & 300.83 & 2.44 & 1.69 & 18.65 & 63.63 & 248.43 &  1.45 & 342.31 & --4.63 & 356.23\\
7\tablefootmark{a} & 2016-Nov-04 & 309.60 & 2.44 & 1.78 & 20.61 & 64.73 & 248.31 &  1.18 & 342.83 & --4.36 & 358.79\\
8 & 2016-Nov-13 & 318.42 & 2.44 & 1.88 & 22.06 & 65.50 & 248.08 &  0.90 & 343.87 & --4.10 & 1.35\\
9 & 2016-Dec-14 & 349.50 & 2.44 & 2.27 & 23.74 & 66.83 & 246.86 &  0.01 & 351.00 & --3.25 & 10.14\\
10 & 2016-Dec-26 & 361.40 & 2.45 & 2.42 & 23.29 & 67.12 & 246.40 & --0.27 & 354.83 & --2.97 & 13.52 \\
11 & 2017-Jan-17 & 383.46 & 2.46 & 2.70 & 21.35 & 67.74 & 245.83 & --0.66 &   2.92 & --2.54 & 19.69\\
12 & 2017-Jan-30 & 396.23 & 2.47 & 2.85 & 19.70 & 68.26 & 245.74 & --0.82 &   8.16 & --2.31 & 23.30\\
13 & 2017-Aug-24 & 602.63 & 2.78 & 3.23 & 17.33 & 272.62 & 269.11 & 0.94 & 96.43 & --0.20 & 74.73\\
14 & 2017-Nov-10 & 680.19 & 2.93 & 2.38 & 17.84 & 279.25 & 276.38 & 0.78 & 113.15 & 0.82 & 90.88\\
16\tablefootmark{b} & 2018-Jan-31 & 762.50 & 3.10 & 2.21 & 9.22 & 98.41 & 271.81 & --1.08 & 101.42 & 2.08 & 106.11\\
17 & 2018-Feb-27 & 789.50 & 3.15 & 2.52 & 15.60 & 95.35 & 270.86 & --1.23 & 99.52 & 2.13 & 110.78\\
18 & 2018-May-07 & 858.31 & 3.27 & 3.59 & 16.04 & 97.62 & 275.68 & --0.54 & 110.32 & 2.02 & 122.07\\
19 & 2018-Nov-13 & 1048.13 & 3.54 & 3.89 & 14.36 & 294.55 & 292.63 & 0.44 & 168.25 & 2.78 & 149.58\\
20 & 2018-Dec-22 & 1087.15 & 3.57 & 3.35 & 15.91 & 293.13 & 293.57 & --0.12 & 175.02 & 3.35 & 154.79\\
21 & 2019-Jan-25 & 1121.73 & 3.60 & 2.89 & 12.18 & 290.19 & 293.84 & --0.74 & 175.86 & 3.96 & 159.25\\
22 & 2019-Feb-16 & 1143.40 & 3.62 & 2.70 & 6.95 & 284.65 & 293.63 & --1.07 & 173.38 & 4.30 & 162.10\\
23 & 2019-May-17 & 1233.55 & 3.65 & 3.24 & 15.41 & 113.70 & 291.92 & --0.46 & 162.51 & 3.64 & 173.64\\
24\tablefootmark{c} & 2020-May-11 & 1593.83 & 3.45 & 2.45 & 2.86 & 124.28 & 289.96 & --0.71 & 222.16 & 2.84 & 220.63\\ 
\noalign{\smallskip}\hline
\end{tabular}
\tablefoot{$N$ is the sequence number of the observation, $r_\mathrm{h}$ and $\Delta$ are the heliocentric and geocentric distances in AU, $\alpha$ is the phase angle, $PA_{-\odot}$ and $PA_\mathrm{-v}$ are the position angle of the anti-solar direction and of the projected negative orbital velocity vector, $\epsilon$ is the angle between the line of sight and the heliocentric orbital plane of 288P, long and lat are the observer-centred ecliptic longitude and latitude, and $\nu$ is the true anomaly angle. All angles are in degrees.
  \tablefoottext{a}{The perihelion was on 2016-Nov-09, between visits 7 and 8.}
  \tablefoottext{b}{Visit 15 suffered from a problem with guide star acquisition with the result that 288P is trailed in the images at a time-dependent rate. The data were excluded from further analysis.}
  \tablefoottext{c}{Visit 24 observations were obtained between UT 2020-May-11 04:07 and 2020-May-12 12:47. The data given in this table refer to UT 2020-May-11 20:00.}
}
\end{table*}

\section{Data analysis by PSF fitting}
\label{sec:data_analysis}

\subsection{PSF-fitting method}
\label{subsec:psf_method}
To carry out PSF fitting, we used the images in the .flt format for visits A-12, and in the .flc format from visit 13 on. In both formats, the images have been flux calibrated while the native pixel dimensions including image distortions have been preserved. In the .flc format, fluxes have additionally been corrected for charge transfer efficiency (CTE) trailing. CTE corrections are most relevant for faint sources. We expect that during the active phase, when 288P was embedded in a bright dust coma, CTE effects played only a minor role.
We minimum-stacked the two (2011: three, 2020: eight) exposures obtained at a given dither station to remove cosmic ray hits and subtracted a constant local background. This resulted in four (2011: two, 2020: six) images per epoch that we analysed independently by PSF fitting. 

We obtained subsampled images of the WFC3 PSF in the F606W filter from the Space Telescope Science Institute (STScI)\footnote{file PSFSTD\_WFC3UV\_F606W.fits from http://www.stsci.edu/hst/wfc3/analysis/PSF}. The file contains PSF images measured at 56 different locations on the WFC3 chips. We interpolated these to the approximate coordinates of the target using an adapted version of the code given in Appendices B-D of \citet{anderson2016}. The coordinates to which we interpolated the PSF were (512,512) for visits 1-23 (corresponding to the centre of the C1K1C sub-frame), (256,256) for visit 24, and (2048,3240) for visits A and B. The pixels in the PSF images are sub-sampled by a factor of 4 (that is to a linear scale of 0.01\arcsec), and the images have 101$\times$101 pixels, corresponding to a linear size of 1\arcsec. We normalised the PSF images such that the sum over all 101$\times$101 pixels equaled unity. 

\begin{figure*}
  \includegraphics[width=\textwidth]{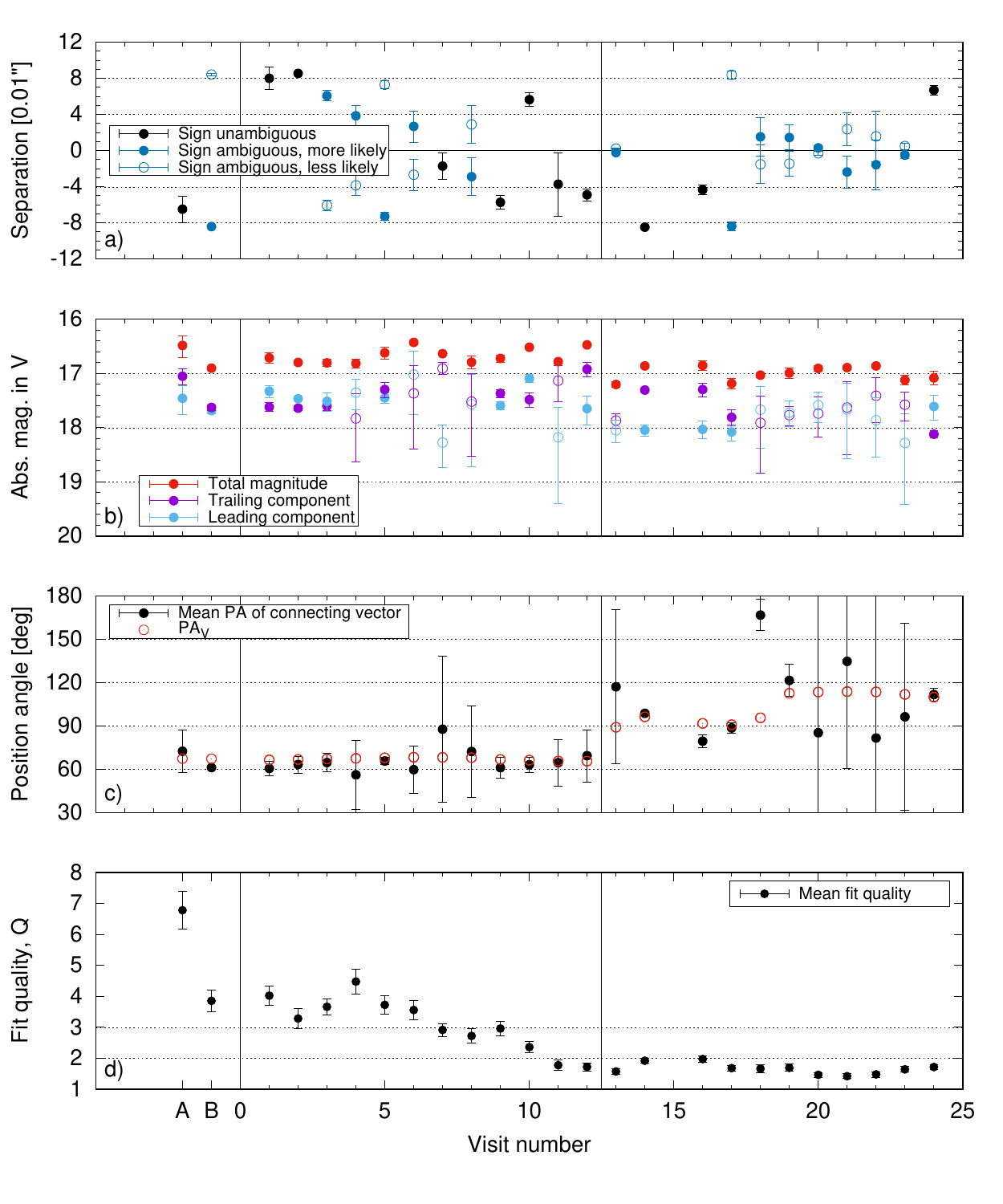}
  \caption{Results of the PSF fitting procedure. a) Component distance $D$ as defined by Eq.~\ref{eq:D}. Open symbols indicate that the brightness relation between the components cannot be determined (overlapping error bars in panel b); b) Absolute magnitudes: Open symbols indicate that $D$<0.04\arcsec, implying significant PSF overlap; The denomination as 'leading' and 'trailing' components does not identify them as unique physical entities between epochs, because their relative position w.r.t. the heliocentric orbital motion changes as they orbit about each other. c) Sky position angles (ccw from north) of the line connecting the components (black) and of the projected heliocentric orbital velocity vector of 288P (red); d) Mean fit quality $Q$ as described by Eq.~\ref{eq:quality}. The horizontal lines indicate fit qualities where the model reproduces the data within 2$\sigma$ and 3$\sigma$ of the expected photon noise on average. Error bars on the data points are 3$\sigma$ in panels a) and c), 2$\sigma$ in panel b), and 1$\sigma$ in panel d).
}
\label{fig:psf_fitting_results}
\end{figure*}

To obtain the best fit to a given image, we added two PSF-images, scaled by factors $f_j$ and centred at coordinates ($x_j$,$y_j$) where $j \in [1,2]$ refers to the components of the binary system. We varied each of these six parameters independently over ranges of possible values constrained from visual inspection of the image. The stepsize for the factors $f_j$ was 100 e$^{-}$ for visits 1-12, and 10 e$^{-}$ for all other visits. For each parameter set, we calculated the sum of squared differences between model and observations, $S$, in a box of 7$\times$7 native pixels or smaller (blue boxes in Figs.~\ref{fig:obsmod_vA} to \ref{fig:obsmod_v24b}). We interpreted the parameter set ($X_j^i, Y_j^i, F_j^i$) that minimised the quantity $S$ at dither station $i$ as the best fit to this observation, and calculated the distance, $D$, and fluxes, $F_1$ and $F_2$, as the averages over all $N$ dither stations obtained at a given epoch:
\begin{equation}
D = \frac{1}{N} \sum_{i=1}^{N} \sqrt{(X_2^i-X_1^i)^2 + (Y_2^i-Y_1^i)^2},
\label{eq:D}
\end{equation}
and
\begin{equation}
F_j = \frac{1}{N} \sum_{i=1}^{N} F_j^i.
\label{eq:Fj}
\end{equation}
Within an image set from the same epoch, we identified the components by their relative position on the sky, rather than by their brightness, assuming that the relative position does not change on the $\sim$1h timescale of the observations, while the relative brightness is more likely to change both intrinsically due to rotation and due to image noise.
We defined $D>$\,0 if the brighter component was preceding the fainter one in the sense of their revolution about the Sun. This definition is based on the assumption that the brightness relation between the two components is always the same. This is not a straightforward assumption if both bodies are of similar sizes and elongated. The measured separations are shown in Fig.~\ref{fig:psf_fitting_results}a). In visits with unclear brightness ratio (overlapping magnitude error bars), we could not decide on the sign of the distance.

For epochs during the active period 2016-17 (visits 1-12) we also tried to add a coma model with the surface brightness dropping inversely proportional to the aperture radius, variable absolute brightness, and centred on either component 1 or 2. This did not significantly improve the fit, mainly because the dust forms a thin linear tail already close to the nucleus rather than a radially symmetric coma. We did not include this coma model into the subsequent analysis.

For visits 6 -- 12, the quantity $S$ decreased when the tailward component in the model was moved away from the central condensation along the tail direction. Our interpretation is that the fitting process was 'locking' on representing the tail with one of the model components. To prevent this, we artificially decreased the area contributing to $S$ on the tailward side for visits 1-12 (blue boxes in Figs.~\ref{fig:obsmod_v1} to \ref{fig:obsmod_v12}). Still, we cannot exclude that the best-fit distances, in particular for visits 6-12, are influenced by the presence of the bright dust tail.

We found component separations below the native pixel scale of 0.04\arcsec\ for visits 13 and 18-23 (Fig.~\ref{fig:psf_fitting_results}a). For these visits, and for A and B (when only two dither stations were used), we repeated the fitting procedure at a 5$\times$ finer pixel scale, using an interpolated PSF as described in \citet{anderson2016}, and a flux stepsize of 10 e$^{-}$. 

The shape of the PSF is known to depend on time, for instance due to thermal breathing of the focal length. We studied this effect by fitting the visit 14 image set (having particularly large residuals) with PSFs obtained at different focus parameters (priv. communication by J. Anderson). While the value of $S$ showed a significant dependency on the focus parameters, the results obtained for $D$ and $F_i$ and their uncertainties were comparable for different focus models.
For the same data set, we also found that the results did not significanly change when we fitted single exposures rather than minimum-stacks of images pairs.

The best fitting PSF models we used for subsequent analysis are shown in Figs.~\ref{fig:obsmod_vA} to \ref{fig:obsmod_v24a}. Panel a) of Fig.~\ref{fig:psf_fitting_results} shows the measured distances $D$, including the sign inferred from the brightness relation (panel b). Panel c) shows the sky position angle of the line connecting the components (cf.~Sec.~\ref{sec:plane_angle}), while panel d) shows the mean quality $Q$ of the fit. We define the quality of a fit to an image obtained at dither station $i$ as follows:
\begin{equation}
  Q_i = \sum_{k=1}^{N_\mathrm{px}} q_k = \sum_{k=1}^{N_\mathrm{px}} \frac{|f_k^{(sim)} - f_k^{(obs)}|}{\sqrt{f_k^{(sim)}}},
  \label{eq:quality}
\end{equation}
where the index $k$ counts all pixels in the fitting area, and $f_k$ is the flux in electrons per pixel. Assuming that the noise is photon dominated, $q_k$ describes the per-pixel difference between observation and model in units of the noise expected from the simulation. The quantity $Q$ shown in Fig.~\ref{fig:psf_fitting_results} describes therefore the average $q_k$ over all pixels and dither stations. For all epochs during the inactive phase (visit 13+) the simulation reproduces the data within 2$\sigma$. Larger deviations during the active phases can be explained by our not including the dust coma into the model.

\subsection{Photometry}
\label{sec:photometry}
Panel b) of Fig.~\ref{fig:psf_fitting_results} shows the absolute magnitudes corresponding to the measured component fluxes. Before converting electrons to absolute magnitudes, we corrected the counts, $n_{{\mathrm e}^-}$, measured at time $t_2$ for the temporal sensitivity loss of WFC3 as described in \citet{khandrika-deustua2018}:
\begin{equation}
n_{{\mathrm e}^-,0} = n_{{\mathrm e}^-} \left(1 - \frac{l (t_2-t_1)}{100} \right),
\end{equation}
where $n_{{\mathrm e}^-,0}$ is the number of counts as it would have been measured at the time, $t_1$, of our first observation (visit A), $l$=-0.1744\% per year describes the sensitivity loss of chip 2 in the F606W filter, and $(t_2-t_1)$ is expressed in years. The corrected flux is up to 1.5\% larger than the measured flux, which is small compared to the uncertainty of the photometry derived from comparing the different methods (Appendix~\ref{app:photometry}).

We converted electrons to apparent magnitudes $V_\mathrm{app}$ using the relation
\begin{equation}
  V_\mathrm{app} (n_{{\mathrm e}^-,0},T_\mathrm{exp}) = 20. - 2.5\log_{10} \left( \frac{n_{{\mathrm e}^-,0}}{T_\mathrm{exp} Cr} \right),
\end{equation}
where $T_\mathrm{exp}$ is the exposure time in seconds. $Cr$= 288.674 e$^-$ s$^{-1}$ is the count rate obtained in F606W filter from a source with a sun-like (Kurucz G2V) spectrum renormalised to Vega magnitude 20 in Johnson/V filter as obtained from the WFC3 Exposure Time Calculator\footnote{ETC Request ID WFC3UVIS.im.1411381}.

\begin{figure}
  \includegraphics[width=\columnwidth]{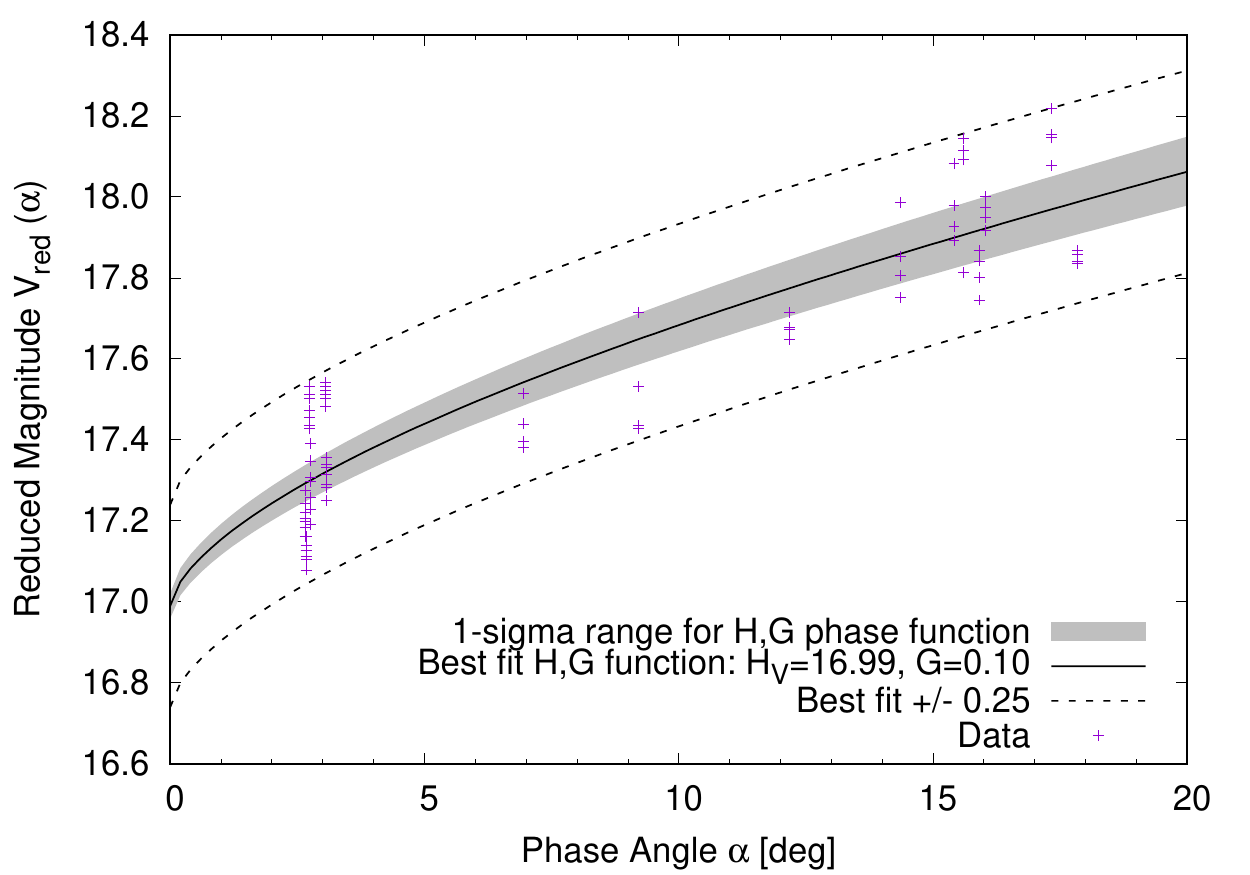}
  \caption{Phase function of the combined, inactive 288P system. The data points are from PSF-fitting of two point sources to individual exposures for visit 24 ($\alpha \sim$ 3$^\circ$), and from fitting minimum-stacked pairs as described in Sec.~\ref{subsec:psf_method} for visits 13-23. The scatter of the data reflects the rotation-induced variability of the system's brightness. The solid line shows the best-fit $H-G$ phase function with $H_\mathrm{V}$=(16.99 $\pm$ 0.03) and $G$=(0.10 $\pm$ 0.04). The shaded area reflects the 1$\sigma$ uncertainty of the fit, and the dashed lines indicate the interval $\Delta V_\mathrm{red} = \pm$0.25 covered by the rotational lightcurve. 
}
\label{fig:phase_fn}
\end{figure}

We next converted the apparent magnitudes to reduced magnitudes valid for unit heliocentric and geocentric distances
\begin{equation}
V_\mathrm{red}(1,1,\alpha) = V_\mathrm{app} - 5 \log_{10}(r_\mathrm{h} \Delta).
\end{equation}
In Fig.~\ref{fig:phase_fn} we plotted the reduced magnitudes of the combined, inactive system (visits 13 -- 24) as a function of the phase angle during observation, and fitted the data with a phase function $\Phi(H_\mathrm{V},G,\alpha)$ as defined in \citet{bowell-hapke1989}.
We found an average absolute magnitude of $H_\mathrm{V}$ = $V_\mathrm{red}$ + 2.5 $\log_{10} \Phi(G,\alpha)$ = (16.99 $\pm$ 0.03) for the combined, inactive system, with a rotation-induced amplitude of at least $\Delta H_\mathrm{V} = \pm$0.25, and $G$=(0.10 $\pm$ 0.04). 
This result is consistent with the reported $H_\mathrm{V}$=17.0$\pm$0.1\,mag in \citet{agarwal-jewitt2016} and $H_\mathrm{R}$=16.80$\pm$0.12\,mag (corresponding to $H_\mathrm{V}$=17.20$\pm$0.12 for solar colours) in \citet{hsieh-ishiguro2018}, and to some extent with the lightcurve amplitude of 0.4\,mag reported preliminarily by \citet{waniak_288P_dps}.

\begin{figure}
  \includegraphics[width=\columnwidth]{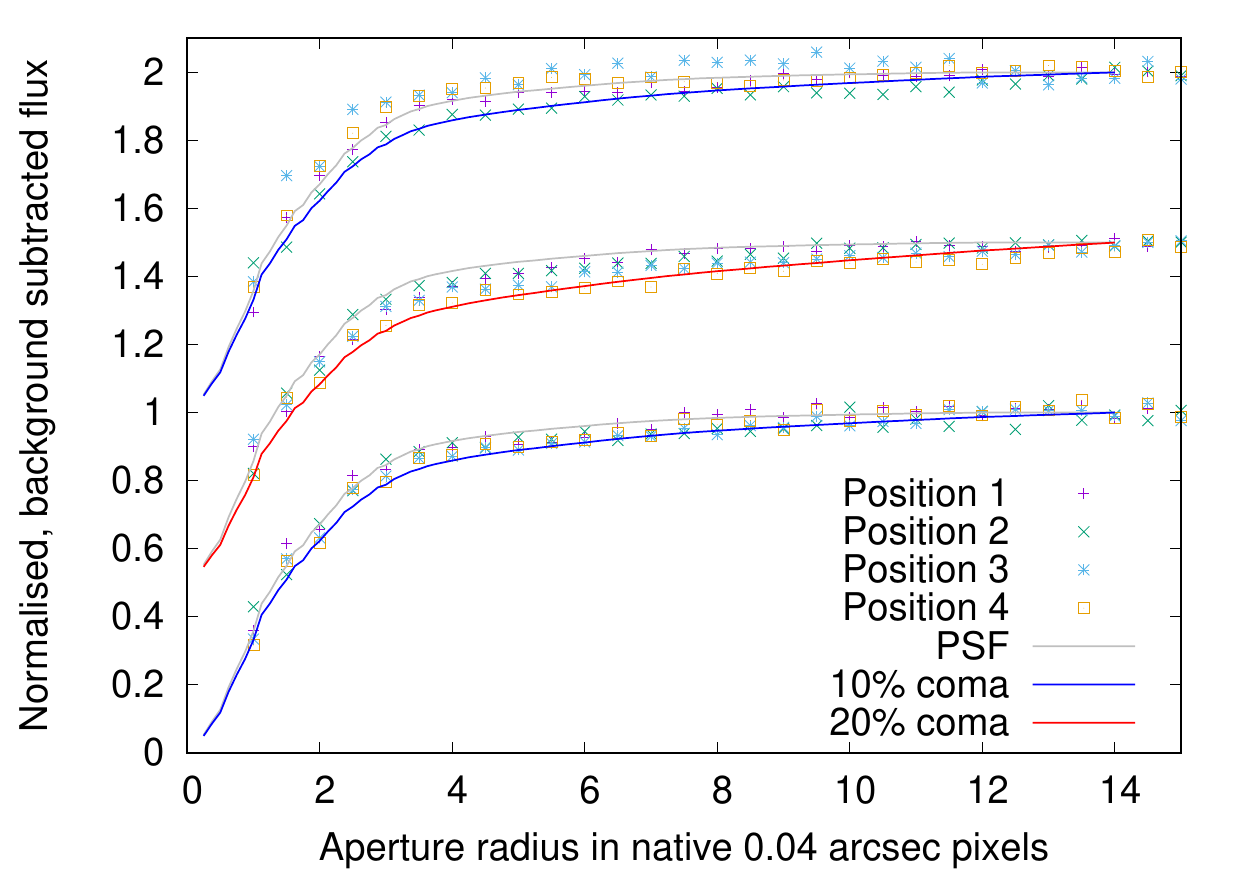}
  \caption{Radial profiles $F_\mathrm{n}(r)$ of the nucleus at four dither stations and of the PSF for visits 13 (bottom), 20 (shifted upwards by 0.5), and 23 (shifted by 1). We do not find a systematic broadening of the asteroid PSF that could be indicative of dust. Deviations between dither stations can be related to thermal breathing of the PSF. The blue and red lines show the PSF profile combined with a steady state coma reaching 10\% and 20\% of the nucleus cross-section at 14 pixels (Eq.~\ref{eq:PSF_plus_coma}).
}
\label{fig:rad_prof}
\end{figure}

To search for possible faint dust remaining near the nucleus from the last perihelion passage, we compared the radial brightness profiles of 288P to that of the PSF. For this analysis we used visits 13, 20, and 23, which have the smallest component separation ($<$0.005\arcsec) such that broadening of the profile due to the binary nature is expected to be negligible. In the minimum-stacked images, we measured the total flux, $F(r)$, in circular apertures centred on the nucleus and having radii, $r$, increasing from 1 to 26 native pixels, and for $r$>13\,px fitted the result with the function
\begin{equation}
F(r) = b \pi r^2 + F_\mathrm{rp},
\end{equation}
where $b$ is the background and $F_\mathrm{rp}$ the nucleus flux. Fig.~\ref{fig:rad_prof} shows the normalised, background-subtracted nucleus flux $F_\mathrm{n}(r) = (F(r) - b \pi r^2) / F_\mathrm{rp}$ together with the PSF profile.

To estimate the possible contribution of a faint coma, we also plotted the PSF profile combined with a function $F_\mathrm{c}(r) = F_0 (r/r_0)$, describing a coma in steady state that reaches a total flux $F_0$ (in units of the brightness of the central point source) in an aperture of radius $r_0$. The combined brightness normalised at $r_0$ is then described by
\begin{equation}
  F_\mathrm{comb} (r) = [ F_\mathrm{PSF} (r) + F_\mathrm{c} (r) ] / (1+F_0).
  \label{eq:PSF_plus_coma}
\end{equation}
Using $r_0$=14 pixels ($\sim$1300\,km), we find that the profile from visit 13 is technically compatible with a steady state coma with a total cross-section inside $r_0$ of 10\% of the combined asteroid cross-sections, while visit 20 tolerates up to 20\%.
However, during visit 23 one of the profiles also tolerates 10\% of coma, while other profiles (especially from position 3) are even steeper than the PSF, which cannot be explained by dust but only by a temporal variation of the PSF itself, such as by thermal breathing. Hence it is possible that the observed deviations during visits 13 and 20 are also intrinsic to the PSF, such that the numbers given for the dust cross-sections are only upper limits but do not prove the presence of dust.

We also compared the fluxes measured through PSF-fitting with results from aperture photometry and radial profile fitting, finding an inter-method variability of about 10\%, but no strong systematic trends (Appendix~\ref{app:photometry}).
The following analysis is based exclusively on the photometry derived from PSF fitting, as shown in panel b) of Fig.~\ref{fig:psf_fitting_results}.
The brightness and sizes of the individual components for both the inactive and the active system will be discussed in Secs.~\ref{sec:component_sizes} and \ref{sec:active_component}.

\subsection{Orbit plane orientation}
\label{sec:plane_angle}
Panel c) of Fig.~\ref{fig:psf_fitting_results} shows the sky position angle of the line connecting the components, together with the position angle of the negative heliocentric velocity vector. The latter corresponds approximately to the projected heliocentric orbit plane of 288P, as the angle between the line of sight and the orbit plane was $\epsilon\!\!<$2.3$^\circ$ during all observations. In particular for situations with $D\!\!\gg$0.04\arcsec, the two position angles are reasonably consistent. Given that the observations cover a wide range of ecliptic latitudes, we interpret this as a strong indication that the mutual and the heliocentric orbits are roughly aligned.

\section{Binary orbit}
\label{sec:orbit_fitting}

\subsection{Method}
\label{sec:orbit_fit_method}
We used the distances shown in Fig.~\ref{fig:psf_fitting_results}a) as constraints to fit a model of a Keplerian mutual orbit with the semi-major axis, $a$, the eccentricity, $e$, the period, $T$, the time of periapsis, $t_{\rm per}$, and the angle, $\alpha_0$, between the major axis and the line of sight from Earth to 288P at $t_0$ = 2016 August 22, as free parameters. Details of this model are described in \citet{agarwal-jewitt2017}. Our fit is based on the assumption that no additional components exist in the 288P system.
We treated the inclination of the binary orbit as fixed to be parallel to the heliocentric orbit plane (Sec.~\ref{sec:plane_angle}). We studied both prograde and retrograde orbits with 70$\leq a$/km$<$300 (stepsize 2\,km), 0$\leq e<$0.94 (stepsize 0.02), 20$\leq T$/days$<$300 (stepsize 0.2 days), 0\degr$\leq\alpha_0<$180\degr, and $t_{\rm per}$ in 50 steps of 1/50 of the orbital period starting from $t_0$. We considered a given model orbit consistent with an observation if the model distance for the time of observation matched the magnitude of the distance $D$ (Fig.~\ref{fig:psf_fitting_results}a) within 3$\sigma$, hence we did not take into account the sign of $D$. The error bars for visit 14 in \ref{fig:psf_fitting_results}a) are formally consistent with zero, and we substituted them with $\sigma_\mathrm{17}$=0.15$\times$10$^{-2}$\arcsec\ to enable the fitting to return results. Also in visit 13, $D$ has a formally small error $D_\mathrm{13}$=(0.23$\pm$0.02)$\times$10$^{-2}$\arcsec. We substituted this by $D_\mathrm{13}$=(0.0$\pm$0.25)$\times$10$^{-2}$\arcsec, because for any of the individual frames (Fig.~\ref{fig:obsmod_v13}) neither $X_2-X_1$ nor $Y_2-Y-1$ exceeded the resolution of the twice sub-sampled PSF image (0.002\arcsec), and, in addition, both the relative position of the brighter and fainter components and the orientation of the line connecting them seem randomly distributed. We conclude that the component separation during visit 13 was below our resolution and therefore set it to zero with an error bar sufficiently large to include the formal solution.
Of all possible parameter combinations, we read out those that at least either matched visits 1, 2, 3, and 5 simultaneously, or visits 13-18.

\subsection{Results}
\label{sec:orbit_fit_results}
The most comprehensive orbit solutions we found reproduce 21 out of the 25 data points (block I in Table~\ref{tab:solutions_overview}).
\begin{table*}
  \includegraphics[width=\textwidth]{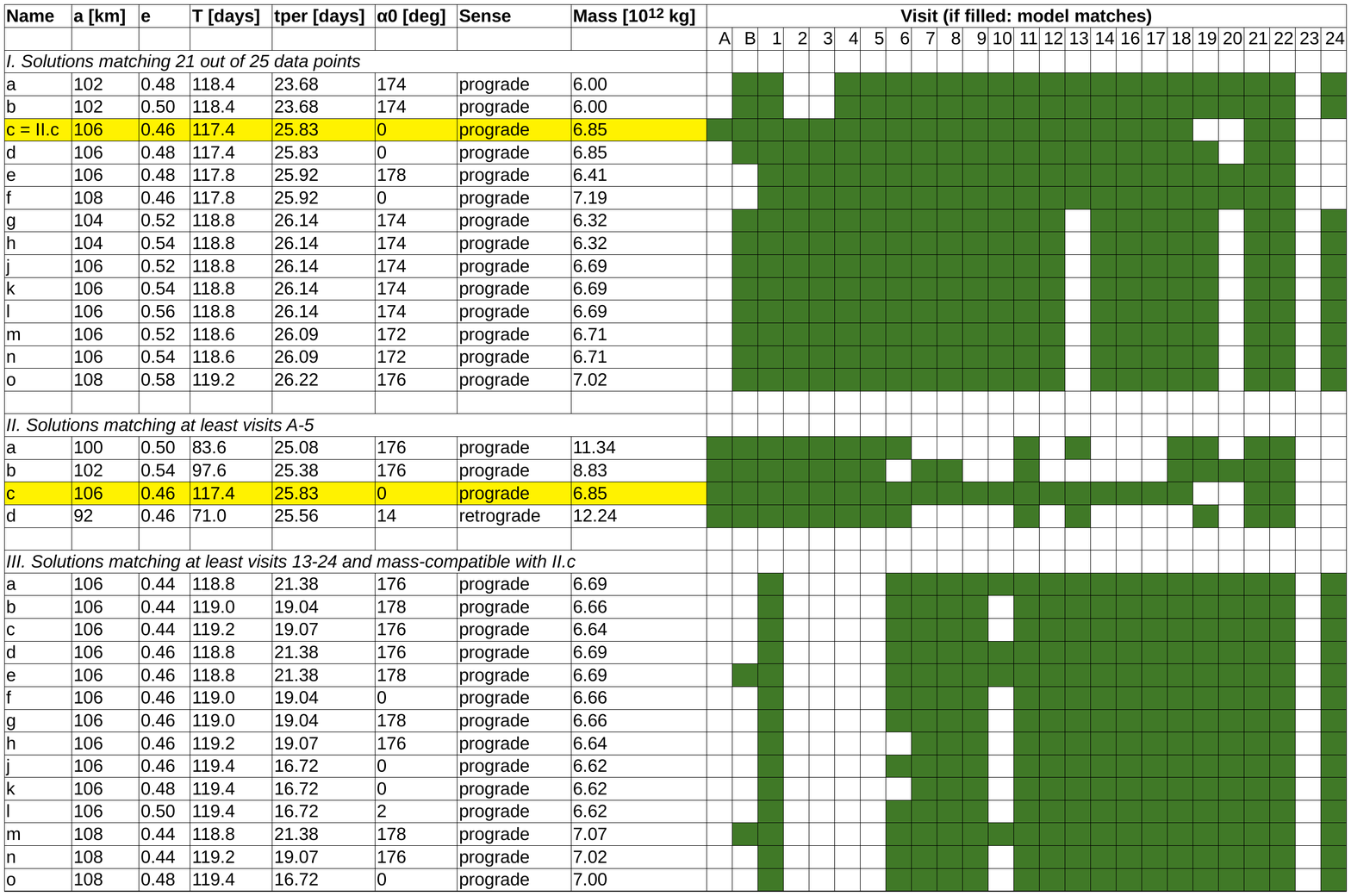}
  \caption{Overview of best-fitting orbit solutions.}
  \label{tab:solutions_overview}
  \tablefoot{The solutions are listed in three blocks (I to III) as discussed in the text. The first column identifies the orbits with letters to facilitate the discussion in the text. Columns 2-8 list the orbit's Keplerian elements (semi-major axis, $a$, eccentricity, $e$, mutual orbital period, $T$, time of perihelion relative to UT 2016 August 22, $t_\mathbf{per}$, and the angle between the line of sight on UT 2016 August 22 and the line of apsides, $\alpha_0$), the sense of rotation and the system mass given by Kepler's third law $M_\mathrm{S} = (4 \pi^2 a^3) / (G T^2)$. The following 24 columns show whether an orbit solution reproduces the measured distance within its error bars during the concerned epoch regardless of its sign (green if true). Marked in yellow is the solution (II.c) that we adopt as the best fit to our data (Fig.~\ref{fig:orbit_solutions}), because it best reproduces the pre-perihelion data and indeed all datasets from 2011 to mid-2018. It is also the only solution with a mass-compatible equivalent matching all data from the inactive phase (block III and refined parameter grid in Fig.~\ref{fig:mass_compatible_solutions}).}
\end{table*}
The only solution (I.c) reproducing both visits A and B is incompatible with visits 19, 20, 23, and 24, while the solutions reproducing visit 24 are incompatible with at least visit A. Some parameter sets reproduce neither A nor 24. Solutions I.g--o fail to reproduce visit 13 by a large amount, which disqualifies them. The visit 23 data point is not matched by any of the block I solutions, which may be related to the comparatively small errorbar of this measurement. Also visit 20 has a possibly too small errorbar.

Mismatches between model and observation typically occur either with the very early, or the very late observations. Possible reasons include the very long (5-years) separation in time between visits B and 1 and the poor spatial resolution during Cycle 26 (visits 19--23), although the latter argument is to some extent invalidated by the much higher spatial resolution achieved during visit 24.

Since the mismatches cluster towards the ends of the data set, it is conceivable that the mutual orbit changed as a consequence of the outgassing during the perihelion passage in 2016/17. We therefore searched separately for orbit solutions that reproduce either the pre-perihelion data (minimum visits A -- 5), or the (post-perihelion) data from the inactive phase (visits 13-24), and tried to identify pairs of solutions with similar masses.

This mass-constraint is derived from the following reasoning. We assumed that the dust mass loss during the 2016/17 perihelion passage was of order 10$^8$ kg from \citet{hsieh-ishiguro2018} reporting a dust production rate of 5.6\,kg\,s$^{-1}$ and an active period of about 200 days. \citet{licandro_288P} infer a total dust production of 2$\times$10$^{6}$\,kg for the 2011 perihelion passage, and \citet{agarwal-jewitt2016} report an instantaneous measurement of 10$^7$\,kg. Typical dust to gas mass ratios assumed for comets and active asteroids range between 1 and 10 \citep{choukroun-altwegg2020}, hence the total system mass loss per perihelion passage likely is of order 10$^{8}$\,kg or less, which is four orders of magnitude smaller than the system mass. We therefore expect the system mass pre- and post-perihelion to be similar.
  
We find four different parameter sets reproducing at least visits A -- 5 (II.a--d in Table~\ref{tab:solutions_overview}).
 The solutions matching all data from visits 13 to 24 (not listed in  Table~\ref{tab:solutions_overview}) have orbital periods in the range of 50 -- 53 days, and semi-major axes between 80\,km and 110\,km. The corresponding masses are in the range (16 -- 35) $\times$ 10$^{12}$ kg. The typical time interval between epochs in 2017/18 (when the data quality was highest) was 40\,days, such that orbital periods $T<$80\,days were not Nyquist sampled. The 2018/19 data are better sampled but suffer from larger uncertainties. Hence we do not consider the solutions with periods of order 50 days reliable.
With the component volumes to be derived in Sec.~\ref{sec:component_sizes}, the derived masses correspond to densities in the range (2900 -- 26000)\,kg\,m$^{-3}$, which are inconsistent with C-type densities. We found a strong anticorrelation between mass and eccentricity for these orbits, such that the lowest (least unlikely) densities corresponded to $e$>0.9, rendering this model even less probable.
We conclude that we did not find a plausible solution reproducing all measurements from the inactive phase within their error bars.

Hence we searched for solutions that formally reproduced only 10 out of these 11 measurements. We further limited the results to those parameter sets that failed to reproduce either visit 20 or 23, because the formal errorbars on these two measurements are very small and may be underestimating the true uncertainty. In the large majority of solutions fulfilling these criteria, the mismatch occurs for visit 23. We find three groups of solutions, prograde ones with periods around 119 days and 140 days, and retrograde ones with periods of about 144 days. To further constrain these solutions, we compared their masses to the four orbits compatible with visits A--5. All visit 13--24 solutions have masses $<$7.1$\times$10$^{12}$ kg, hence they are incompatible with solutions II.a, b and d. The visit 13--24 solutions with masses most similar to the 6.85$\times$10$^{12}$ kg of II.c have masses clustering around 6.65$\times$10$^{12}$ kg and 7.02$\times$10$^{12}$ kg (block III in Table~\ref{tab:solutions_overview}).

\begin{figure}
  \includegraphics[width=\columnwidth]{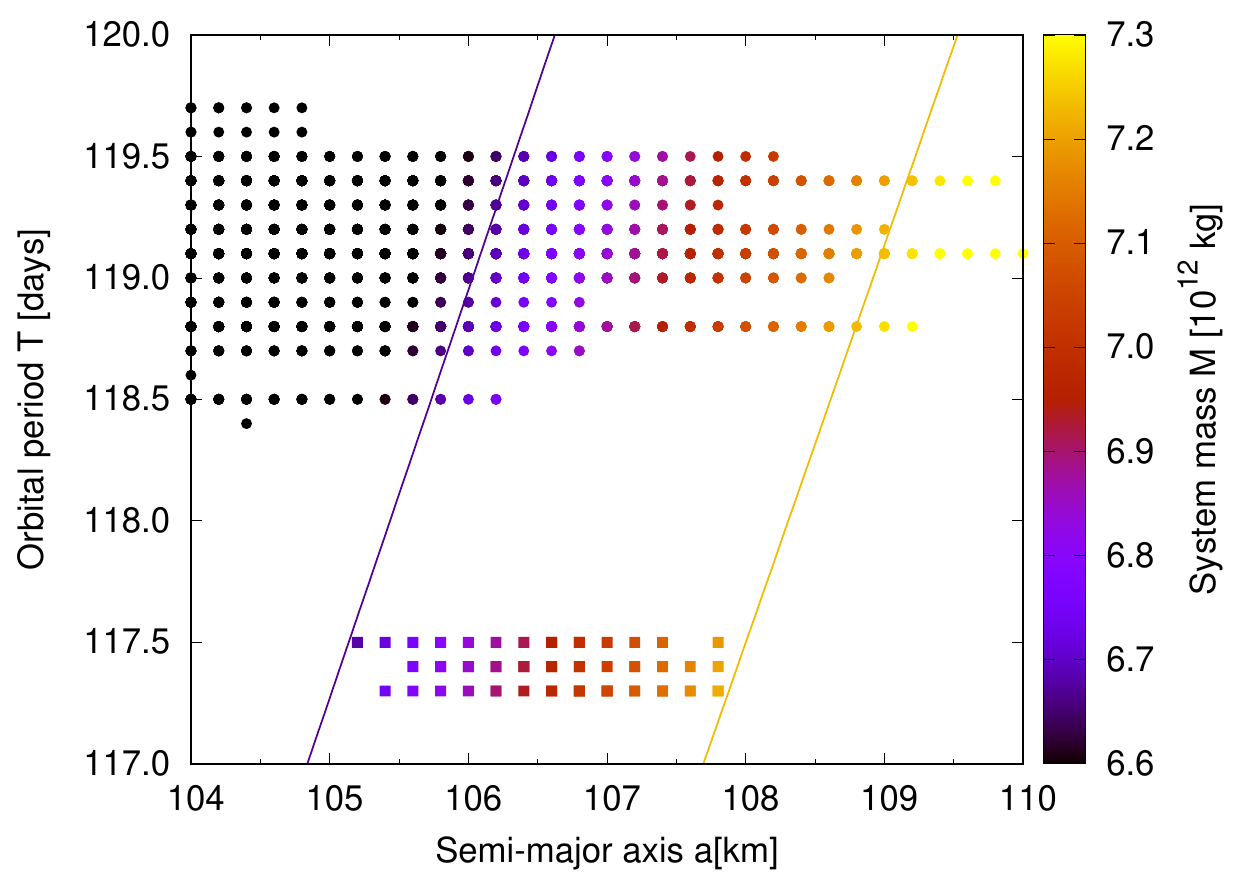}
  \caption{Orbital periods $T$ and semi-major $a$ axes of possible orbit solutions to either visits A--5 (squared symbols) or visits 13--22+24 (circular symbols). The system mass is colour-coded. The lines show the relationship $T(a,M)$ given by Kepler's third law for $M_\mathrm{min}=6.67\times10^{12}$ kg (violet) and $M_\mathrm{max}=7.23\times10^{12}$ kg (yellow).} 
\label{fig:mass_compatible_solutions}
\end{figure}

We suspected that the clustering around masses embracing the value of solution II.c was due to our discretisation of the parameter space and hence ran the search for mass-compatible solutions again with finer stepsizes for some of the parameters: 104$\leq a$/km$<$115 (stepsize 0.2\,km), 0.40$\leq e<$0.51 (stepsize 0.01), 117$\leq T$/days$<$123 (stepsize 0.1 days), while keeping the earlier stepsizes for $\alpha_0$ and $t_{\rm per}$. The resulting possible solutions in $a$-$e$-space are shown in Fig.~\ref{fig:mass_compatible_solutions}.
Solutions matching the pre-perihelion data all have periods 117.3 days $<T_\mathrm{pre}<$ 117.5 days. The narrow range is dictated by the long, almost five year time interval between visits B and 1, and is a consequence of our assumption that the orbit did not change during this period of presumed inactivity. The period $T_\mathrm{pre}$ is incompatible with data from visits 13-24, also with the finer parameter grid. For the post-activity epochs we find orbital periods in the range 118.5 days $<T_\mathrm{post}<$ 119.5 days that are mass-compatible with the pre-perihelion solutions. The allowable mass-range for the post-perihelion solutions is larger than for the pre-perihelion solutions, hence we consider orbits that reproduce post-perihelion data but do not have a mass-equivalent orbit reproducing pre-perihelion data as improbable. The derived system mass is hence in the range 6.67$\times$10$^{12}$ kg $<M<$ 7.23$\times$10$^{12}$ kg, while the orbital period likely increased by (1.0 -- 2.2) days. To ensure mass conservation, the semi-major axis must have increased accordingly by (0.6 -- 1.4) km. 

The range of possible eccentricities remained almost unchanged from 0.41(pre) or 0.42(post) to 0.51. The time of the fall-2016 periapsis $t_\mathrm{per}$ changed from September-16 to the interval between September-07 and September-14, where a stronger shift in periapsis is correlated with a stronger shift in orbital period, such that longer final orbital periods require earlier dates of periapsis to reproduce the data. The range of possible orientations of the lines of apsides, $\alpha_0$ remains largely unchanged (176$^\circ$ -- 184$^\circ$ pre, 175$^\circ$ -- 182$^\circ$ post). 

\begin{figure*}
  \includegraphics[width=\textwidth]{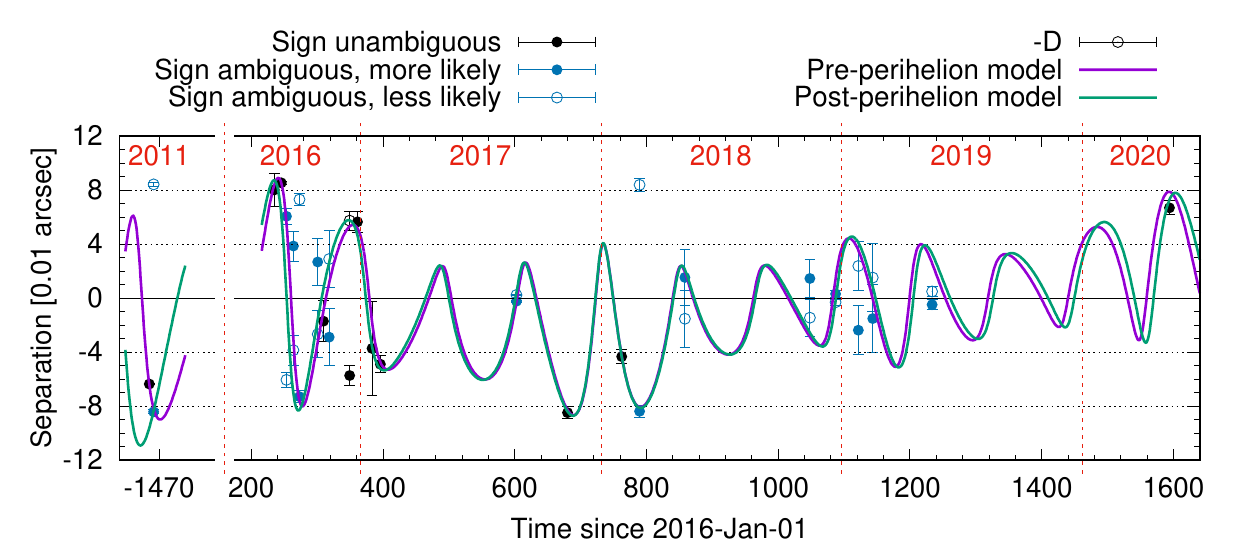}
  \caption{Measured distances including their sign and possible orbit fits. The violet curve shows a representative orbit solution that was required to at least reproduce visits A--5 ($a$=106.5\,km, $e$=0.46, $t_\mathrm{p}$=25.83\,d,  $T$=117.4\,d, $\alpha_0$=0.0\degr, $M_\mathrm{S}$=6.94$\times$10$^{12}$\,kg). The green curve shows a representative solution that was required to reproduce visits 13--22 and 24 ($a$=107.5\,km, $e$=0.46, $t_\mathrm{p}$=19.00\,d,  $T$=119.0\,d, $\alpha_0$=0.0\degr, $M_\mathrm{S}$=6.95$\times$10$^{12}$\,kg). The perihelion passage took place between visits 7 and 8 on DOY=314.
}
\label{fig:orbit_solutions}
\end{figure*}

Representative solutions for the pre- and post-perihelion orbit solutions are shown in Figs.~\ref{fig:orbit_solutions} and \ref{fig:plane_view}. The key difference between the two solutions is whether or not they reproduce visit 24 data. The difference is small but significant given the high spatial resolution and deep sensitivity of this measurement. We note that the post-perihelion data from the active phase (visits 8--12) are also better fitted by the post-perihelion solution, although both solutions reproduce these measurements within their error bars.

\section{Discussion}
\label{sec:discussion}

\subsection{Visit 23}
\label{sec:V23}
Our adopted best-fitting solution (Fig.~\ref{fig:orbit_solutions}) does not reproduce the measured distance during visit 23 within its error bars. Moreover, none of the solutions shown in Table~\ref{tab:solutions_overview} match this particular data point. The orbital elements given in Table~\ref{tab:solutions_overview} are those of a system where one component (of reduced mass $\mu$, see below) orbits about a central mass corresponding to the system mass $M_\mathrm{S}$. The length of the radius vector in this system corresponds to the true objects' mutual distance. Hence the typical 3d component separation is of order 100\,km (Fig.~\ref{fig:plane_view}). At geocentric distances beyond 3.2\,AU (visits 13, 18--20, 23) this is comparable to the linear size of a native WFC3 pixel (93\,km at 3.2\,AU), such that projected distances are typically sub-pixel-sized. The error bars covering the full range of a native WFC3 pixel for visits 18, 19, 21, and 22, imply that the distance is essentially unconstrained during these visits. Visits 13, 20, and 23 have distances consistent with zero but very small error bars. Given their comparable observation geometry, it is possible that the distances from these visits have error bars similar to those of their neighbours and that for some reason these error bars are underestimated by our procedure. While our solution is compatible with measurements for visits 13 and 20, it is outside the error bars for visit 23. A possible reason could be that the fainter component was not detected by the PSF fitting due to a strong instantaneous difference in component brightness. This might be related to an unfavourable combination of relative rotation phases (Sec.~\ref{sec:component_sizes}) or to a mutual event with a potential third component.

\subsection{Component sizes}
\label{sec:component_sizes}
\begin{figure}
  \includegraphics[width=\columnwidth]{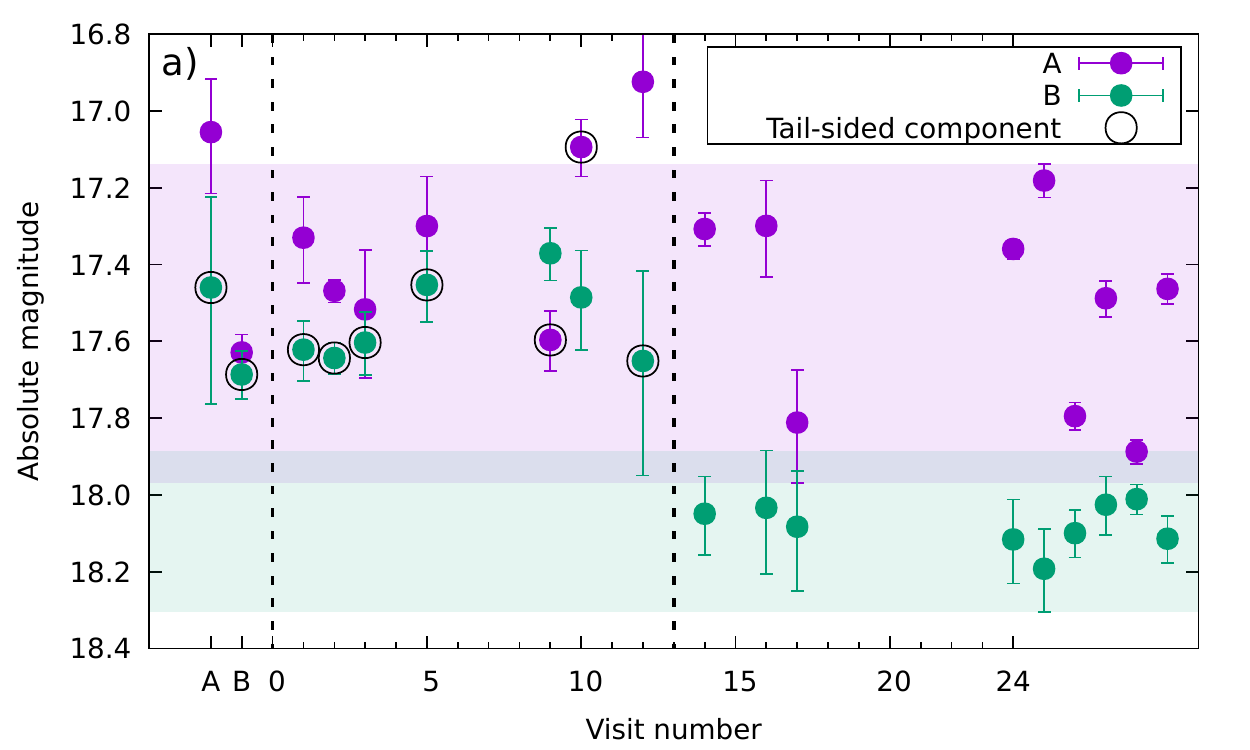}
  \includegraphics[width=\columnwidth]{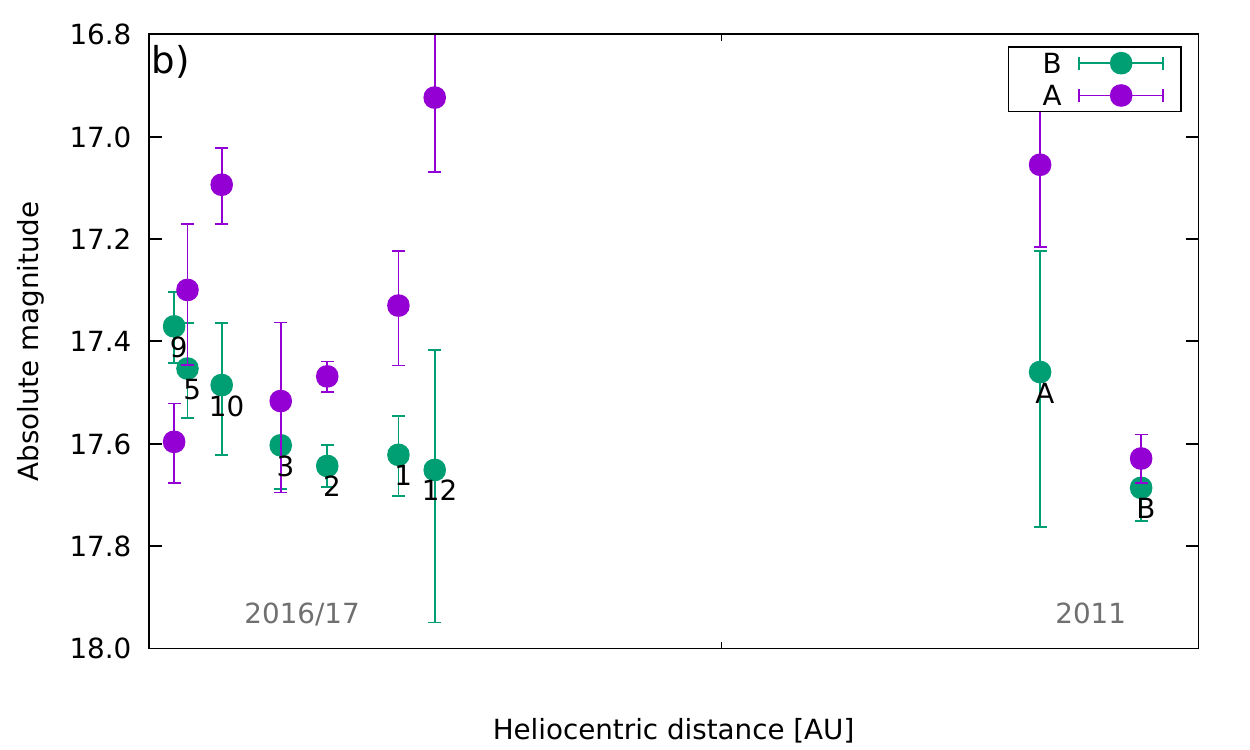} 
  \caption{Photometry of individual components of the 288P system. The components are identified based on their relative positions according to the orbit models shown in Fig.~\ref{fig:orbit_solutions}. The component that is brighter in the inactive state is labelled 'A', the fainter one 'B'. Panel a) shows the absolute magnitudes of the individual components for those visits where $D>$0.04\arcsec. Data from visit 24 are plotted separately for each HST orbit. The areas shaded in violet and green mark the brightness ranges compatible with visits 14--17 + 24, assuming that these are the minimum ranges of their rotational lightcurves. Black circles indicate the tail-sided component for any given time, hence the non-circled components were located in the direction of the Sun. Vertical dashed lines separate 2011 from 2016 visits, and the active from the inactive phase. Panel b) shows the absolute magnitudes during the active phases as functions of heliocentric distance $r_\mathrm{h}$. Numbers label the visit. Within the 2016/17 data set, component B shows a systematic dependency between magnitude and $r_\mathrm{h}$, while A does not.
}
\label{fig:active_component}
\end{figure}

With the additional knowledge of their mutual orbit we could then identify the components as physical objects, rather than merely from their brightness or relative positions as we did in Fig.~\ref{fig:psf_fitting_results}. This allowed us to study the photometry of the individual components in more detail, and to derive constraints on their sizes, shapes, and activity. Fig.~\ref{fig:active_component}a) shows the absolute magnitudes of the individual components for all epochs with $D$>0.04\arcsec.
We first studied the sizes and elongations of the individual components from their absolute magnitudes during visits 14, 16, 17, and 24, when the system was presumably free of dust.
We find 17.1$<H_\mathrm{A}<$18.0 and 17.9$<H_\mathrm{B}<$18.3, corresponding to cross-sections of 1.3$<C_\mathrm{A}$/km$^2<$2.8 and 0.9$<C_\mathrm{B}$/km$^2<$1.4 for a family-average albedo of $p_\mathrm{V}$=0.07 \citep{novakovic-hsieh2012}. The main uncertainty results from the uncertainty of the albedo ($\Delta p_\mathrm{V} = \pm$0.03) and amounts to 40\%. For prolate spheroids with semi-axis lengths $a<b$ rotating about one of their short axes and viewed equator on, this corresponds to $a_\mathrm{A}\leq$0.6\,km, $b_\mathrm{A}\geq$1.4\,km, $a_\mathrm{B}\leq$0.5\,km, and $b_\mathrm{B}\geq$0.8\,km, with typical uncertainties of 20\%.
The numbers given correspond to volumes of $V_\mathrm{A}$=2.4\,km$^3$ and $V_\mathrm{B}$=1.0\,km$^3$, with an uncertainty of 60\% induced by the uncertainty of the albedo, and a further, hard to quantify uncertainty arising from the possibly not fully sampled lightcurve ranges and the unknown details of the shapes.

With these volumes and the system mass range derived from the orbit solutions shown in Fig.~\ref{fig:mass_compatible_solutions}, we calculated a system bulk density of 2000 kg m$^{-3}$. The main uncertainty of this density stems from the volume and is at least 60\%. Typical C-type densities (albeit measured only for objects having diameters $>$100\,km) range between 1000 and 2000\,kg m$^{-3}$ \citep{hanus-viikinkoski2017}. Our derived value is consistent with this range within its uncertainties.

\subsection{Active component}
\label{sec:active_component}
We tried to gain some insight on which component(s) may have been active during 2016/17 from their brightness.
Fig.~\ref{fig:active_component}a) shows that during the active phase (visis A -- 12), the magnitude range of component A was elevated by about 0.3 mag compared to the inactive phase, although most individual measurements are compatible with the brightness range derived from the inactive phase.
Component B was systematically brighter by about 0.5\,mag, with all measurements from the active phase being incompatible with the range derived from the inactive phase. This may suggest that more dust may have been present in the central pixel of B than of A, which in turn may indicate that the dust was emerging from B.

However, it is also possible that radiation pressure drove the dust towards the component further from the Sun, enhancing selectively the brightness of the tail-sided component, which we identify by black circles. Indeed, B was in the tail during most epochs. However, during visits 9 and 10, B was closer to the Sun than A but still showed enhanced brightness. We tentatively conclude that enhanced brightness due to dust in the central pixel is more likely associated with component B.

We note that the measured brightness of B during the 2016/17 active phase anti-correlates with heliocentric distance (Fig.~\ref{fig:active_component}b), as would be expected for activity driven by a thermal process. The systematic magnitude-$r_\mathrm{h}$ relation cannot be extended between data sets from different perihelion passages, but our data do not allow us to conclude if this means that the relation seen within the 2016/17 data set is coincidence (as a result of the randomly sampled rotational lightcurve), or if the dust production rate differed between the two apparitions. Regardless of the $r_\mathrm{h}$-dependence, we find that also during 2011, it was mainly component B that showed enhanced brightness. 

\subsection{Energy and angular momentum considerations}
The total energy, $E$, and angular momentum, $L$, of a Keplerian binary system are
\begin{equation}
  E=\frac{2\pi^2 a^2 e^2 \mu}{(1-e^2) T^2}
\end{equation}
and
\begin{equation}
  L=\frac{2\pi a^2 \sqrt{1-e^2} \mu}{T^2},
\end{equation}
where $\mu=(M_\mathrm{A}M_\mathrm{B})/M_\mathrm{S} = M_\mathrm{A}/M_\mathrm{S} (1+f)$ is the reduced mass of the system, and $f=M_\mathrm{B}/M_\mathrm{A} \leq 1$ the component mass ratio.
From the estimated volumes and assuming similar densities and conservative volume uncertainties of 60\% (Sec.~\ref{sec:photometry}), we estimate 0.1$<f<$1, or 0.08$M_\mathrm{S}<\mu<$0.25$M_\mathrm{S}$. 

The changes in orbital period and semi-major axis suggested by Fig.~\ref{fig:mass_compatible_solutions} correspond to a change of specific angular momentum of 3.4$\times$ 10$^{-6}$ m$^2$s$^{-1} < \Delta L/\mu <$6.6$\times$ 10$^{-6}$ m$^2$s$^{-1}$. Including the uncertainty of the reduced mass, we find a possible angular momentum change in the range of 2$\times$ 10$^{6}$ kg m$^2$s$^{-1} < \Delta L <$12$\times$ 10$^{6}$ kg m$^2$s$^{-1}$.

To estimate if such a change in angular momentum can be achieved by outgassing forces during a single perihelion passage, we used the relation $\Delta L = k Q_\mathrm{gas} v_\mathrm{th} a \Delta T$ derived in \citet{agarwal-jewitt2017}, where 0$<k<$1 describes the degree of focussing of the gas stream (k=1 corresponding to a parallel stream), $v_\mathrm{th}$=500\,m s$^{-1}$ is the thermal speed of the gas, $\Delta T$ is the duration of the activity, and $Q_\mathrm{gas}$ the gas production rate. We assumed $k$=0.1, $Q_\mathrm{gas}$ = $Q_\mathrm{dust}$ = 5\,kg\,s$^{-1}$, and $\Delta T$ = 200\,days \citep{hsieh-ishiguro2018}. Using $a$=120\,km, we obtain $\Delta L$ = 520 $\times$ 10$^{12}$\,kg\,m$^2$\,s$^{-1}$. Given the uncertainties of especially $k$ and $Q_\mathrm{gas}$, the uncertainty of $\Delta L$ is easily a factor 10. This theoretically estimated $\Delta L$ is much larger than the $\Delta L$ corresponding to Fig.~\ref{fig:mass_compatible_solutions}, such that an outgassing-induced orbit change during perihelion seems possible from the angular momentum point of view.

We also estimate the possible change in system energy due to outgassing. The change of linear momentum of the active component, $\Delta p$, is given by the momentum carried by the gas: $\Delta p$ = $k v_\mathrm{gas} Q_\mathrm{gas} \Delta T$ = 4 $\times$ 10$^{9}$ kg\,m\,s$^{-1}$. With a component mass of order 2$\times$10$^{12}$\,kg, the resulting increase in velocity of the active component is $\Delta v_\mathrm{n}$= 2$\times$10$^{-3}$ m\,s$^{-1}$, and its increase in kinetic energy $\Delta E_\mathrm{n}$= 4$\times$10$^6$ J, or $\Delta E_\mathrm{n} / \mu$=4$\times$10$^{-6}$ m$^{2}$\,s$^{-2}$. This value probably has an uncertainty of at least a factor 100.
  The change of specific system energy corresponding to the orbits shown in Fig.~\ref{fig:mass_compatible_solutions} ranges between 300 and 700 m$^{2}$\,s$^{-2}$, which is again small compared to the estimated change above.
  Hence, even a much weaker or less focussed outgassing than assumed above would be sufficient to explain the suspected orbit change.

\subsection{Comparison with earlier results}
\begin{figure}
  \includegraphics[width=\columnwidth]{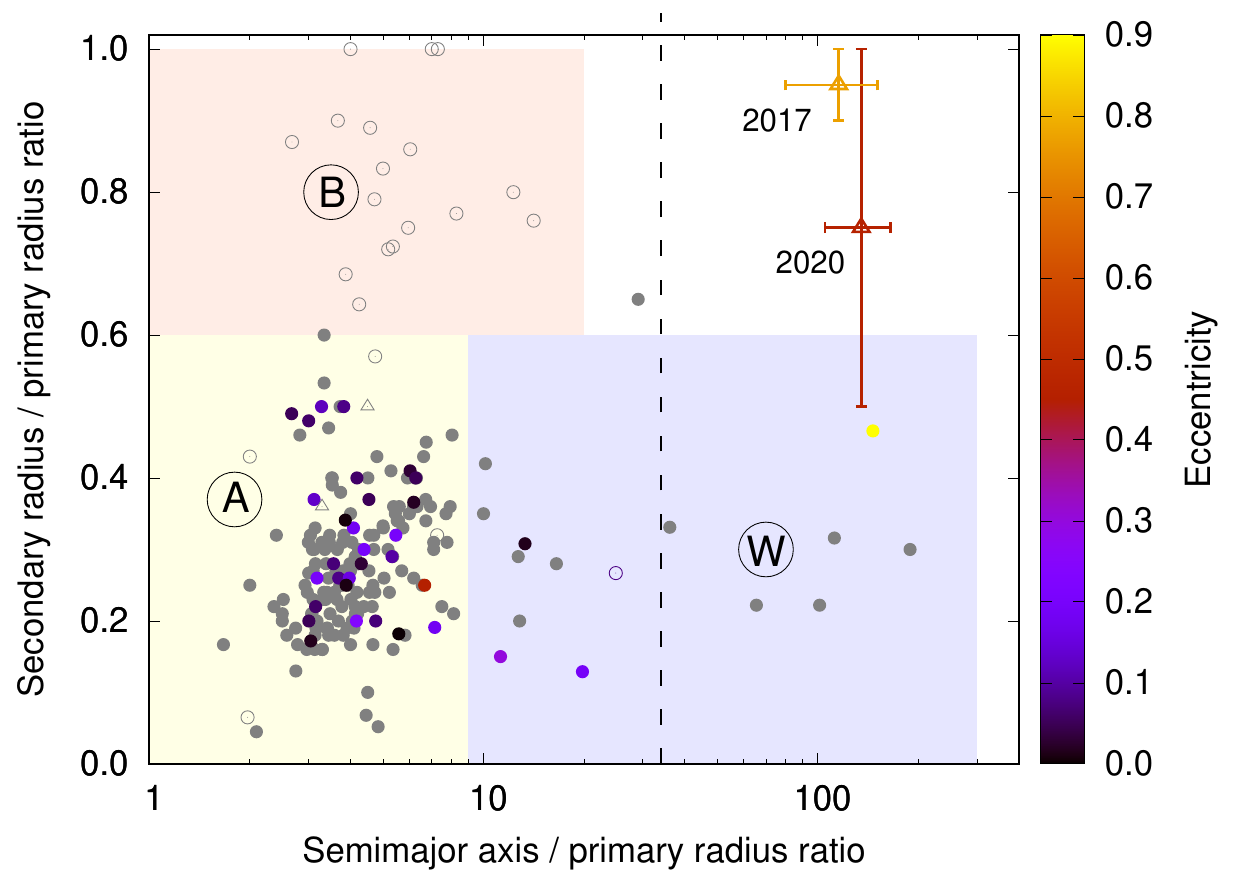}
  \caption{Comparison of 288P with the known binary systems from \citet{johnston2019}. Filled circles represent binaries with a primary rotation period $T_\mathrm{p}<$5h, open circles represent those with $T_\mathrm{p}>$5h, while triangles represent systems with unknown $T_\mathrm{p}$. Grey colour indicates unknown eccentricity.  Letters A, B, and W refer to the three main binary populations defined in \citet{pravec-harris2007}, where group A has small size ratios and fast-rotating primaries, B has doubly synchronous systems with a high size ratio, and W are wide, asynchronous binaries. The vertical dashed line indicates the 34\,$r_\mathrm{p}$ limit out to which binaries can form directly from rotational fission. The error bars show the solutions obtained in this work and in \citet{agarwal-jewitt2017} from visit 1-12 data for comparison. As component radii we used those of volume-equivalent spheres with a 20\% uncertainty. This plot has been updated from Fig.~3 in \citet{agarwal-jewitt2017}.
}
\label{fig:binary_population}
\end{figure}

The distances measured using PSF-fitting (Fig.~\ref{fig:psf_fitting_results}) are systematically smaller than those we obtained by visual examination in \citet{agarwal-jewitt2017}, although the majority of the measurements are consistent within their error bars. The orbit solution shown in Fig.~\ref{fig:orbit_solutions} was not found in our earlier analysis, and neither were the high-frequency solutions. All solutions that seemed to match the data from visits 1-12 during the earlier analysis are incompatible with the data from visits 13--24, that were not available at the time of publication of \citet{agarwal-jewitt2017}. However, we confirm the general ranges of $T>$100\,days and $a\sim$100\,km, and the mass range (1.3$\times$10$^{12}$kg$<M_\mathrm{S}<$1.1$\times 10^{13}$kg previously). We also confirm a considerable, although lower than previously derived eccentricity (now $e$=0.46). Fig.~\ref{fig:binary_population} shows 288P in the context of the known binary population from \citet{johnston2019}. 288P remains the only known binary combining similarly sized components with a wide orbital separation, and has one of the highest eccentricities, although for many systems (especially those with wide separation or high mass ratio) the eccentricity remains unmeasured. The mass ratio we derive from the new data (0.5 -- 1.0) is lower than the value of 0.9 -- 1.0 we found from the earlier data obtained while 288P was active. The earlier measured brightness thus included a certain amount of dust near at least one of the components, which led us to overestimate its brightness, size, and therefore mass (Fig.~\ref{fig:active_component}a).

\subsection{Formation and evolution of the 288P system}
\label{sec:evolution}
One of the key questions arising from our results is whether there is a causal connection between the activity and the unusual mutual orbit that cannot have formed directly from rotational fission \citep{jacobson-scheeres2011}. 
Starting from the assumption that 288P is a collisional fragment from a precursor that disrupted during an event, E0, $\sim$10$^7$ years ago \citep{novakovic-hsieh2012}, \citet{agarwal-jewitt2017} outlined several conceivable formation scenarios that we rephrase here as follows. One possibility (A) is that a single fragment from E0 was spun up by the YORP effect and recently disintegrated into two similarly sized fragments, that are now both active and drive the orbit evolution through sublimation torques. Alternatively (B), a single fragment or a contact binary remaining from E0 split and slowly developed into an inactive wide binary by radiation torques, and was subsequently activated by an impact (BI) or fast rotation (BR). In scenario C, the system emerged already from E0 as a (wide) Escaping Ejecta Binary \citep[EEB, ][]{durda-bottke2004a}, and was activated recently and independently of the binary formation.
Finally (D), the system can have already been a (typical) close binary when one component was impacted (DI) or rotationally disrupted (DR), with the resulting activity driving the orbit evolution to a wide binary. 

We consider scenario C as unlikely because in this case the alignment of the mutual and heliocentric orbital planes would be coincidental, as EEBs are not expected to have any preferred orientation of their orbital planes. In addition, the stability of wide binary systems over timescales of millions of years has not been proven. Scenario A is disfavoured by our suspicion that the activity was concentrated on component B. If true, the activation of B was likely independent of the formation of the binary system.
We currently cannot discriminate between scenarios B and D. We see slight evidence for orbit widening during the 2016/17 perihelion, but further modelling of the orbital evolution of a binary system with one active component is required to understand if the past activity can have driven the formation of a wide binary. We also cannot currently discriminate between a collisional or rotational activation scenario.

\section{Summary and conclusions}
\label{sec:conclusion}
We have analysed HST images of the 288P system while apparently inactive from 11 epochs between 2017 and 2019, and re-analysed data from 12 epochs during the 2016/17 perihelion passage and from 2 epochs during the 2011 perihelion. Our key findings are
\begin{itemize}

\item The combined absolute V-band magnitude of the inactive system ranges between 16.75 and 17.25, consistent with earlier findings \citep{agarwal-jewitt2016,hsieh-ishiguro2018,waniak_288P_dps}. The data can be fitted with a phase function having $G$ = (0.1 $\pm$ 0.04). 

\item We confirm the alignment of the mutual and the heliocentric orbital planes \citep{agarwal-jewitt2017}.

\item We confirm \citet{agarwal-jewitt2017} that 288P is unusual among the known binary asteroids due to its combination of high mass ratio ($r_\mathrm{p}/r_\mathrm{s} >$ 0.5) and wide separation $a/r_\mathrm{p} >$100, high eccentricity (0.4$<e<$0.5), and the sublimation-driven activity.

\item The absolute magnitudes of the individual components are in the ranges 17.1$<H_\mathrm{A}<$18.0 and 17.9$<H_\mathrm{B}<$18.3, corresponding to cross-sections of 1.3$<C_\mathrm{A}$/km$^2<$2.8 and 0.9$<C_\mathrm{B}$/km$^2<$1.4 for $p_\mathrm{V}$=0.07, with an albedo-induced uncertainty of 40\%.
Approximating both components by prolate spheroids with semi-axis lengths $a<b$, the derived cross-section ranges translate to $a_\mathrm{A}\leq$0.6\,km, $b_\mathrm{A}\geq$1.4\,km, $a_\mathrm{B}\leq$0.5\,km, and $b_\mathrm{B}\geq$0.8\,km, with typical uncertainties of 20\%. Such spheroids would have volumes $V_\mathrm{A}$=2.4\,km$^3$ and $V_\mathrm{B}$=1.0\,km$^3$, with uncertainties of at least 60\%.

\item Data from the inactive phase do not show any indication of dust, but we cannot exclude the presence of a low-level coma comprising up to 10-20\% of the combined component cross-section inside a 1300\,km radius.

\item Component photometry from the active phase indicates a selective brightening of component B by about $\Delta H_\mathrm{B}$=0.5\,mag in the central PSF, and a possible correlation of $H_\mathrm{B}$ with heliocentric distance during this phase. A potential brightening of component A by $\Delta H_\mathrm{A}$=0.3\,mag is less certain, as it may be a consequence of the incompletely sampled lightcurve. This may indicate that dust activity was concentrated on component B, which is the smaller and less elongated component.

\item To reproduce all data points except one (visit 23), we need to introduce a change of orbital period by 1-2 days around the time of the 2016 perihelion passage from (117.3 -- 117.5) days pre- to (118.5 -- 119.5) days post-perihelion. Assuming that the system mass (in the range (6.67 -- 7.23)$\times$10$^{12}$ kg) remained constant, this requires a corresponding change of semi-major axis of (0.6 -- 1.4) km from (105 -- 108) km pre- to (106 -- 109) km post-perihelion. The system eccentricity ranged between 0.41 and 0.51, but the magnitude of a possible change could not be inferred from our data and model. The derived system mass and volume imply a bulk density of 2000\,kg\,m$^{-3}$ with a volume-driven uncertainty of at least 60\%.

\item We favour a scenario of formation and evolution where the binary system formed by rotational splitting following YORP spin-up, and where the activation happened independently of the splitting. We currently cannot decide whether evolution to a wide binary was driven by radiation (BYORP) or sublimation forces.
  
\end{itemize}
The data presented here show the 288P system resolved and free of dust, which allowed us to study the properties of the nuclei and to put new constraints on models of the formation of this system. We found that the mutual orbit can be meaningfully studied when the expected maximum projected separation exceeds the linear size of a WFC3 pixel and that separate component photometry is possible when the actual separation is larger than this limit. 
Open questions remain concerning the processes triggering the activity and driving the orbit evolution. We expect that future resolved observations of the system will put additional constraints on a potential outgassing-induced orbit change during the 2021 perihelion passage, and will help to constrain the rotation states of the components.

\begin{acknowledgements}
We thank Jay Anderson (STScI) for sharing his knowledge on the Point Spread Function of HST/WFC3. We also thank Alison Vick and William Januszewski (STScI) for their reliable support in preparing and scheduling the HST observations.

JA and YK were supported by the German Aerospace Centre (DLR) under Grant No. FKZ~50~OR~1703 and by the European Research Council (ERC) Starting Grant No. 757390 (CAstRA). JA was in part funded by the Volkswagen Foundation.

This work is based on observations made with the NASA/ESA \emph{Hubble Space Telescope}, obtained from the archive at the Space Telescope Science Institute (STScI). STScI is operated by the association of Universities for Research in Astronomy, Inc. under NASA contract NAS~5-26555. These observations are associated with GO programs 12597, 14790, 14864, 14884, 15328, 15481, and 16073.

This work made use of the software IRAF, distributed by the National Optical Astronomy Observatory, which is operated by the Association of Universities for Research in Astronomy (AURA) under a cooperative agreement with the National Science Foundation. 

This research has made use of NASA's Astrophysics Data System, of the JPL/Horizons ephemerides service, maintained by the JPL Solar System Dynamics group, and of data and services provided by the International Astronomical Union's Minor Planet Center. 
\end{acknowledgements}

\bibliographystyle{bibtex/aa} 
\bibliography{/home/agarwal/Latex/refs} 

\clearpage
\begin{appendix}
\include{input_appendix_single_frames}

\include{input_photometry}

\include{input_plane_view}

\end{appendix}

\end{document}

%% file: input_appendix_single_frames.tex
\renewcommand{\floatpagefraction}{0.9}
\onecolumn
\section{Observations and Best-fitting Models}
\begin{figure}[H]
\includegraphics[width=0.48\textwidth]{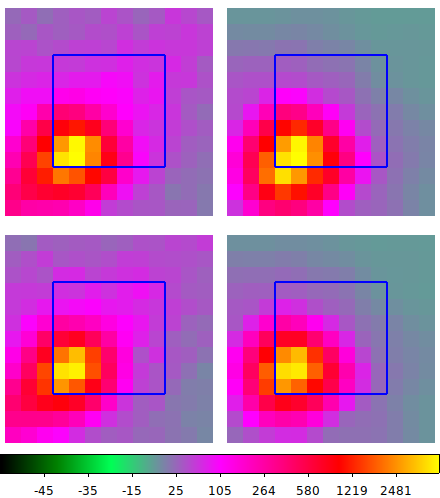}
\includegraphics[width=0.24\textwidth]{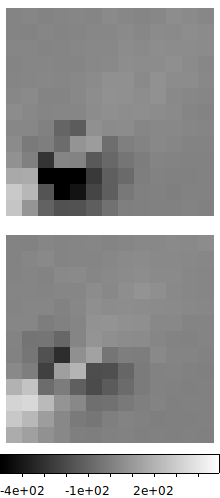}
\includegraphics[width=0.25\textwidth]{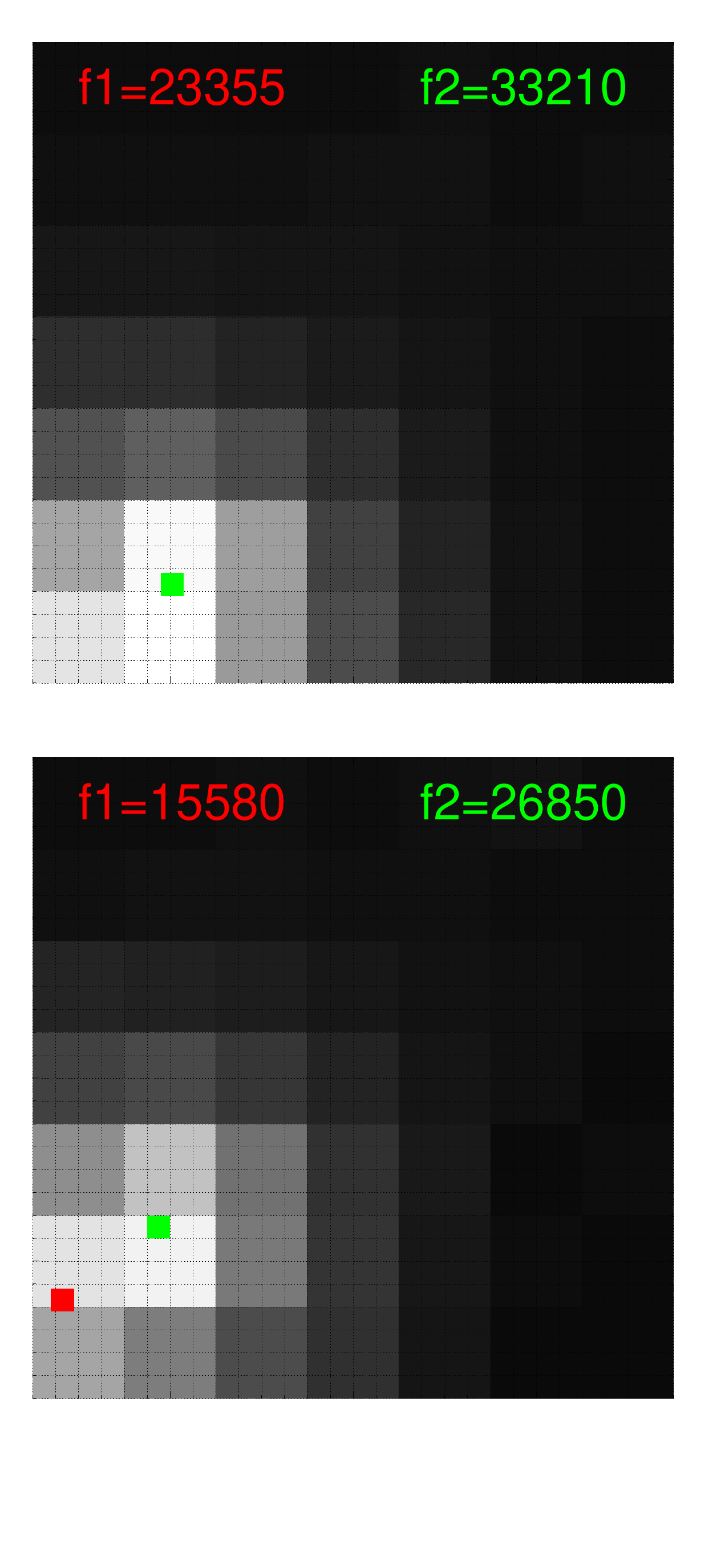}
\includegraphics[width=0.48\textwidth]{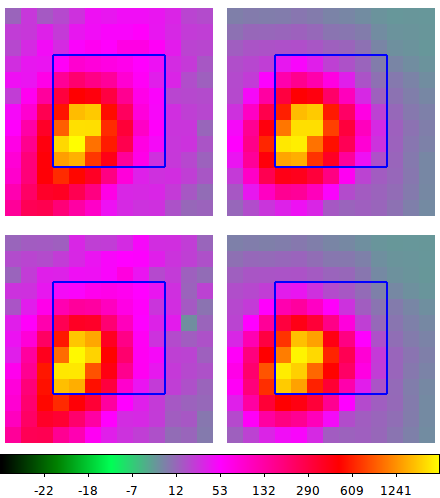}
\includegraphics[width=0.24\textwidth]{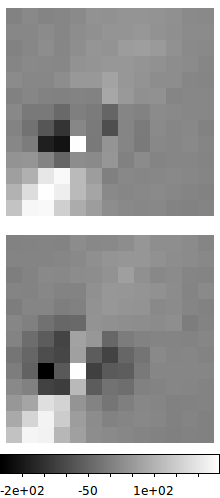}
\includegraphics[width=0.25\textwidth]{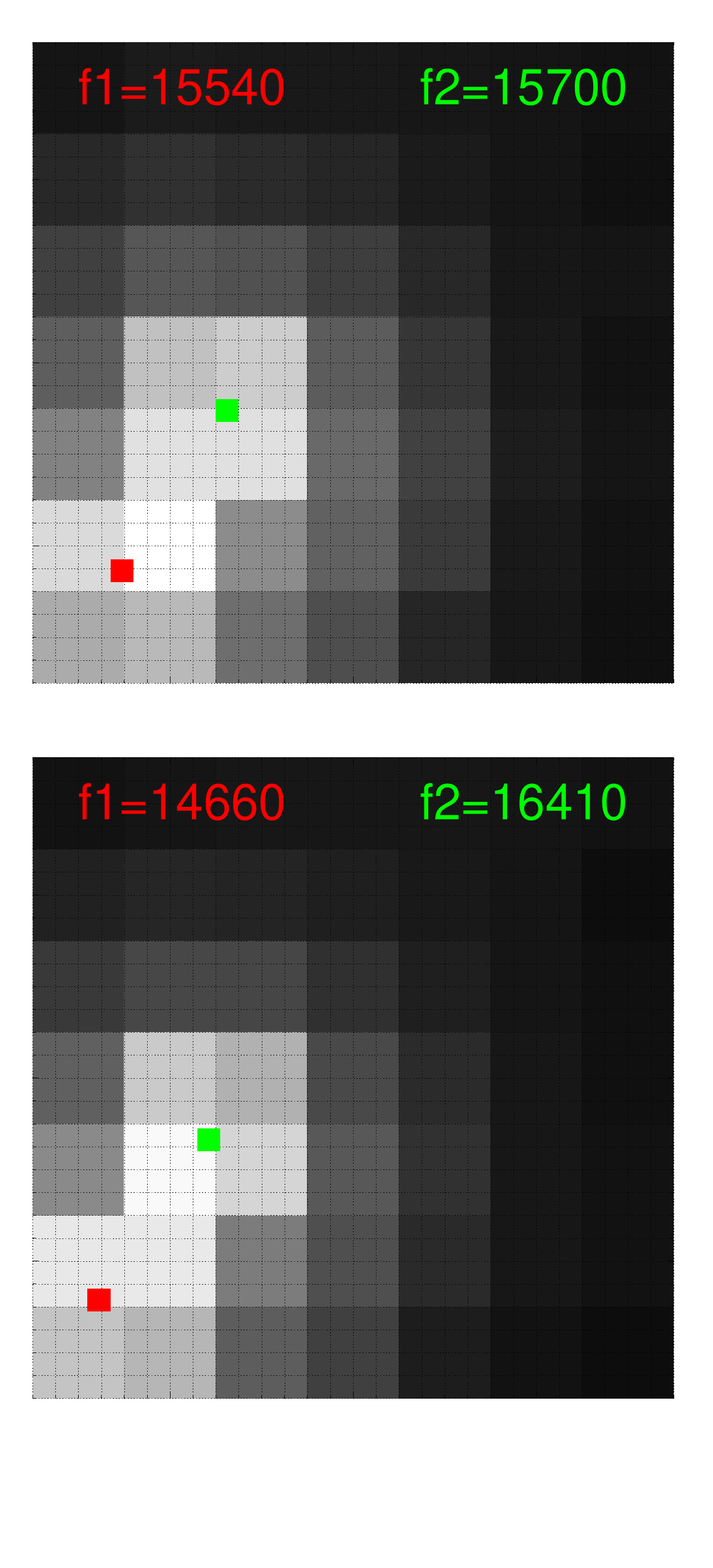}
\caption{{\bf Visits A (upper two rows, brightness scale factor $b$=50) and B (lower two rows, rightness scale factor $b$=25)}, dither stations 1-2 from top to bottom. respectively. 
{\it Left:} Observation (minimum stack of two images obtained at the same dither station, logarithmic brightness scale with range [-$b$, 100$b$]). The blue line marks the area that was included in calculating the sum of squared differences $S$. This area is off-centred in the direction opposite to the tail in order to avoid fitting the tail as best possible. {\it Center left:} Best fitting model at the same brightness scale. {\it Center right:} Difference between observation and model at a linear brightness scale, range [-10$b$, 10$b$]. {\it Right:} Central seven pixels of observation with the sub-pixel positions of the two components marked. The brighter component is marked green, the fainter component is red. The brightness scale is linear in the range [-10, 100$b$]. The quantities f1 and f2 are the total counts from the two components in the simulation.
}
\label{fig:obsmod_vA}
\end{figure}

\twocolumn
\begin{figure*}
\includegraphics[width=0.48\textwidth]{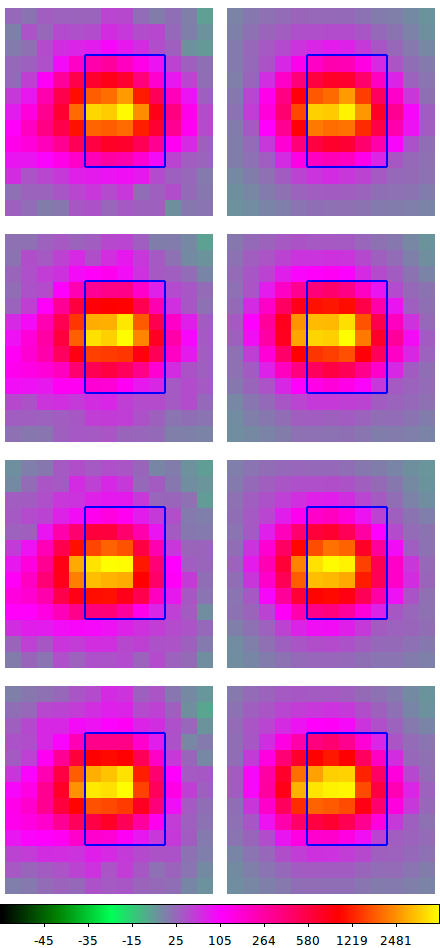}
\includegraphics[width=0.24\textwidth]{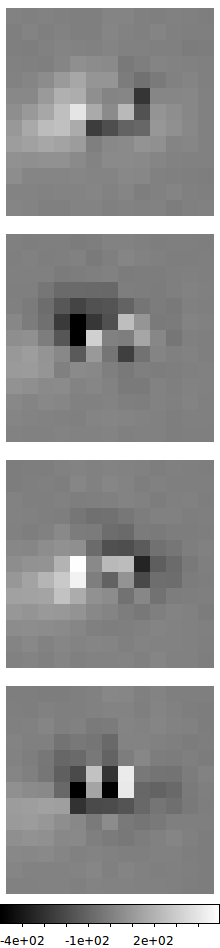}
\includegraphics[width=0.25\textwidth]{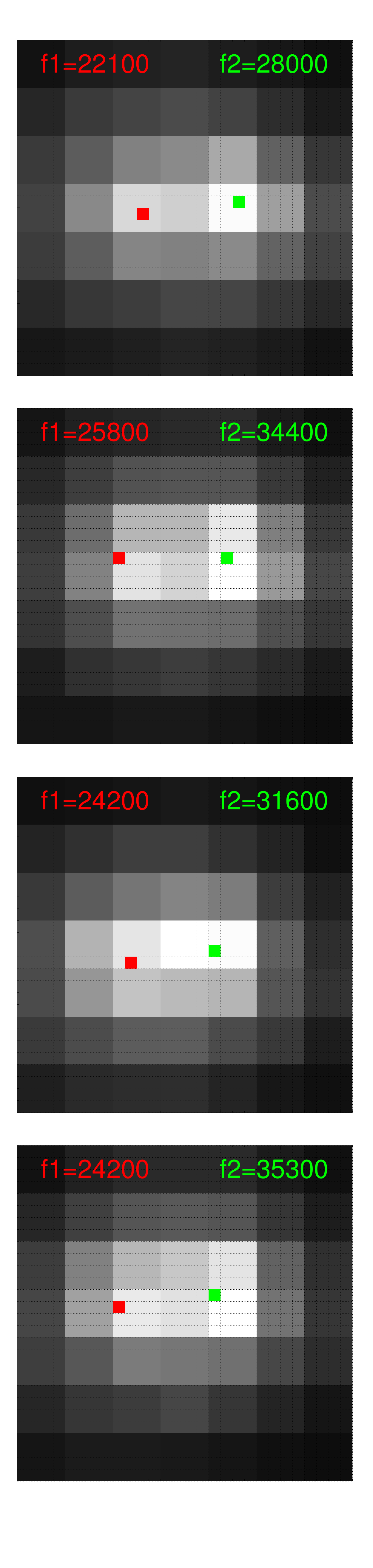}
\caption{{\bf Visit 1}, dither stations 1-4 from top to bottom. {\bf Brightness scale factor $b$=50}. The blue fitting region was reduced in size towards the tail-side to avoid fitting the tail instead of the nucleus. Otherwise same as Figure~\ref{fig:obsmod_vA}.
}
\label{fig:obsmod_v1}
\end{figure*}

\begin{figure*}
\includegraphics[width=0.48\textwidth]{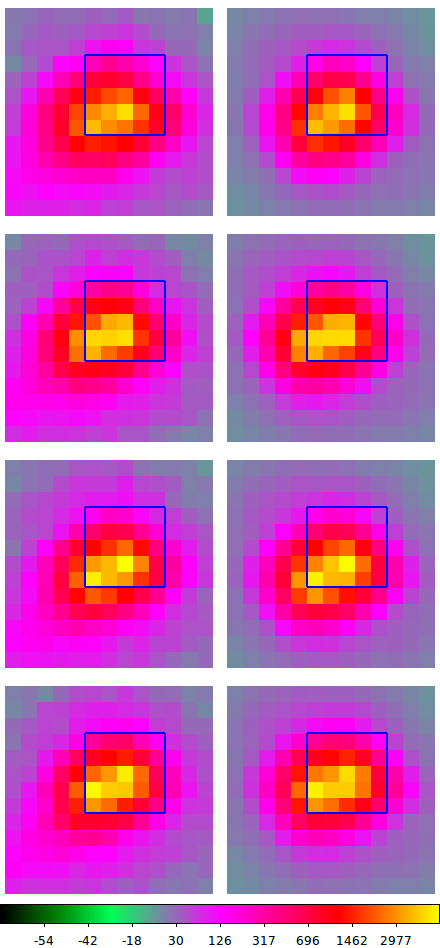}
\includegraphics[width=0.24\textwidth]{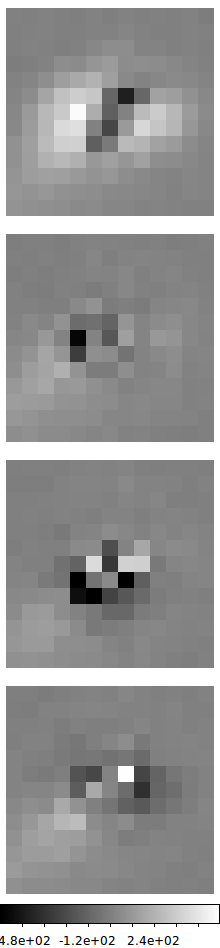}
\includegraphics[width=0.25\textwidth]{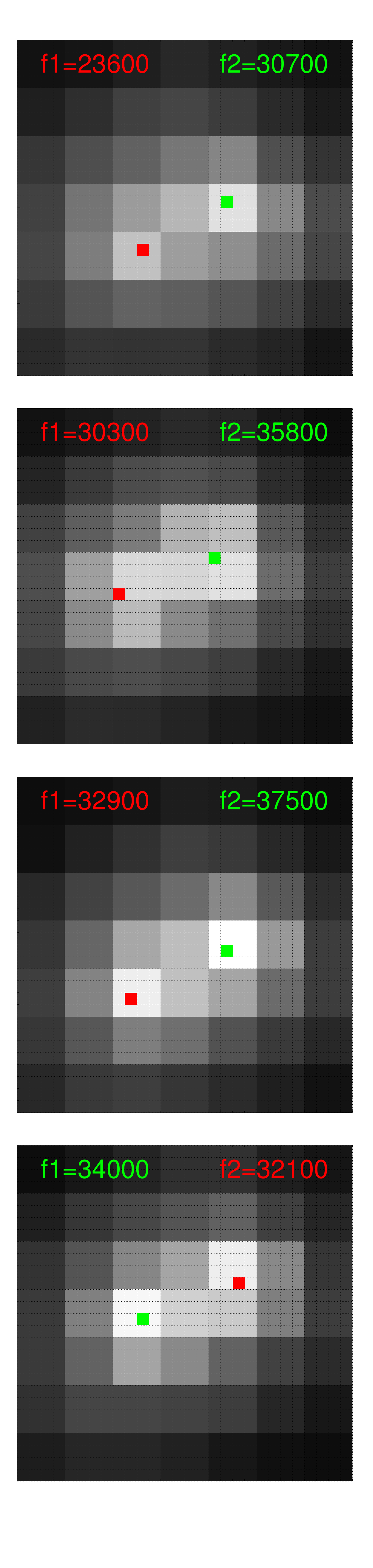}
\caption{Same as Figure~\ref{fig:obsmod_vA}, for {\bf Visit 2}, {\bf Brightness scale factor $b$=60}. While the measured distance is consistent at all dither stations, the flux ratio inverts between stations 3 and 4. This may reflect an intrinsic brightness variation induced by rotation of irregularly shaped components, but can also be due to image noise and/or thermal breathing of the PSF.
}
\label{fig:obsmod_v2}
\end{figure*}

\begin{figure*}
\includegraphics[width=0.48\textwidth]{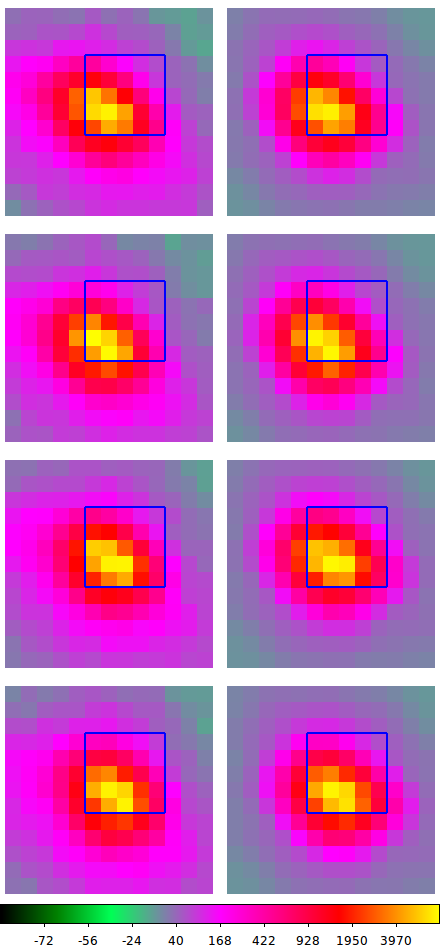}
\includegraphics[width=0.24\textwidth]{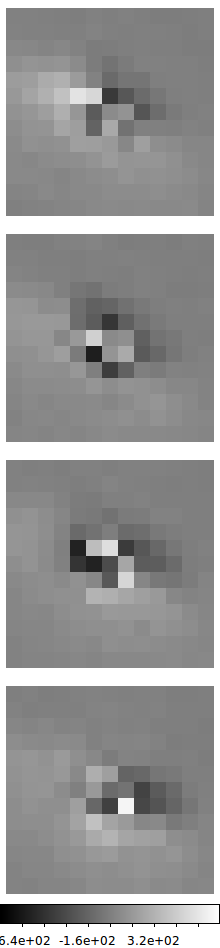}
\includegraphics[width=0.25\textwidth]{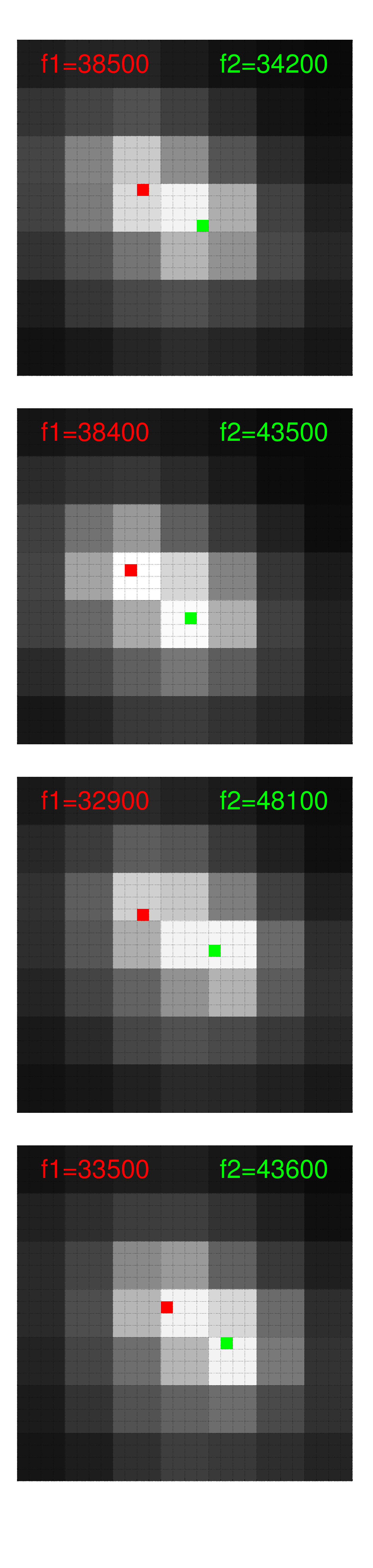}
\caption{Same as Figure~\ref{fig:obsmod_vA}, for {\bf Visit 3}, {\bf Brightness scale factor $b$=80}.
}
\label{fig:obsmod_v3}
\end{figure*}

\begin{figure*}
\includegraphics[width=0.48\textwidth]{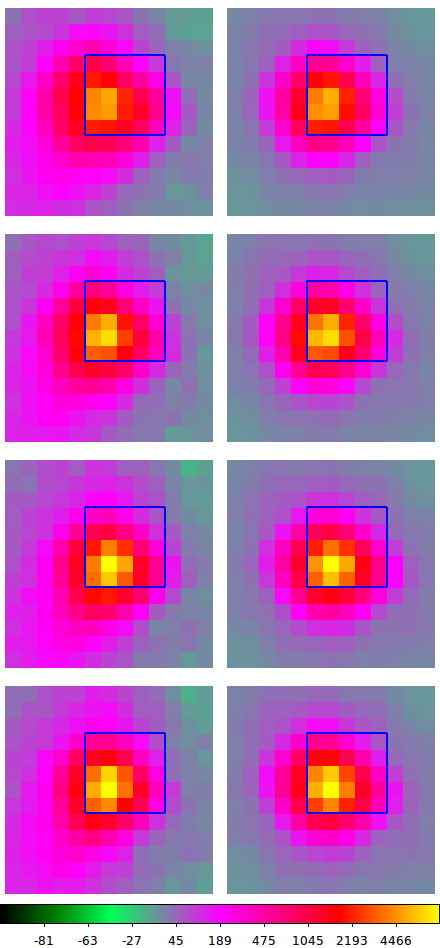}
\includegraphics[width=0.24\textwidth]{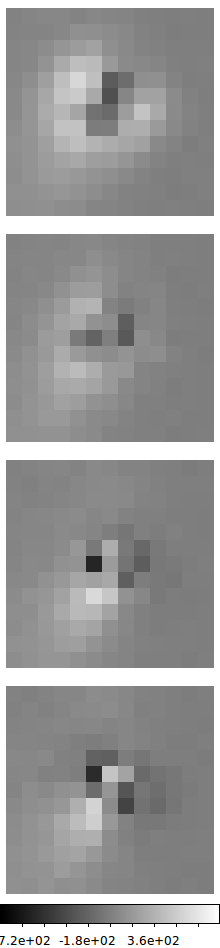}
\includegraphics[width=0.25\textwidth]{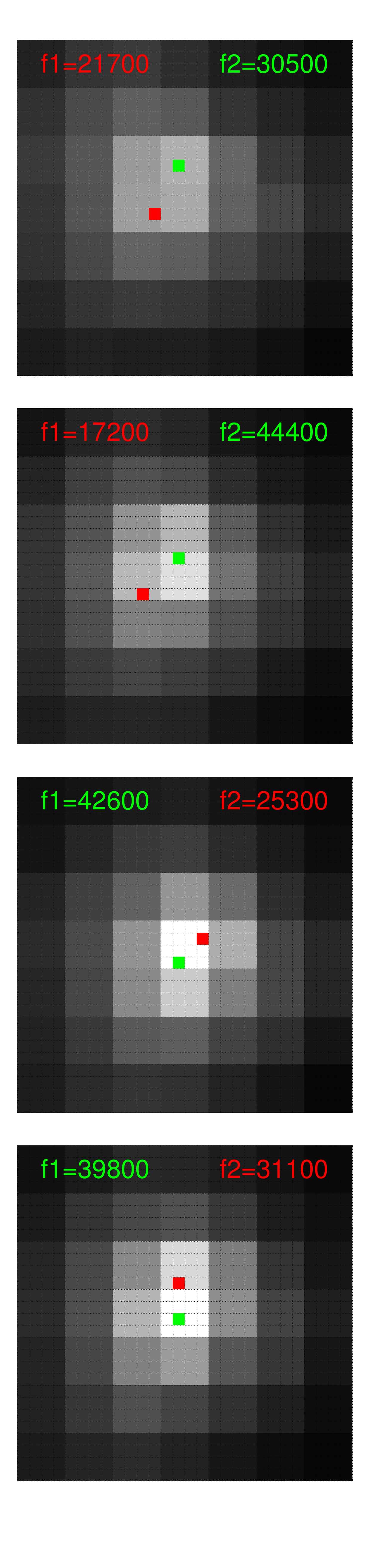}
\caption{Same as Figure~\ref{fig:obsmod_vA}, for {\bf Visit 4}, {\bf Brightness scale factor $b$=90}. The modelled brighness ratio of the components inverts between visits 2 and 3, reflecting either rotational variation, or the uncertainty of the fitting results due to image noise and thermal breathing of the PSF. The conservation of the total flux makes a rotational cause less likely, as it is unlikely that the apparent cross-section of one component increases at the same time and by a similar amount as the other decreases.
}
\label{fig:obsmod_v4}
\end{figure*}

\begin{figure*}
\includegraphics[width=0.48\textwidth]{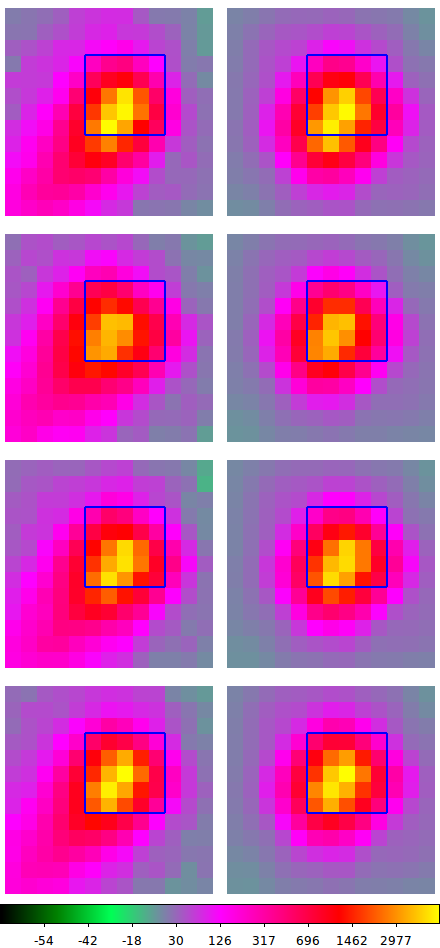}
\includegraphics[width=0.24\textwidth]{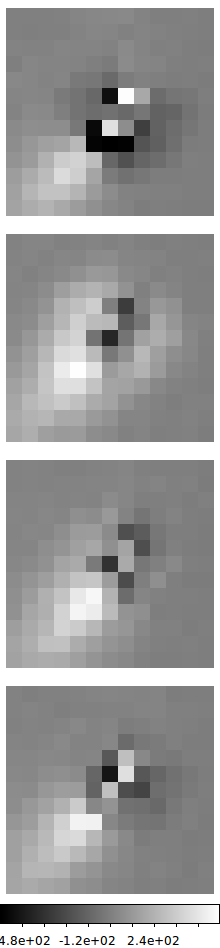}
\includegraphics[width=0.25\textwidth]{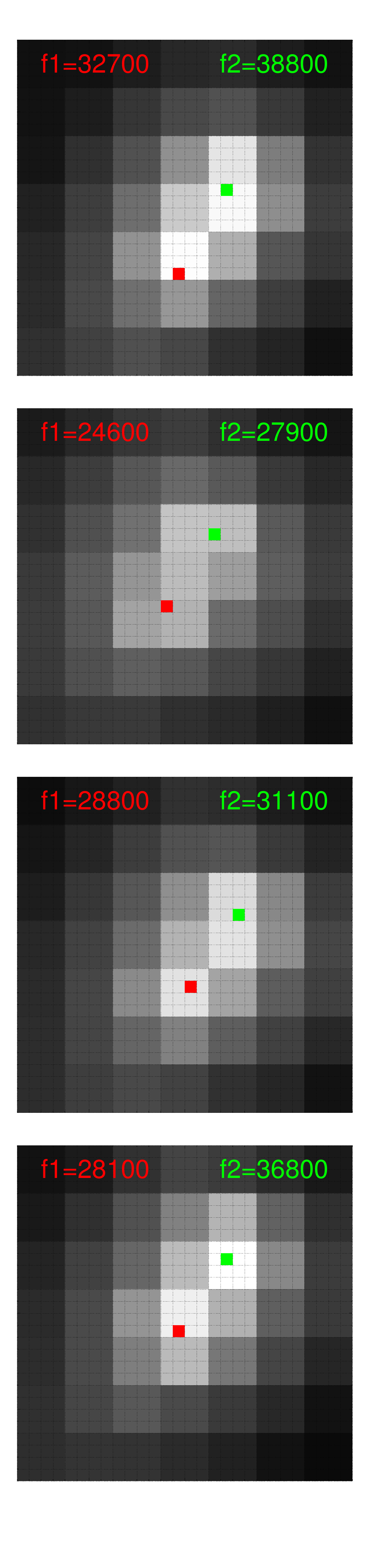}
\caption{Same as Figure~\ref{fig:obsmod_vA}, for {\bf Visit 5}, {\bf Brightness scale factor $b$=60}.
}
\label{fig:obsmod_v5}
\end{figure*}

\begin{figure*}
\includegraphics[width=0.48\textwidth]{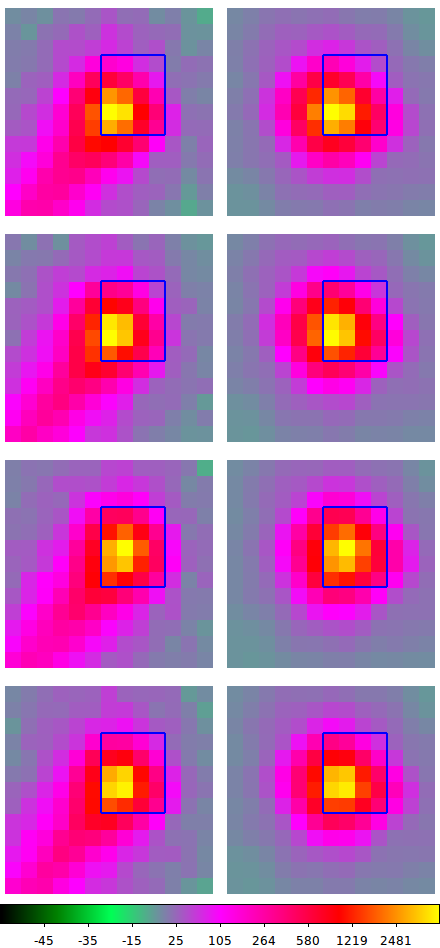}
\includegraphics[width=0.24\textwidth]{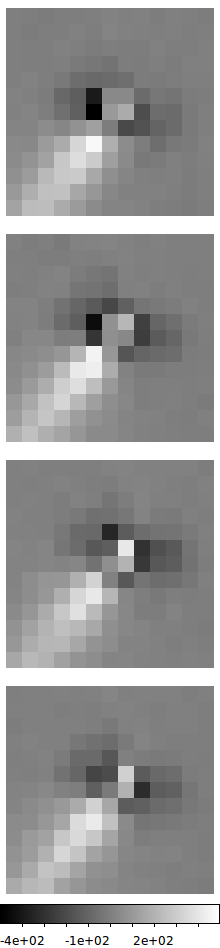}
\includegraphics[width=0.25\textwidth]{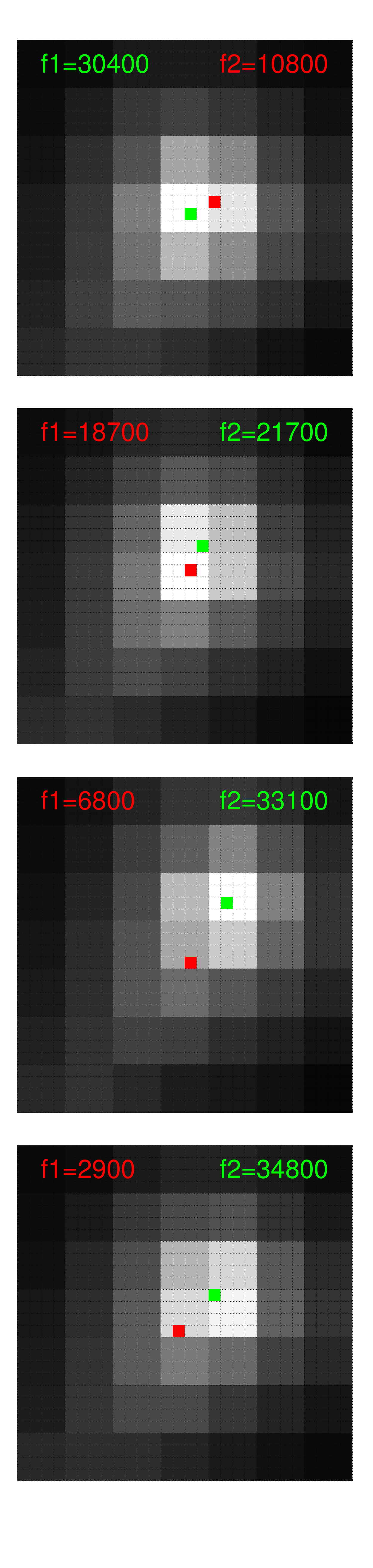}
\caption{Same as Figure~\ref{fig:obsmod_vA}, for {\bf Visit 6}, {\bf Brightness scale factor $b$=50}. The modelled brighness ratio of the components inverts between visits 2 and 3, reflecting either rotational variation (unlikely due to roughly constant total flux), or the uncertainty of the fitting results due to image noise and thermal breathing of the PSF. Since the distance between the model components increases with their brightness swap, it is possible that the reconstructed location of the faint component at stations 3 and 4 is influenced by the dust tail. 
}
\label{fig:obsmod_v6}
\end{figure*}

\begin{figure*}
\includegraphics[width=0.48\textwidth]{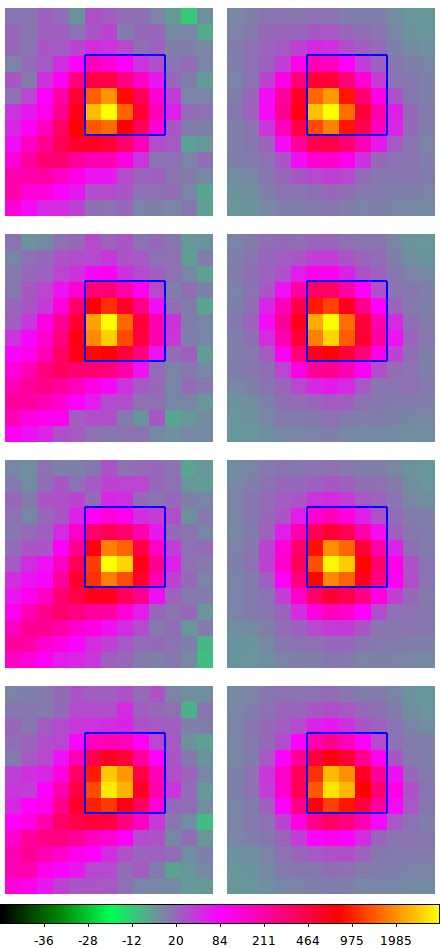}
\includegraphics[width=0.24\textwidth]{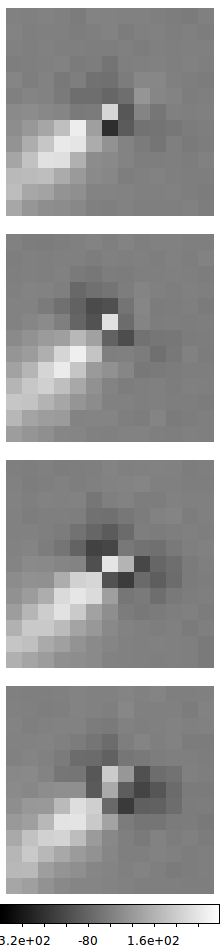}
\includegraphics[width=0.25\textwidth]{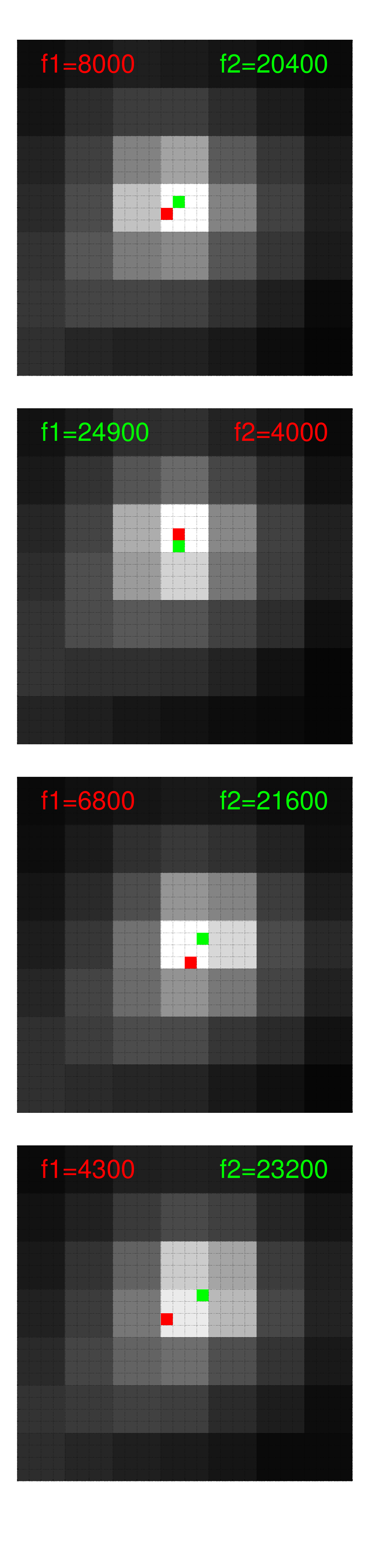}
\caption{Same as Figure~\ref{fig:obsmod_vA}, for {\bf Visit 7}, {\bf Brightness scale factor $b$=40}.
}
\label{fig:obsmod_v7}
\end{figure*}

\begin{figure*}
\includegraphics[width=0.48\textwidth]{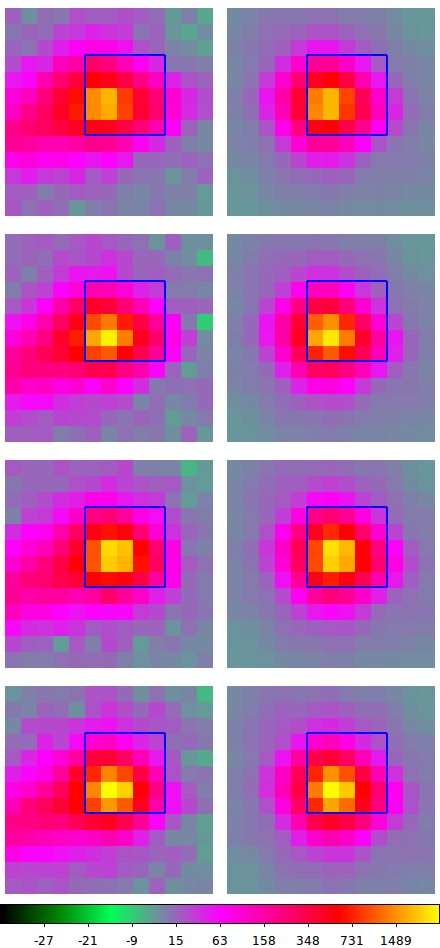}
\includegraphics[width=0.24\textwidth]{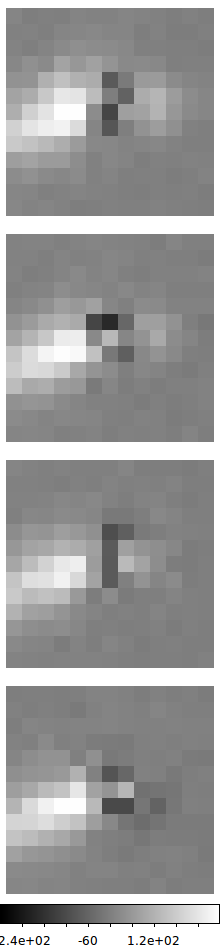}
\includegraphics[width=0.25\textwidth]{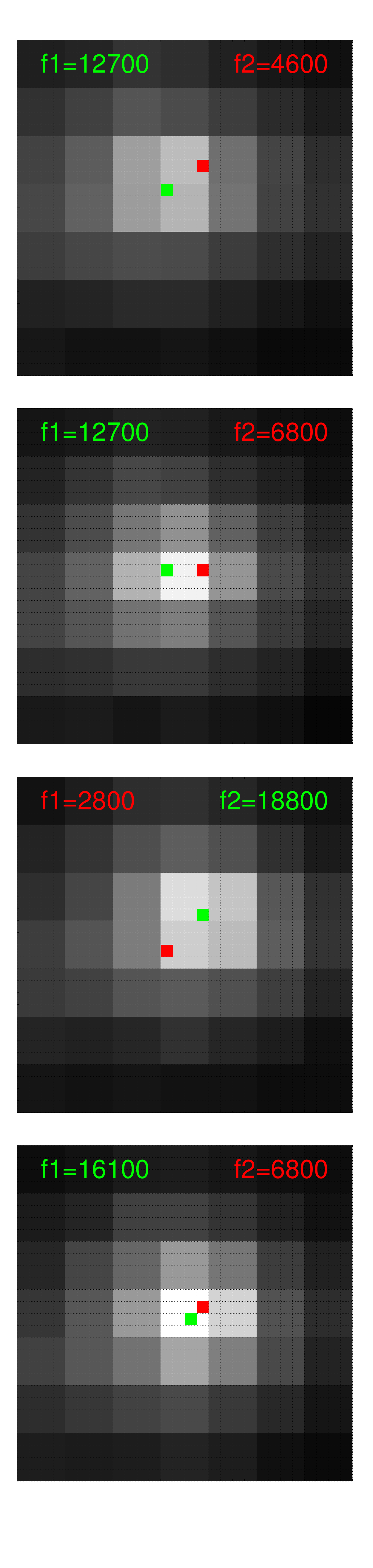}
\caption{Same as Figure~\ref{fig:obsmod_vA}, for {\bf Visit 8}, {\bf Brightness scale factor $b$=30}. The inverted model brightness ratio at station 3 could reflect rotational variation (unlikely due to roughly constant total flux), or the uncertainty of the fitting results due to image noise and thermal breathing of the PSF. 
}
\label{fig:obsmod_v8}
\end{figure*}

\begin{figure*}
\includegraphics[width=0.48\textwidth]{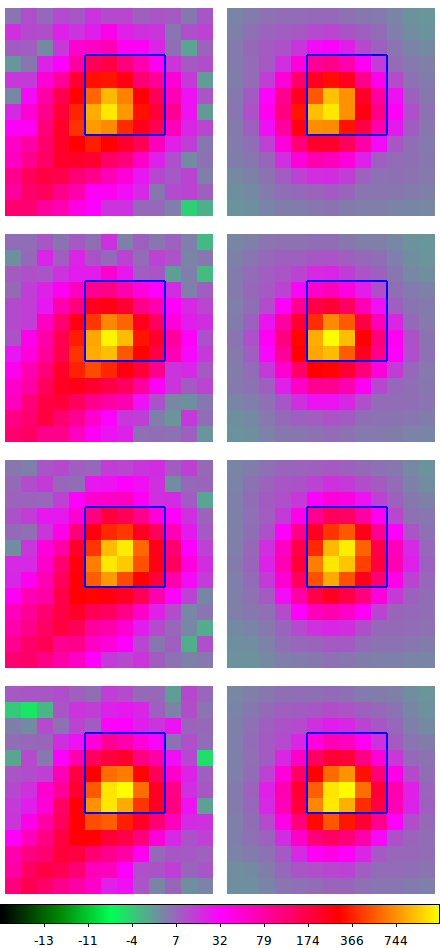}
\includegraphics[width=0.24\textwidth]{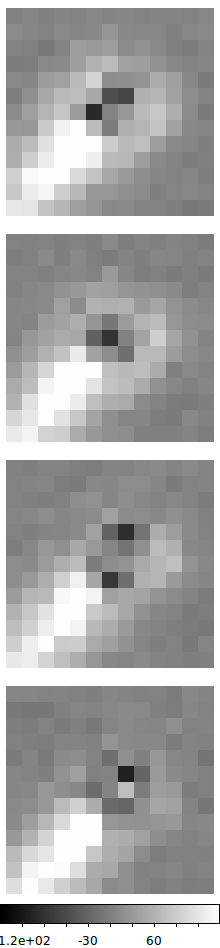}
\includegraphics[width=0.25\textwidth]{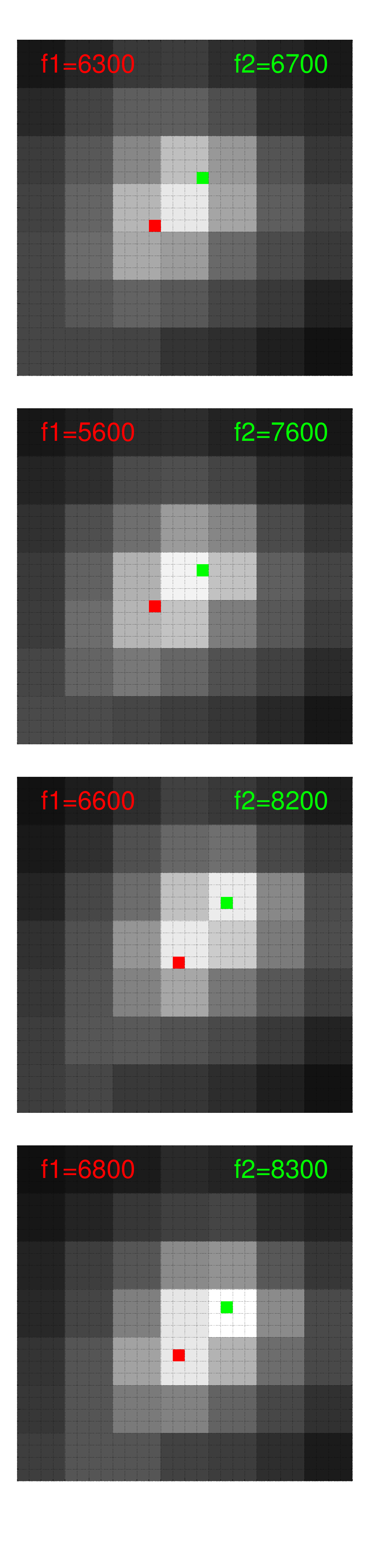}
\caption{Same as Figure~\ref{fig:obsmod_vA}, for {\bf Visit 9}, {\bf Brightness scale factor $b$=15}.
}
\label{fig:obsmod_v9}
\end{figure*}

\begin{figure*}
\includegraphics[width=0.48\textwidth]{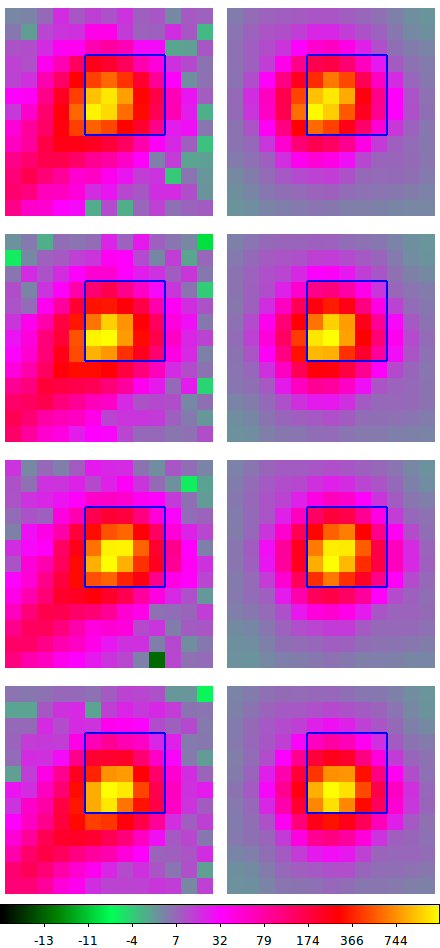}
\includegraphics[width=0.24\textwidth]{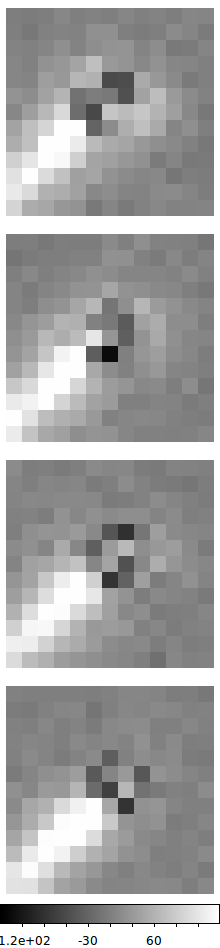}
\includegraphics[width=0.25\textwidth]{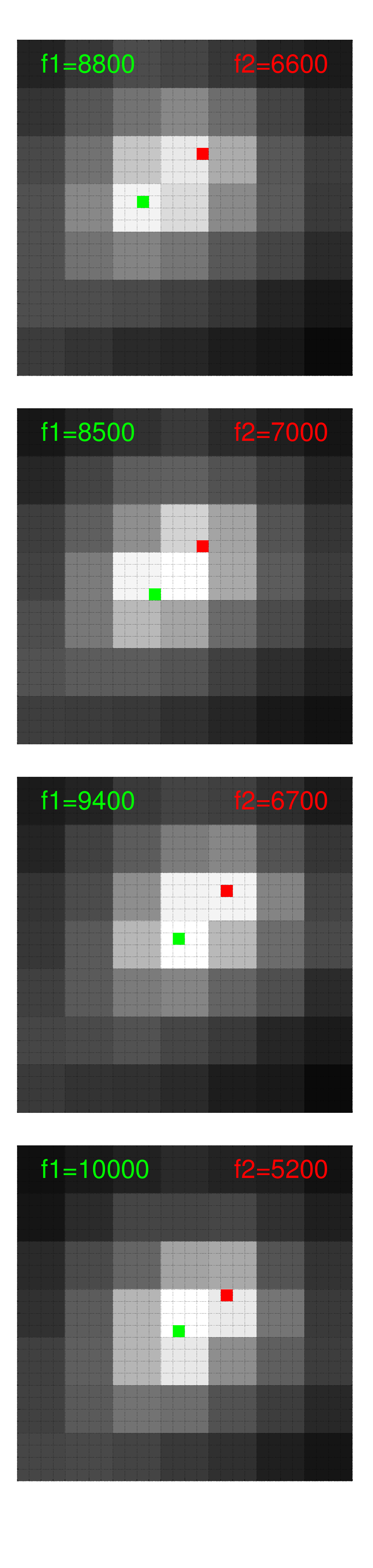}
\caption{Same as Figure~\ref{fig:obsmod_vA}, for {\bf Visit 10}, {\bf Brightness scale factor $b$=15}.
}
\label{fig:obsmod_v10}
\end{figure*}

\begin{figure*}
\includegraphics[width=0.48\textwidth]{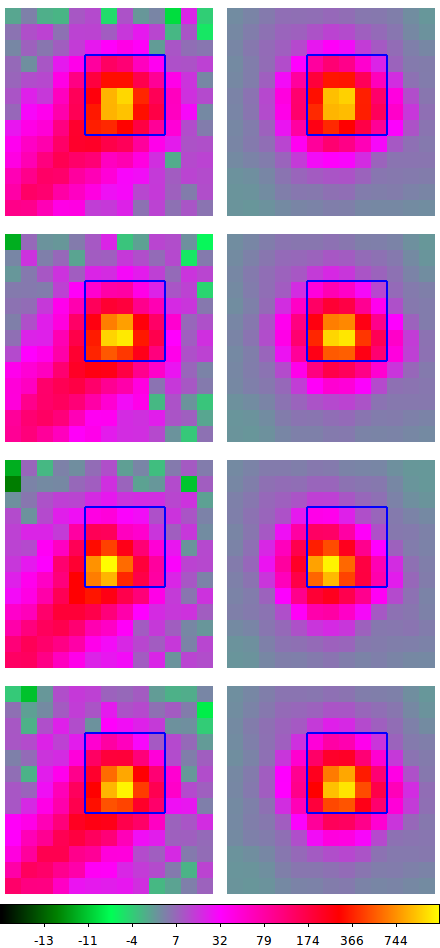}
\includegraphics[width=0.24\textwidth]{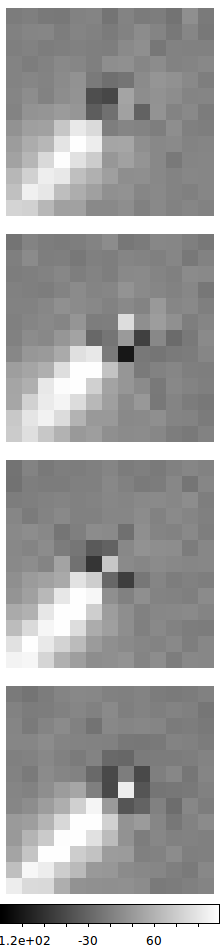}
\includegraphics[width=0.25\textwidth]{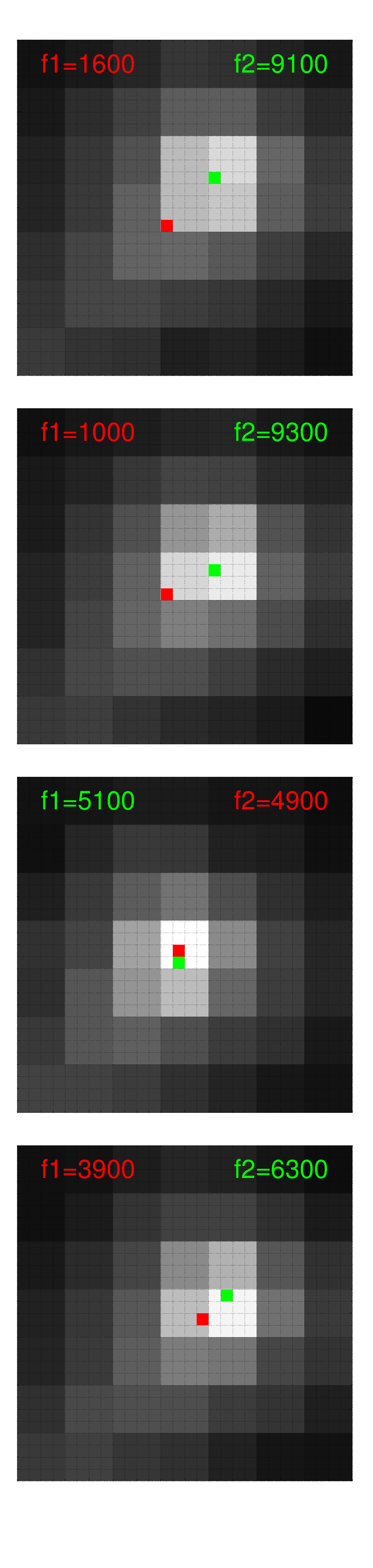}
\caption{Same as Figure~\ref{fig:obsmod_vA}, for {\bf Visit 11}, {\bf Brightness scale factor $b$=15}.
}
\label{fig:obsmod_v11}
\end{figure*}

\begin{figure*}
\includegraphics[width=0.48\textwidth]{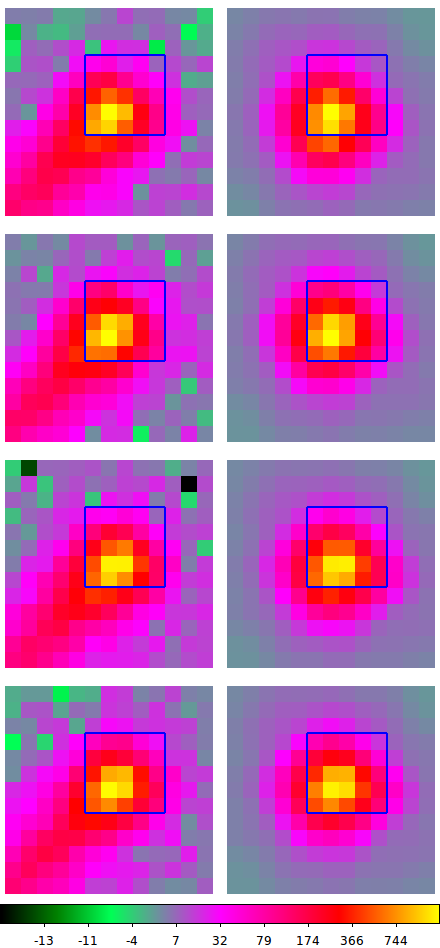}
\includegraphics[width=0.24\textwidth]{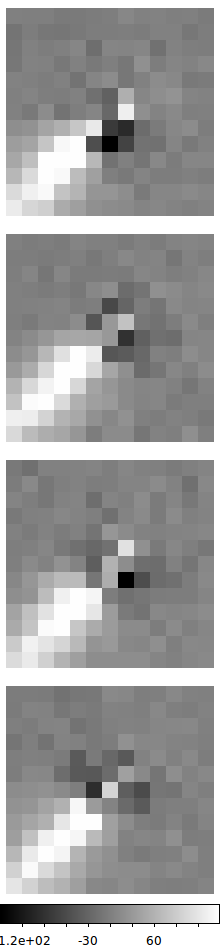}
\includegraphics[width=0.25\textwidth]{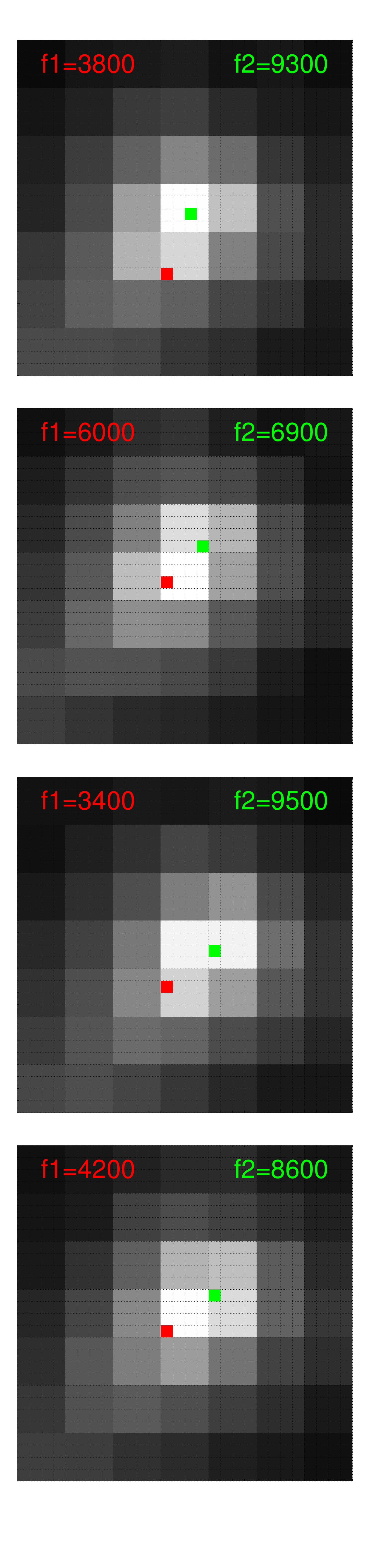}
\caption{Same as Figure~\ref{fig:obsmod_vA}, for {\bf Visit 12}, {\bf Brightness scale factor $b$=15}.
}
\label{fig:obsmod_v12}
\end{figure*}

\begin{figure*}
\includegraphics[width=0.48\textwidth]{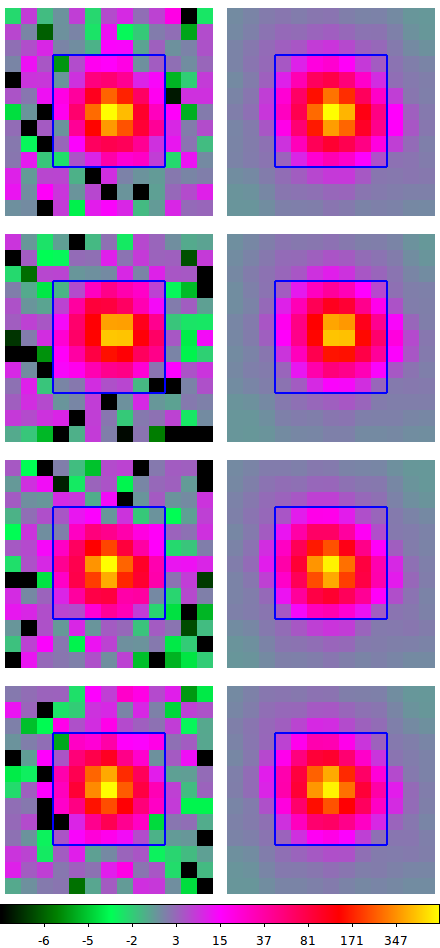}
\includegraphics[width=0.24\textwidth]{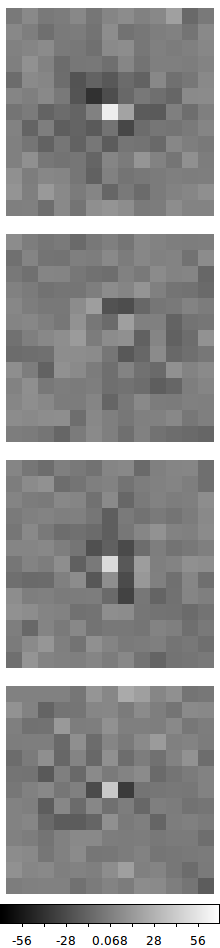}
\includegraphics[width=0.25\textwidth]{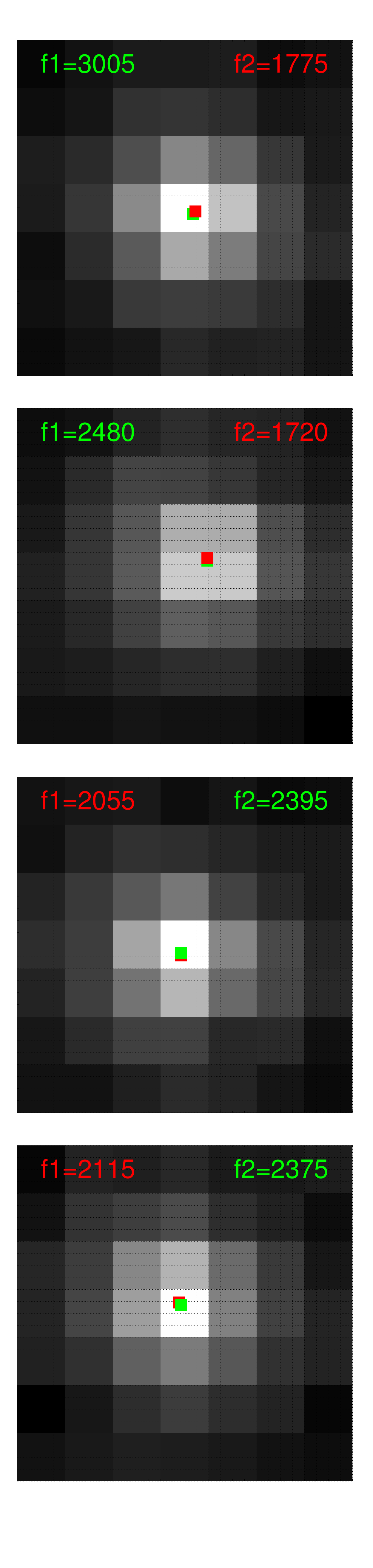}
\caption{Same as Figure~\ref{fig:obsmod_vA}, for {\bf Visit 13}, {\bf Brightness scale factor $b$=7}. The extremely small distance prevents a clear definition of the sign and results in small (absolute) error bars. 
}
\label{fig:obsmod_v13}
\end{figure*}
\pagebreak

\begin{figure*}
\includegraphics[width=0.48\textwidth]{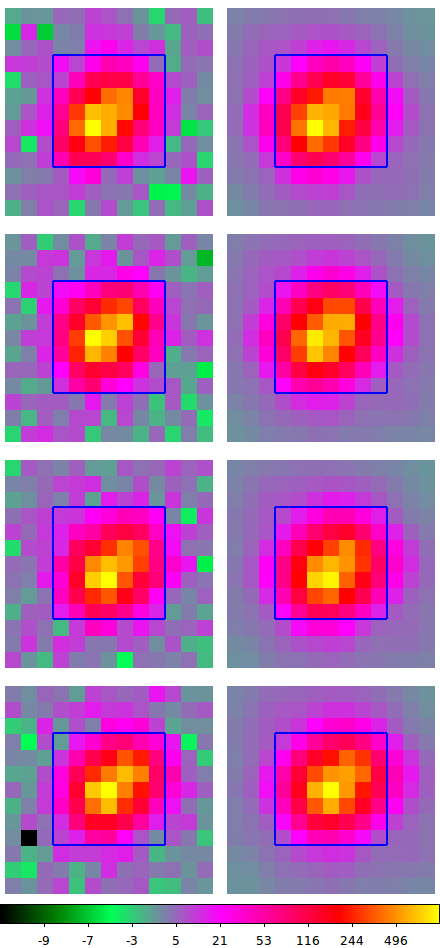}
\includegraphics[width=0.24\textwidth]{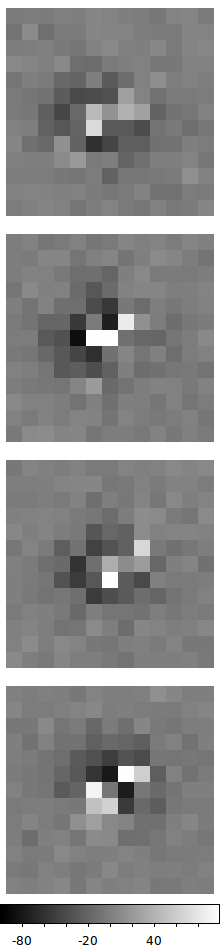}
\includegraphics[width=0.25\textwidth]{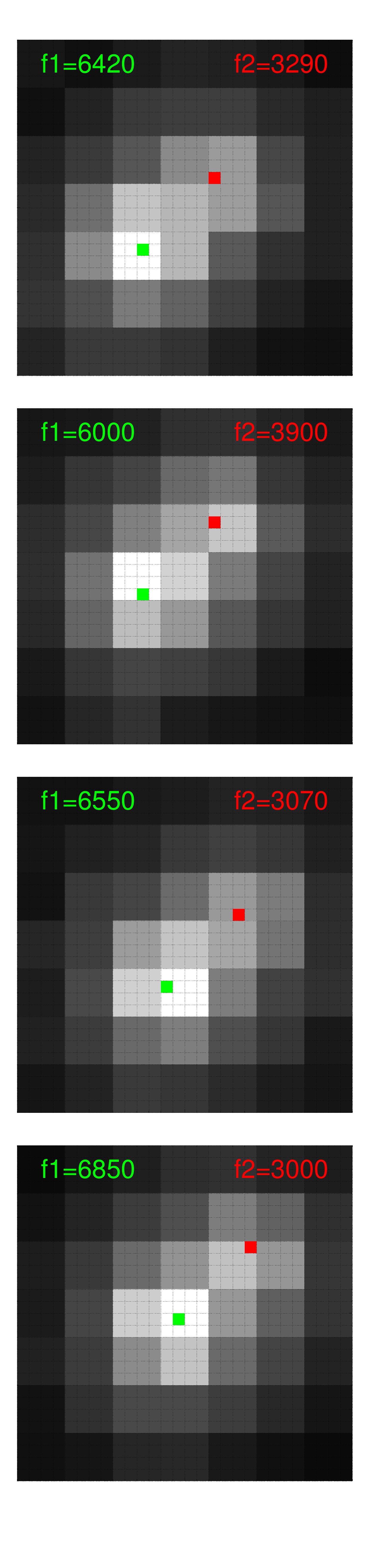}
\caption{Same as Figure~\ref{fig:obsmod_vA}, for {\bf Visit 14}, {\bf Brightness scale factor $b$=10}.
}
\label{fig:obsmod_v14}
\end{figure*}

\begin{figure*}
\includegraphics[width=0.48\textwidth]{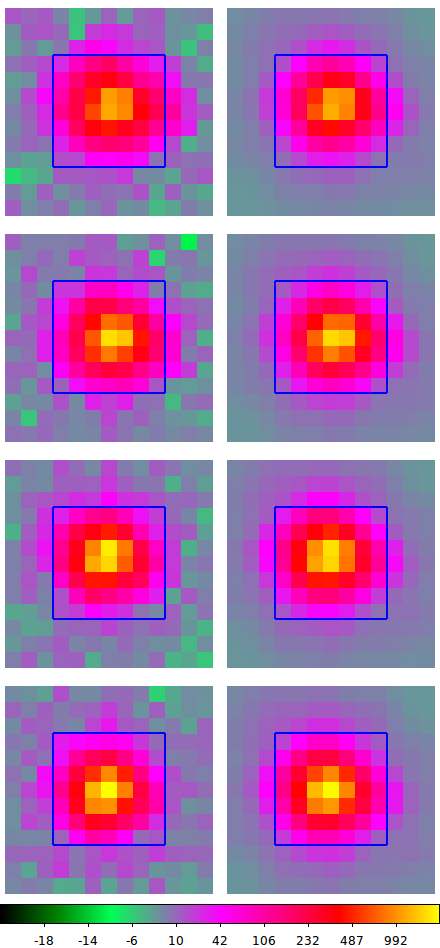}
\includegraphics[width=0.24\textwidth]{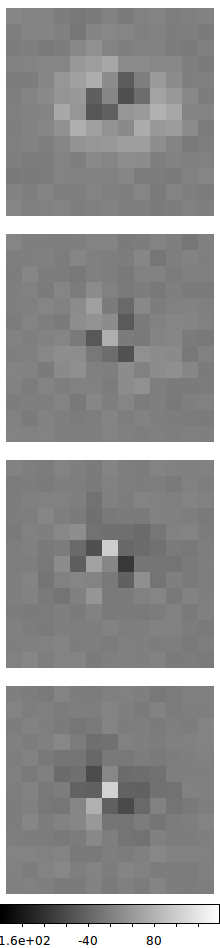}
\includegraphics[width=0.25\textwidth]{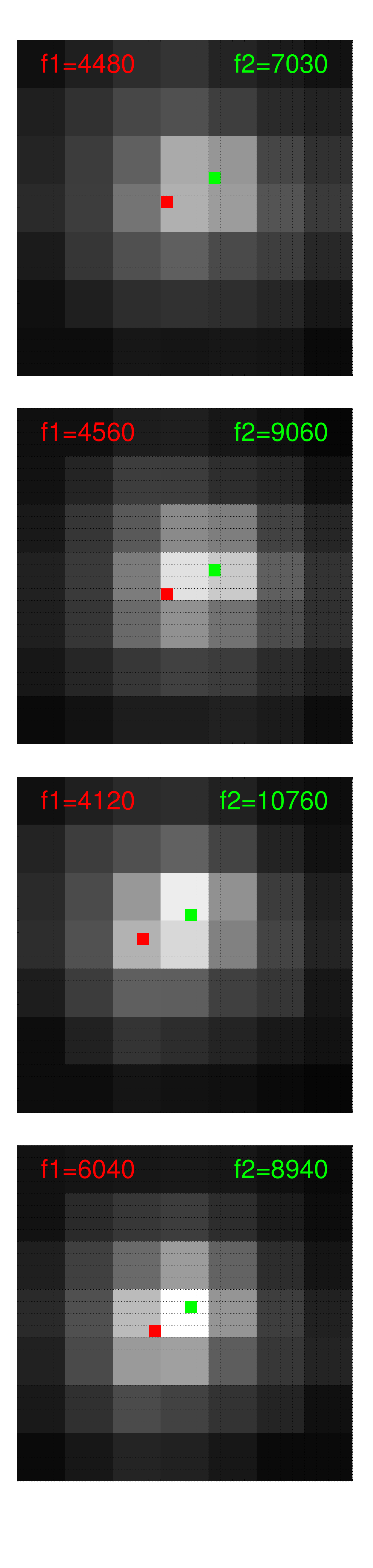}
\caption{Same as Figure~\ref{fig:obsmod_vA}, for {\bf Visit 16}, {\bf Brightness scale factor $b$=20}.
}
\label{fig:obsmod_v16}
\end{figure*}

\begin{figure*}
\includegraphics[width=0.48\textwidth]{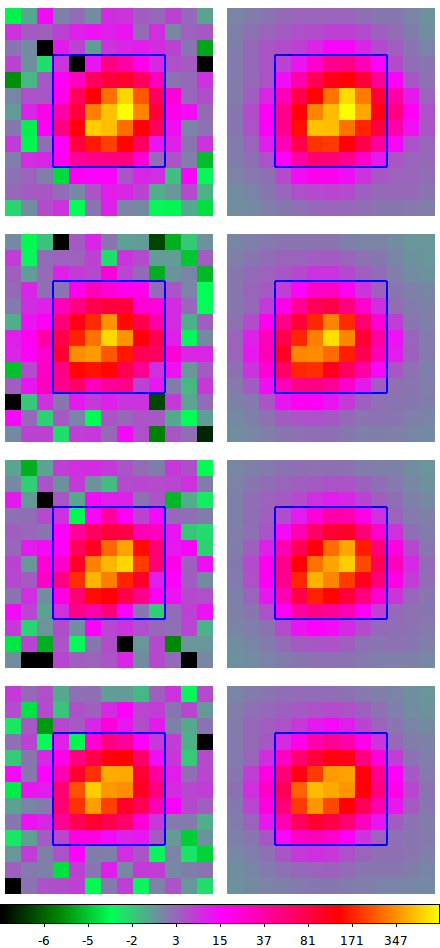}
\includegraphics[width=0.24\textwidth]{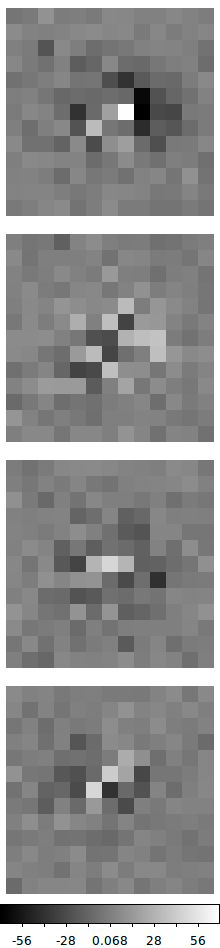}
\includegraphics[width=0.25\textwidth]{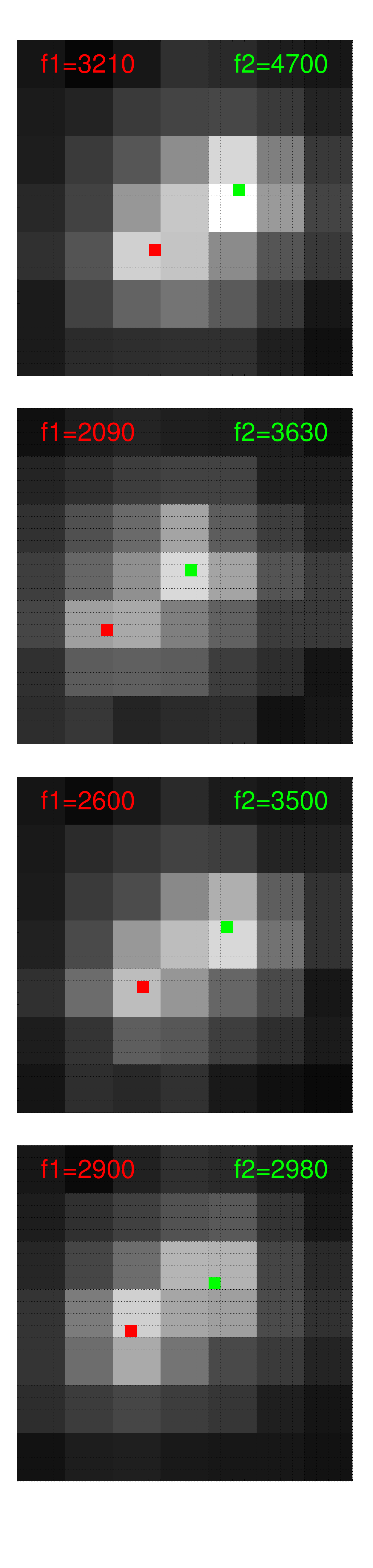}
\caption{Same as Figure~\ref{fig:obsmod_vA}, for {\bf Visit 17}, {\bf Brightness scale factor $b$=7}.
}
\label{fig:obsmod_v17}
\end{figure*}
\pagebreak

\begin{figure*}
\includegraphics[width=0.48\textwidth]{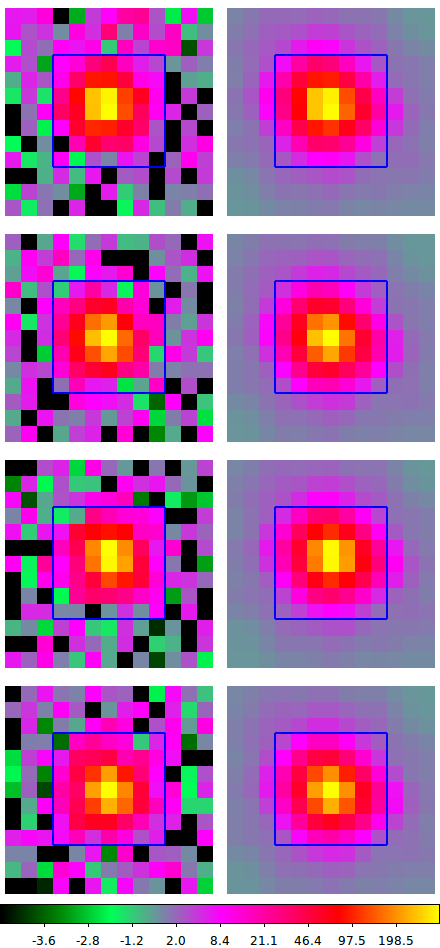}
\includegraphics[width=0.24\textwidth]{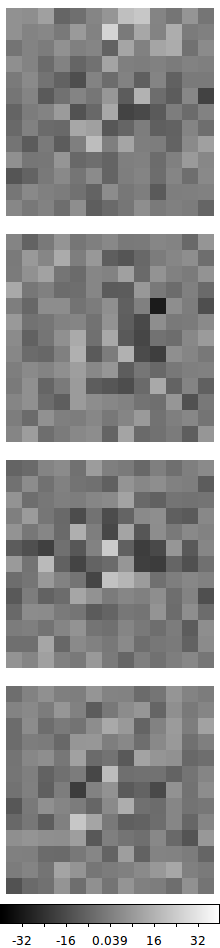}
\includegraphics[width=0.25\textwidth]{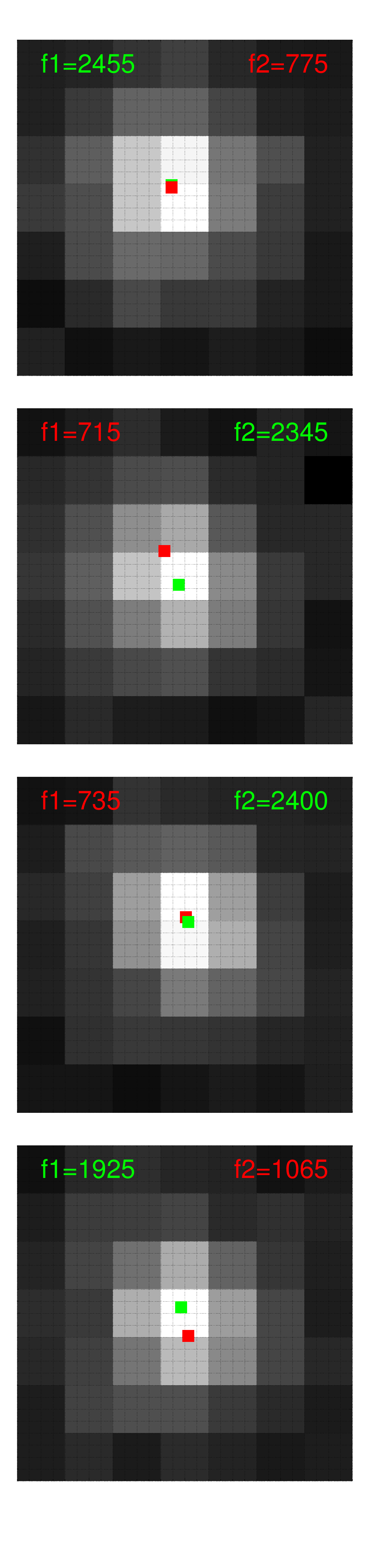}
\caption{Same as Figure~\ref{fig:obsmod_vA}, for {\bf Visit 18}, {\bf Brightness scale factor $b$=4}. Likely due to the small distance and the uncertainty of the fitting due to image noise and PSF breathing, it cannot be identified which component is brighter, and hence the sign of $D$ is undetermined.
}
\label{fig:obsmod_v18}
\end{figure*}

\begin{figure*}
\includegraphics[width=0.48\textwidth]{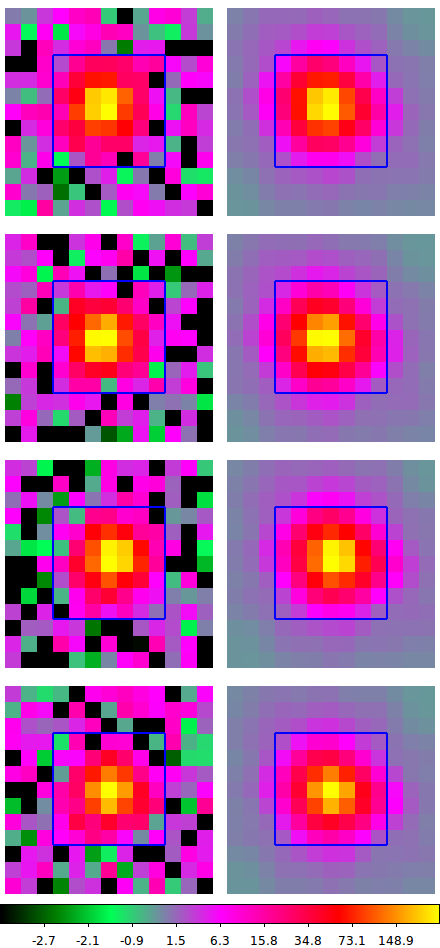}
\includegraphics[width=0.24\textwidth]{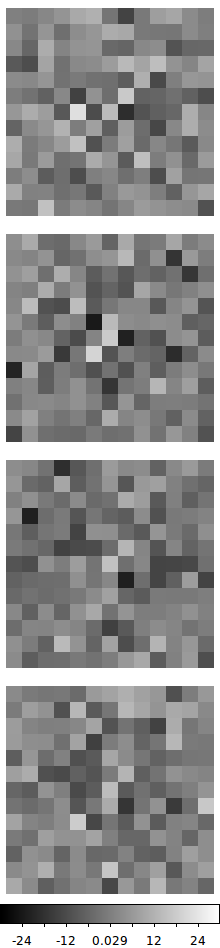}
\includegraphics[width=0.25\textwidth]{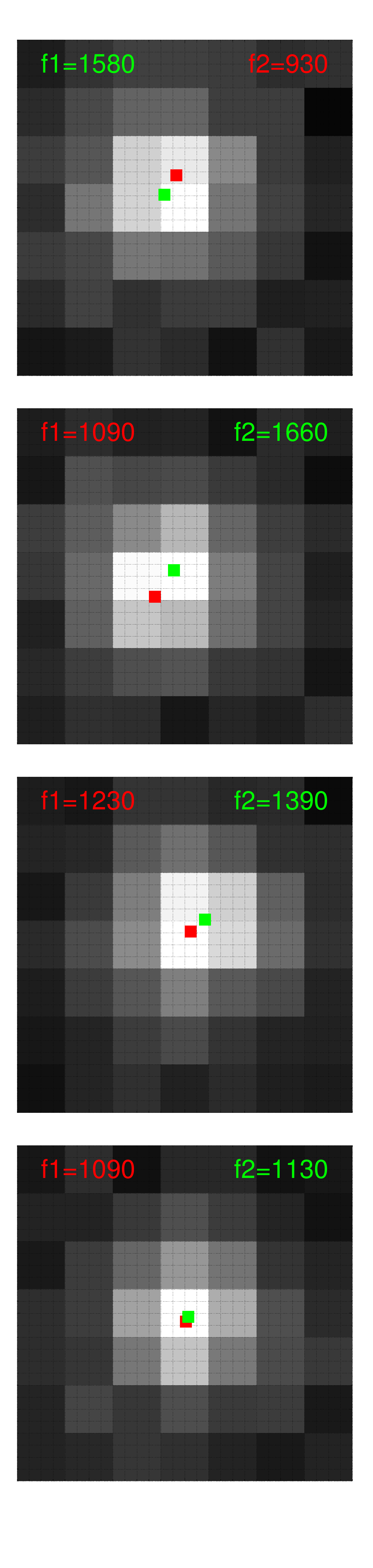}
\caption{Same as Figure~\ref{fig:obsmod_vA}, for {\bf Visit 19}, {\bf Brightness scale factor $b$=3}. Likely due to the small distance and the uncertainty of the fitting due to image noise and PSF breathing, it cannot be identified which component is brighter, and hence the sign of $D$ is undetermined.
}
\label{fig:obsmod_v19}
\end{figure*}

\pagebreak

\begin{figure*}
\includegraphics[width=0.48\textwidth]{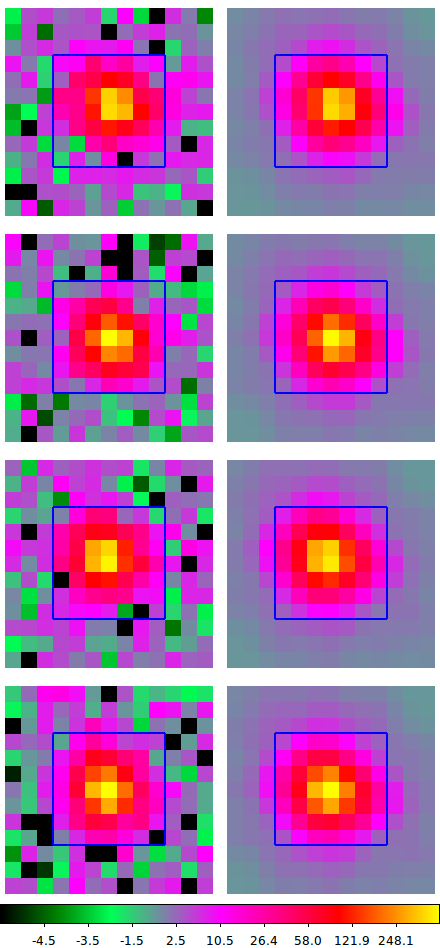}
\includegraphics[width=0.24\textwidth]{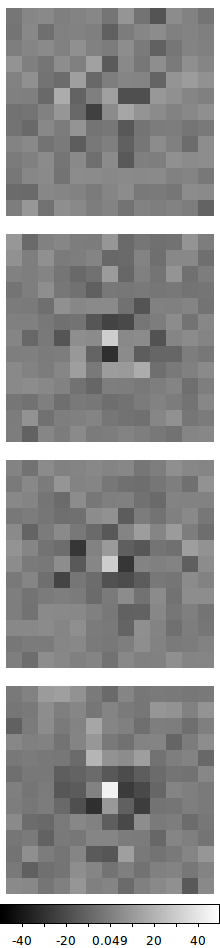}
\includegraphics[width=0.25\textwidth]{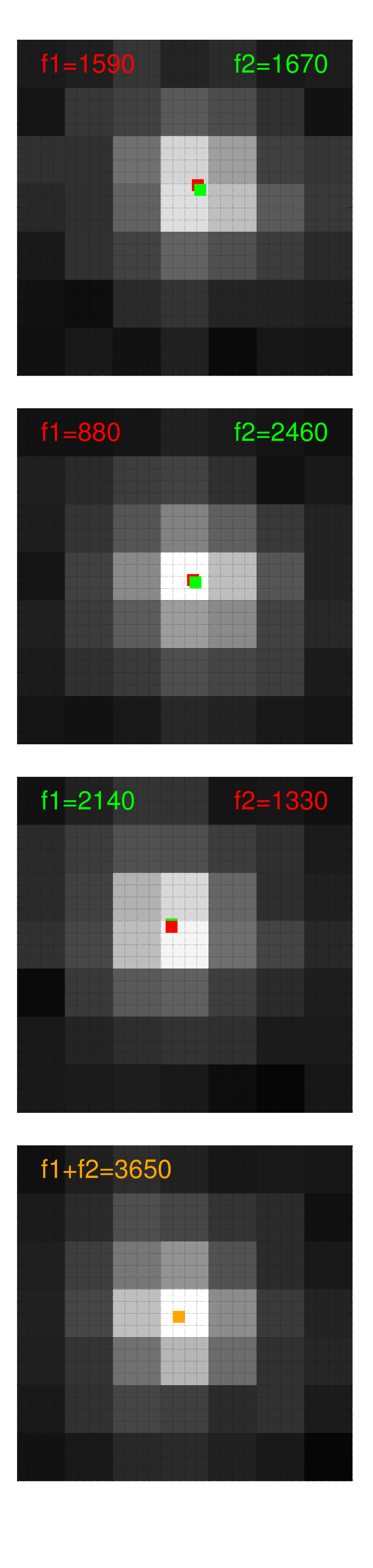}
\caption{Same as Figure~\ref{fig:obsmod_vA}, for {\bf Visit 20}, {\bf Brightness scale factor $b$=5}. Likely due to the small distance and the uncertainty of the fitting due to image noise and PSF breathing, it cannot be identified which component is brighter, and hence the sign of $D$ is undetermined. For station 4, the best fit was obtained with a single point source (both components in the same sub-pixel).
}
\label{fig:obsmod_v20}
\end{figure*}

\begin{figure*}
\includegraphics[width=0.48\textwidth]{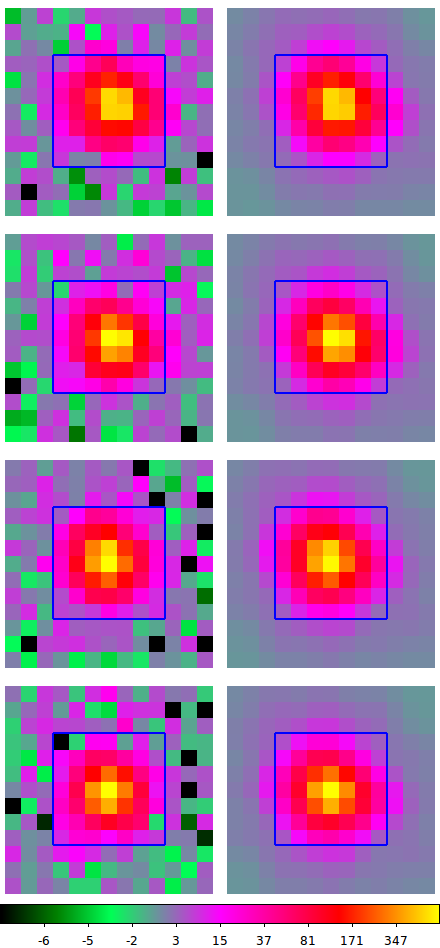}
\includegraphics[width=0.24\textwidth]{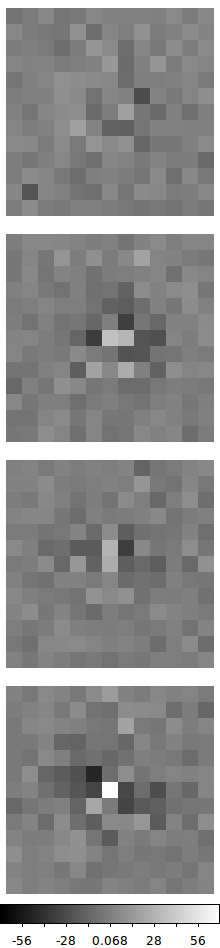}
\includegraphics[width=0.25\textwidth]{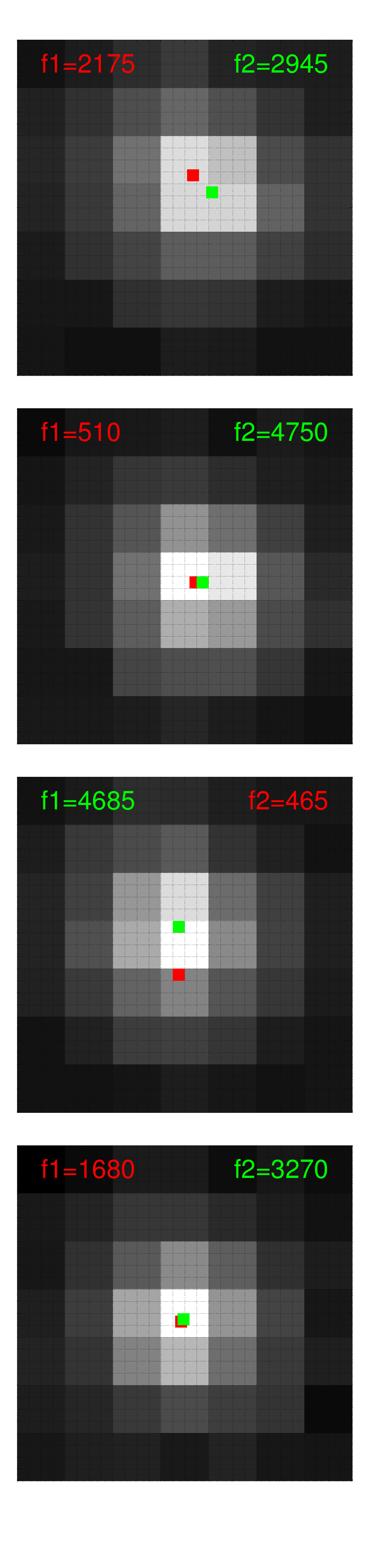}
\caption{Same as Figure~\ref{fig:obsmod_vA}, for {\bf Visit 21}, {\bf Brightness scale factor $b$=7}. Likely due to the small distance and the uncertainty of the fitting due to image noise and PSF breathing, it cannot be identified which component is brighter, and hence the sign of $D$ is undetermined.
}
\label{fig:obsmod_v21}
\end{figure*}

\begin{figure*}
\includegraphics[width=0.48\textwidth]{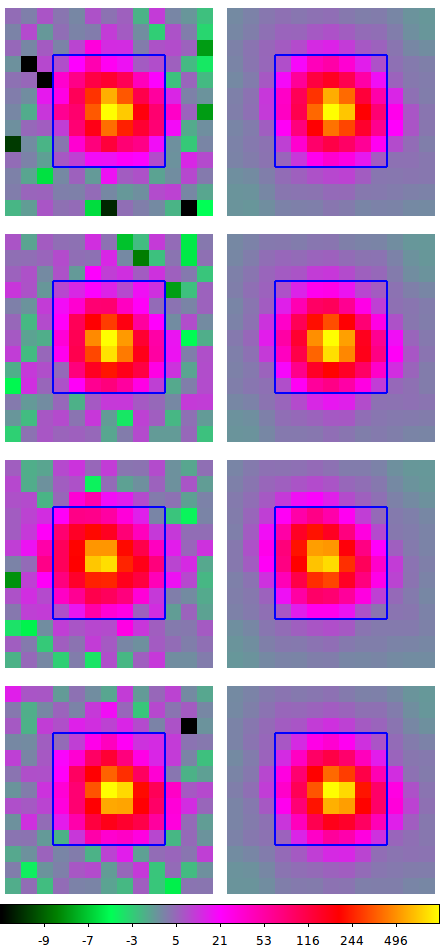}
\includegraphics[width=0.24\textwidth]{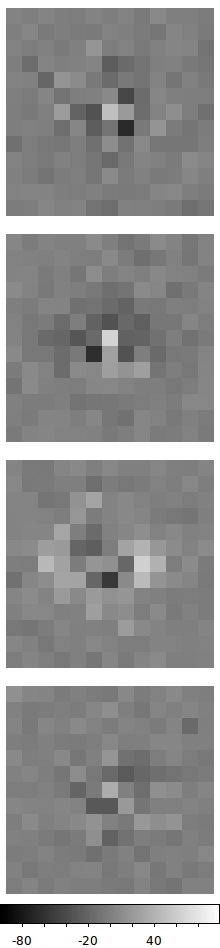}
\includegraphics[width=0.25\textwidth]{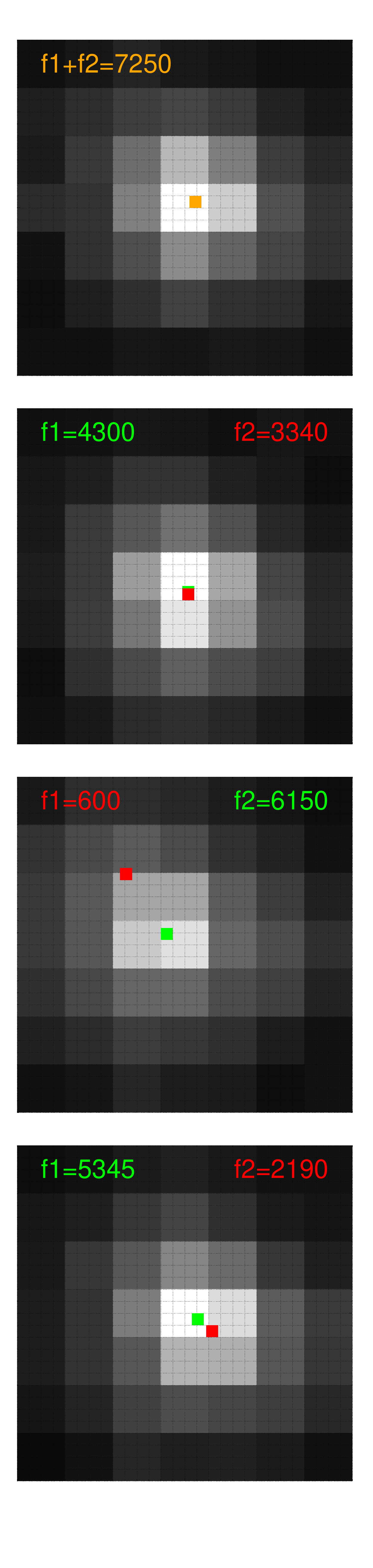}
\caption{Same as Figure~\ref{fig:obsmod_vA}, for {\bf Visit 22}, {\bf Brightness scale factor $b$=10}. Likely due to the small distance and the uncertainty of the fitting due to image noise and PSF breathing, it cannot be identified which component is brighter, and hence the sign of $D$ is undetermined. For station 1, the best fit was obtained with a single point source (both components in the same sub-pixel.)
}
\label{fig:obsmod_v22}
\end{figure*}

\begin{figure*}
\includegraphics[width=0.48\textwidth]{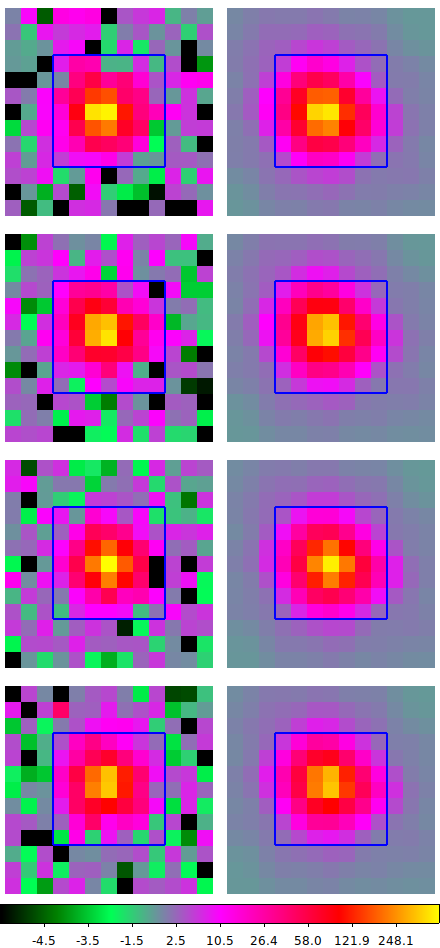}
\includegraphics[width=0.24\textwidth]{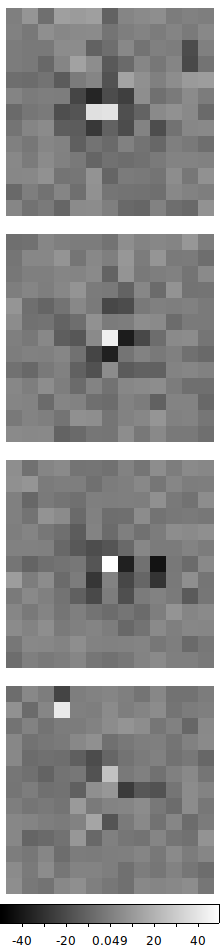}
\includegraphics[width=0.25\textwidth]{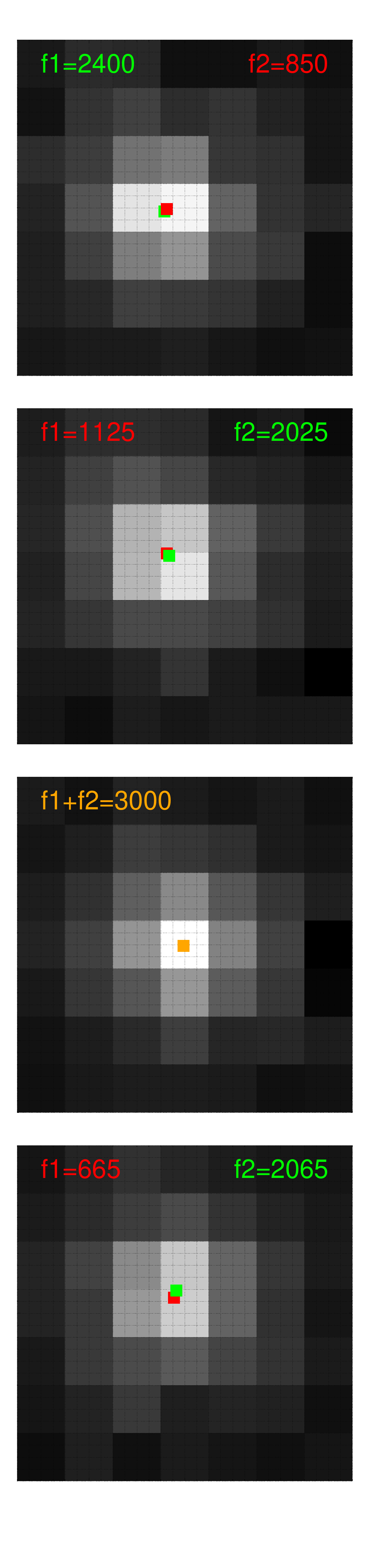}
\caption{Same as Figure~\ref{fig:obsmod_vA}, for {\bf Visit 23}, {\bf Brightness scale factor $b$=5}. Likely due to the small distance and the uncertainty of the fitting due to image noise and PSF breathing, it cannot be identified which component is brighter, and hence the sign of $D$ is undetermined. For station 3, the best fit was obtained with a single point source (both components in the same sub-pixel.)
}
\label{fig:obsmod_v23}
\end{figure*}

\pagebreak

\begin{figure*}
\includegraphics[width=0.48\textwidth]{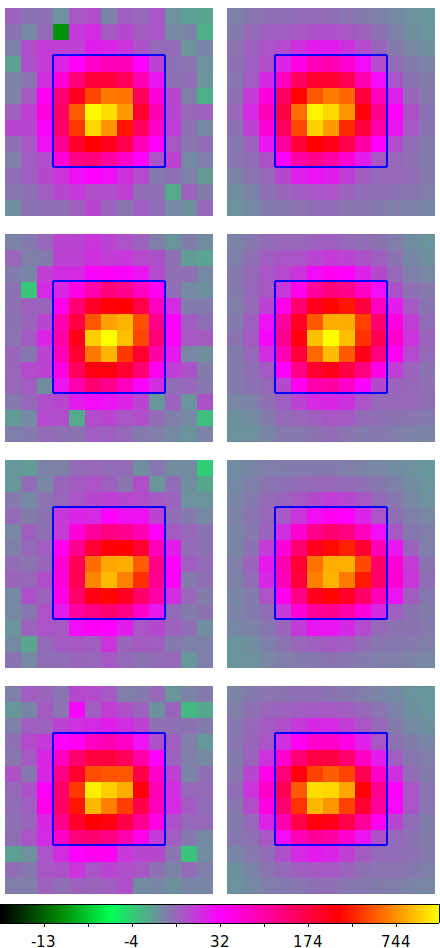}
\includegraphics[width=0.24\textwidth]{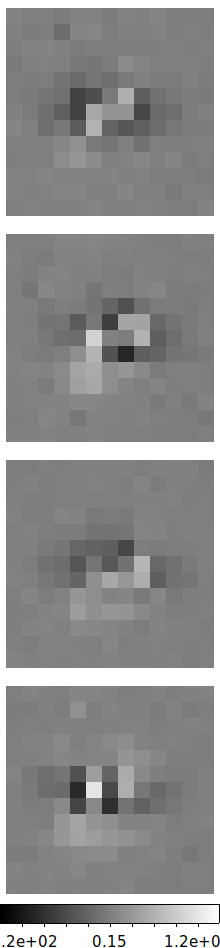}
\includegraphics[width=0.25\textwidth]{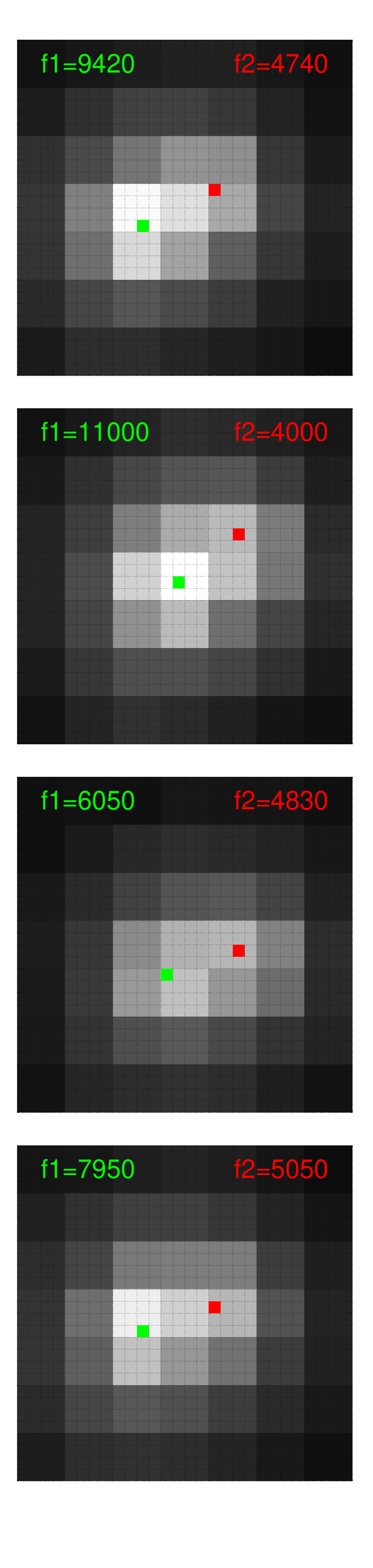}
\caption{Same as Figure~\ref{fig:obsmod_vA}, for the initial four orbits of {\bf Visit 24}, {\bf Brightness scale factor $b$=15}.
}
\label{fig:obsmod_v24a}
\end{figure*}

\begin{figure*}
\includegraphics[width=0.48\textwidth]{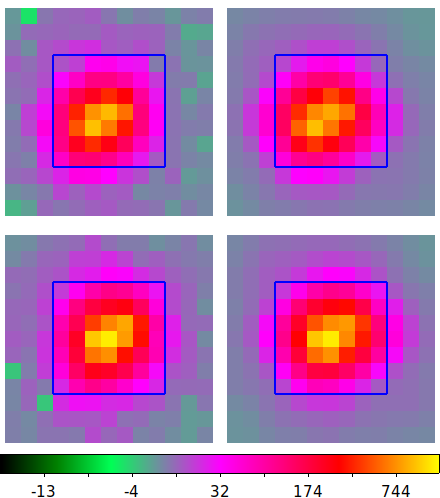}
\includegraphics[width=0.24\textwidth]{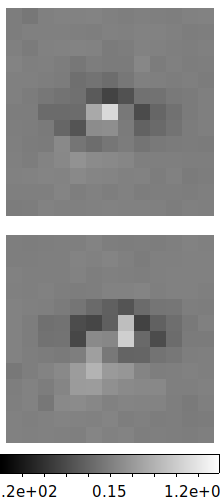}
\includegraphics[width=0.25\textwidth]{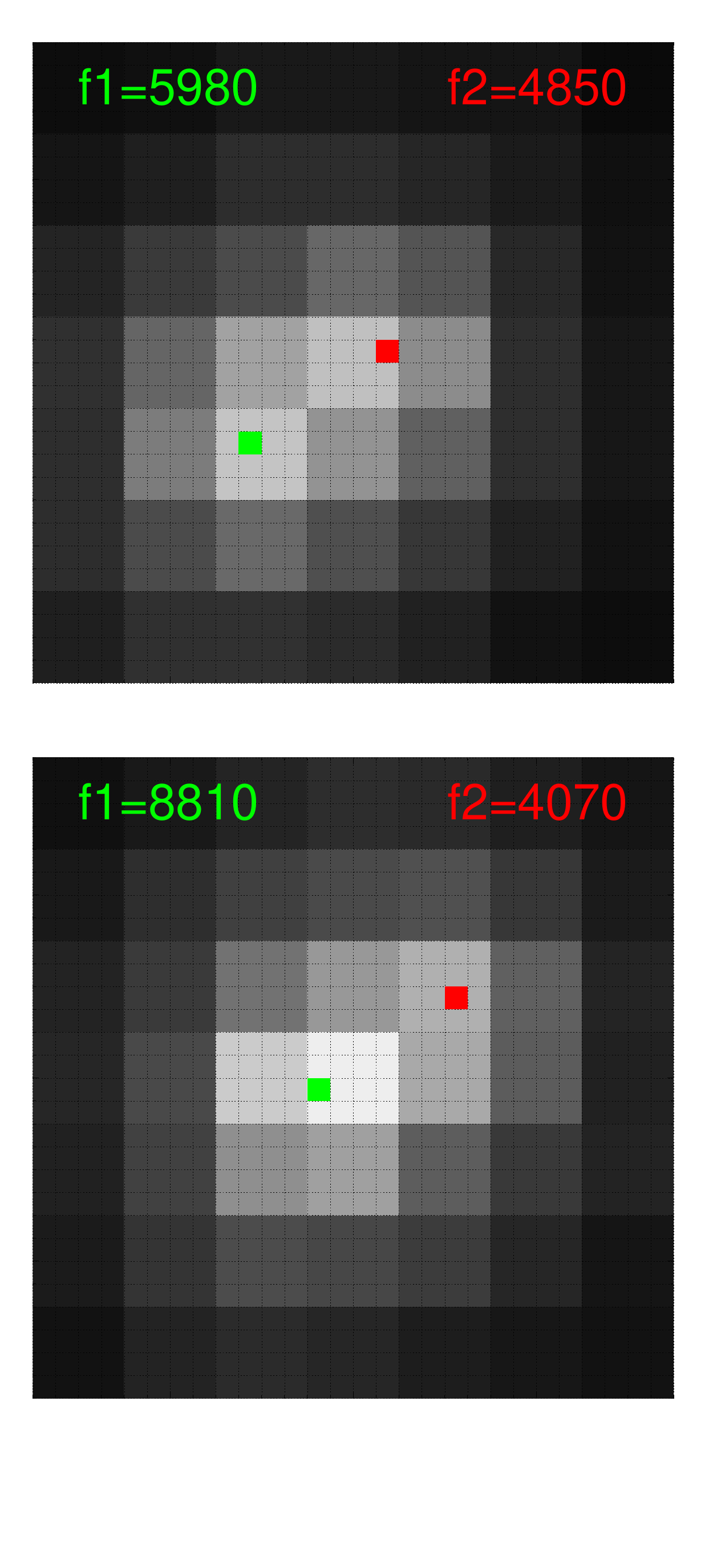}
\caption{Same as Figure~\ref{fig:obsmod_vA}, for the last two orbits of {\bf Visit 24}, {\bf Brightness scale factor $b$=15}.
}
\label{fig:obsmod_v24b}
\end{figure*}

%% file: input_photometry.tex
\section{Photometry}
\label{app:photometry}
For each dither station, $j$, we compared the flux $F_{rp}^j$ with the combined flux from PSF fitting $F_{PSF}^j = F_j^1 + F_j^2$, and from aperture photometry $F_{ap}^j$. We measured the latter in a 5-pixel-radius circular aperture and subtracted the background measured in an annulus 4 pixel wide and separated from the central aperture by 7 pixels. 
\begin{figure}[H]
  \includegraphics[width=\columnwidth]{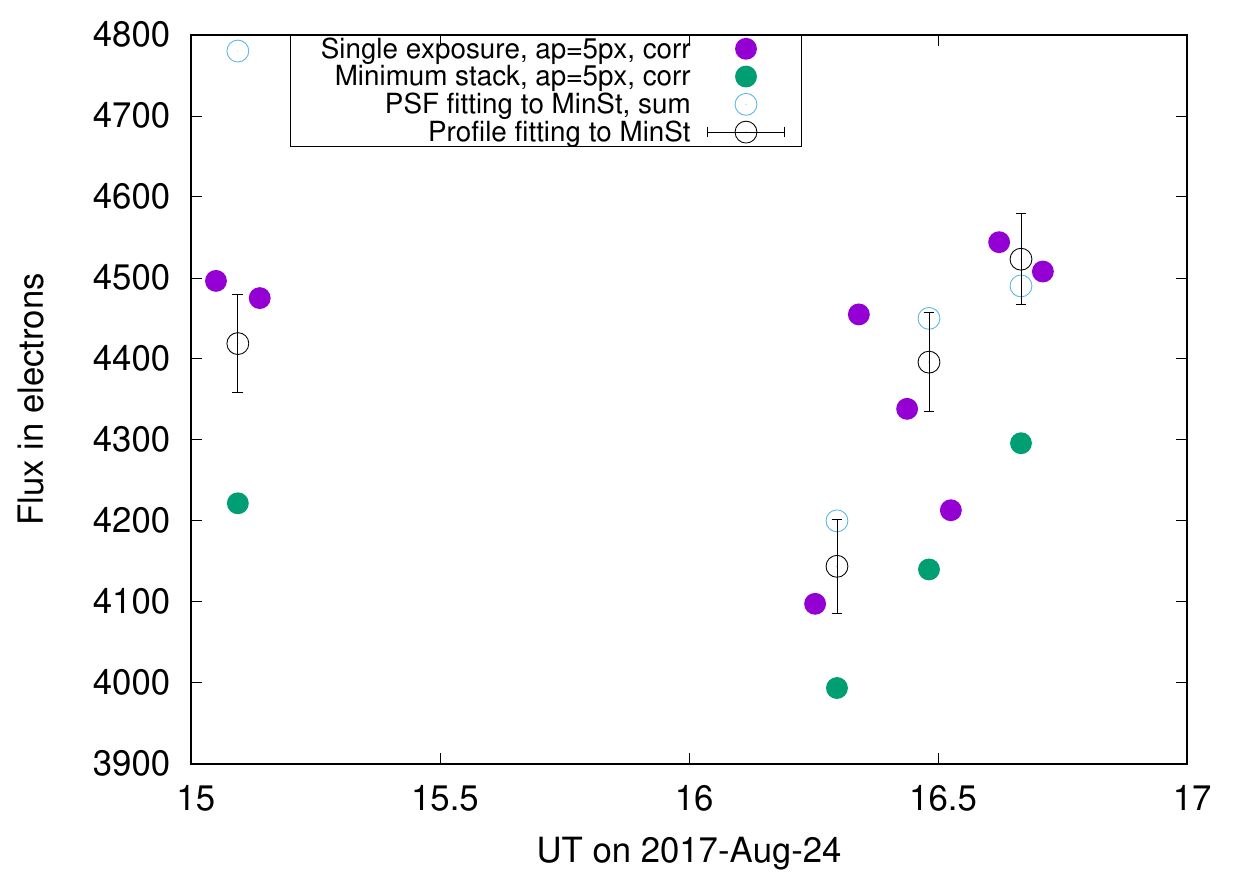}
  \includegraphics[width=\columnwidth]{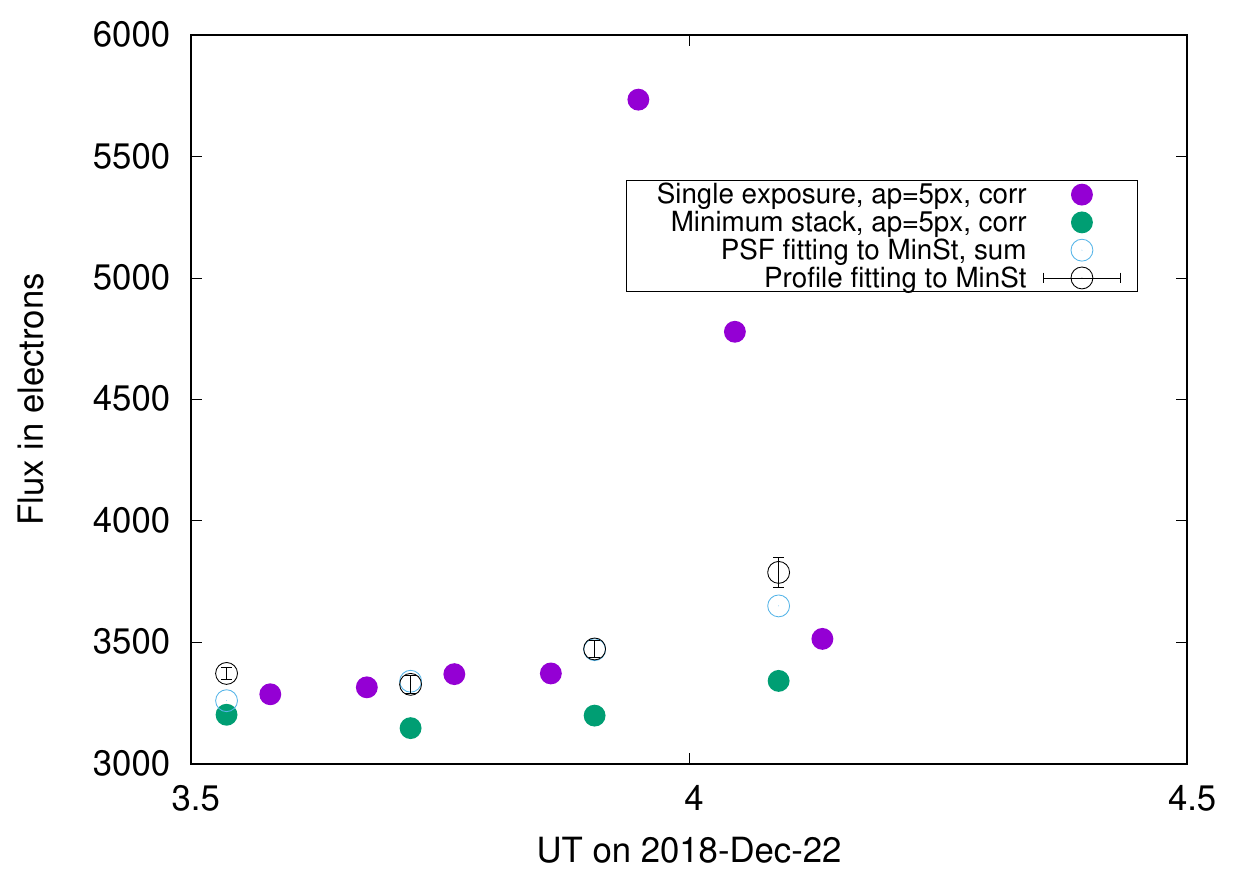}
  \includegraphics[width=\columnwidth]{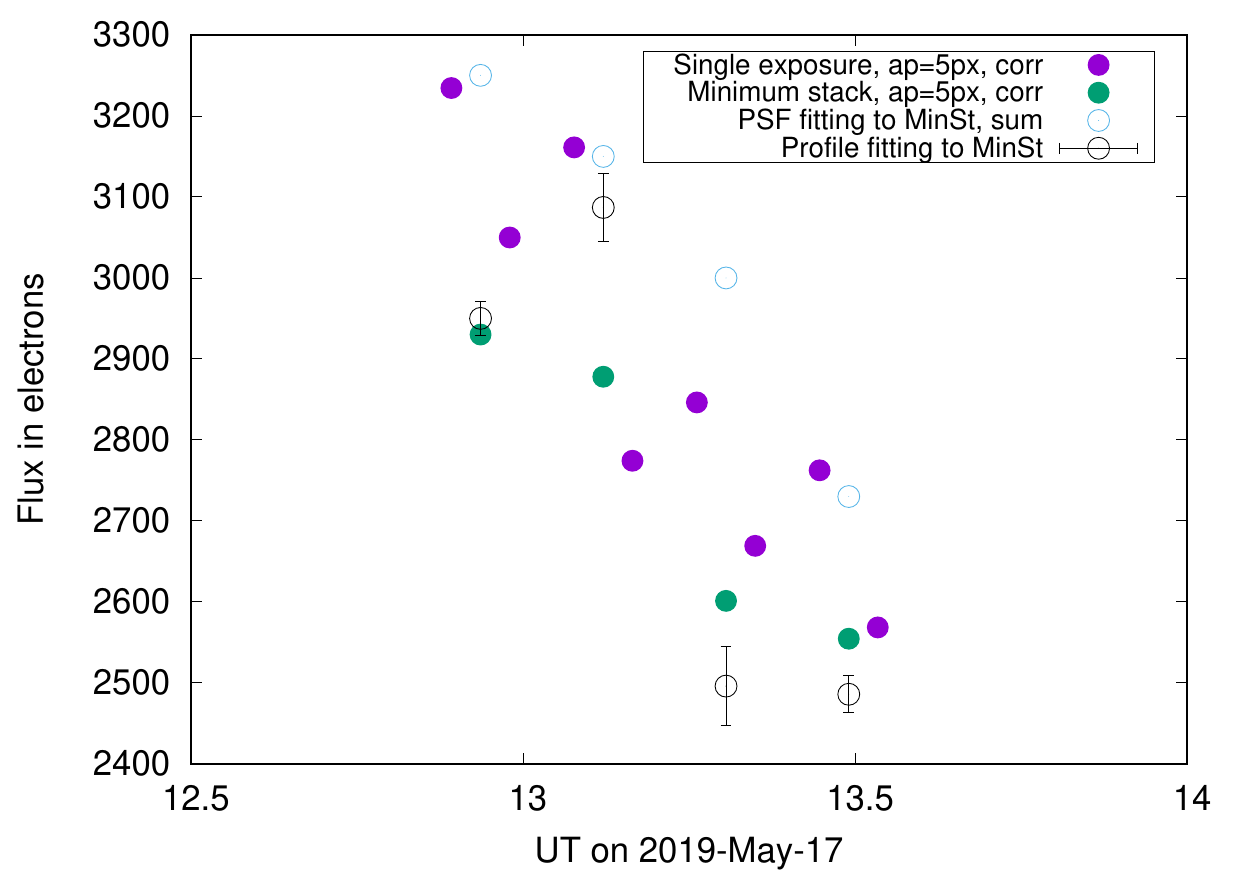}  
 \caption{Flux measured by PSF fitting, profile fitting, and aperture photometry in minimum-stacked images, and by aperture photometry in single exposures for visits 13 (top), 20 (centre) and 23 (bottom). The two strong outliers for single exposures in visit 20 are probably due to cosmic ray hits that were removed by the minimum stacking.
}
\label{fig:photometry_point_source}
\end{figure}

The PSF profile in Fig.~\ref{fig:rad_prof} shows that a 5-pixel aperture contains about 94\% of the total flux. 
Hence we divided the measured value by 0.94 to obtain $F_{ap}$. We repeated this measurement also for the single exposures that are potentially contaminated by cosmic ray hits. Fig.~\ref{fig:photometry_point_source} shows the three measurements for comparison.

The measurements generally vary by up to 10\%, with no obvious systematic trend except that the aperture photometry in the minimum-stacked images typically returns the lowest values. During visits 13 and 23, statistically significant brightness changes corresponding to 0.1 and 0.2\,mag, respectively, may indicate rotational variability. 

\begin{figure}[!b]
  \includegraphics[width=\columnwidth]{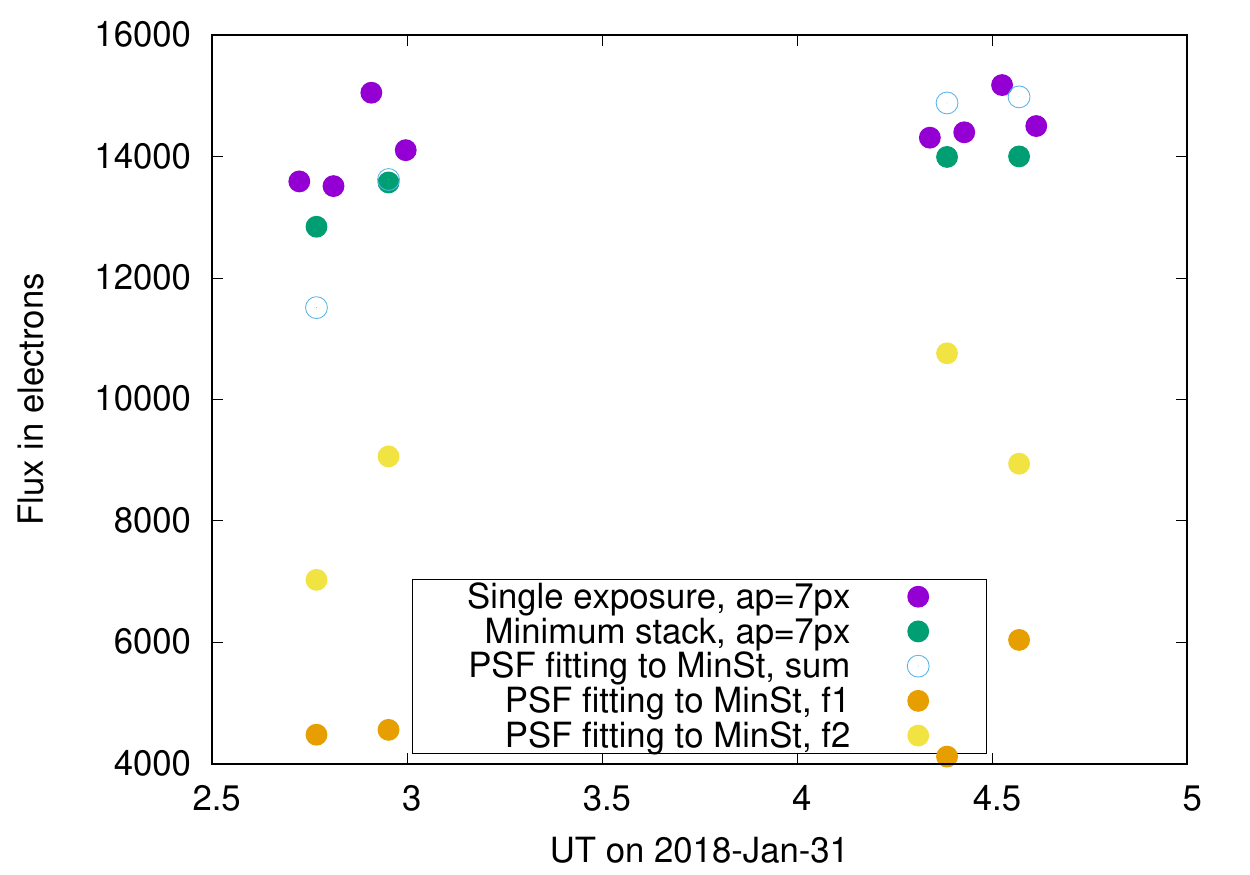}
  \includegraphics[width=\columnwidth]{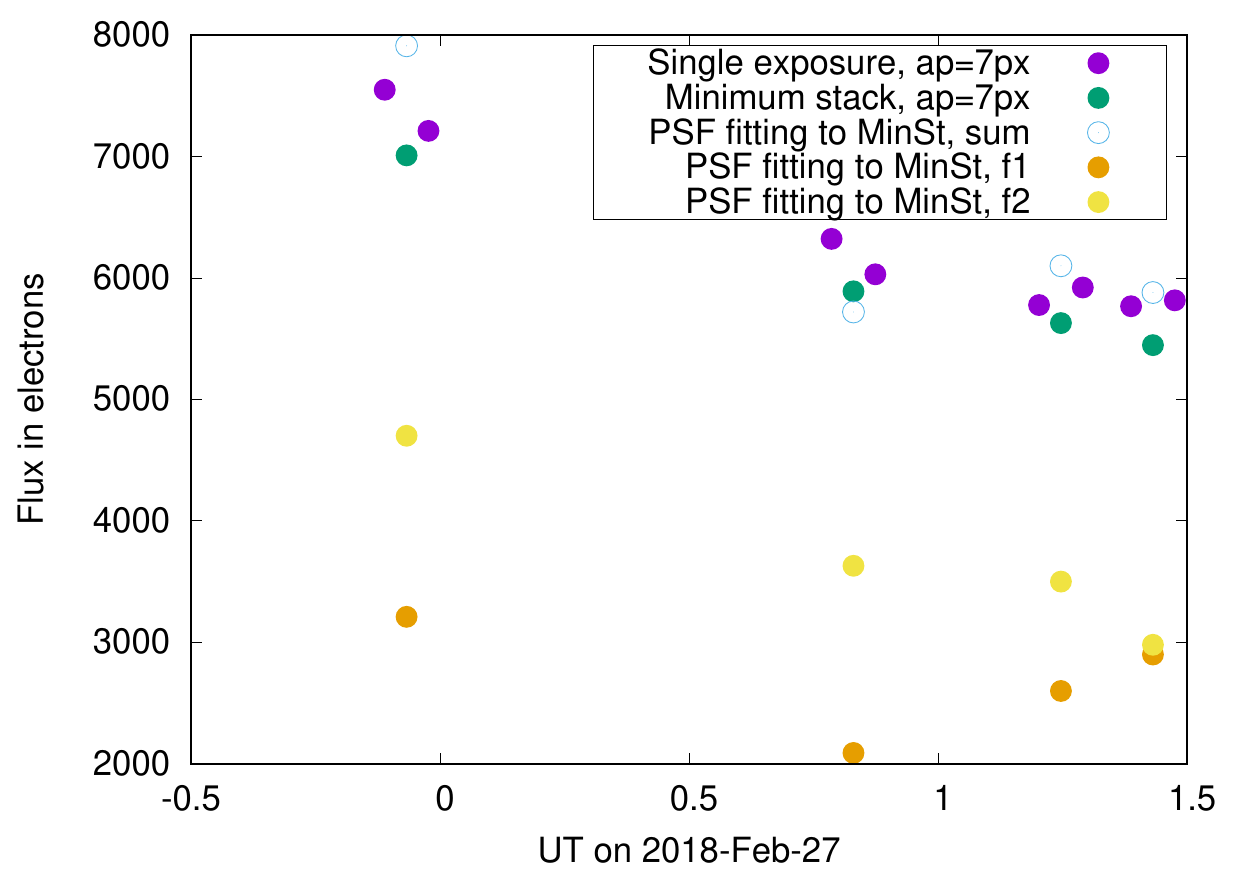}
 \caption{Flux (combined and for the individual components) measured by PSF fitting, and aperture photometry (7 pixels radius) of the combined flux in minimum-stacked images and in single exposures for visits 16 (top) and 17 (bottom), when the components were at maximum separation. We did not correct the aperture flux because of the larger radius used here, and did also not use radial profile fitting due to the non-circular shape of the combined targets. 
}
\label{fig:photometry_binary_separated}
\end{figure}

Fig.~\ref{fig:photometry_binary_separated} shows a comparison of aperture fluxes and PSF fitting results for two situations, when the components were at maximum separation of about 2 native pixels. Also here, an inter-method variability of $\sim$10\% can be observed, and a statistically significant decrease of the total brightness by about 0.2\,mag in visit 17. These data are too sparse to judge the time variability of the individual components.

%% file: input_plane_view.tex
\onecolumn
\section{Orbital geometry}
\label{sec:plane_view}
\begin{figure}[h]
\includegraphics[width=\textwidth]{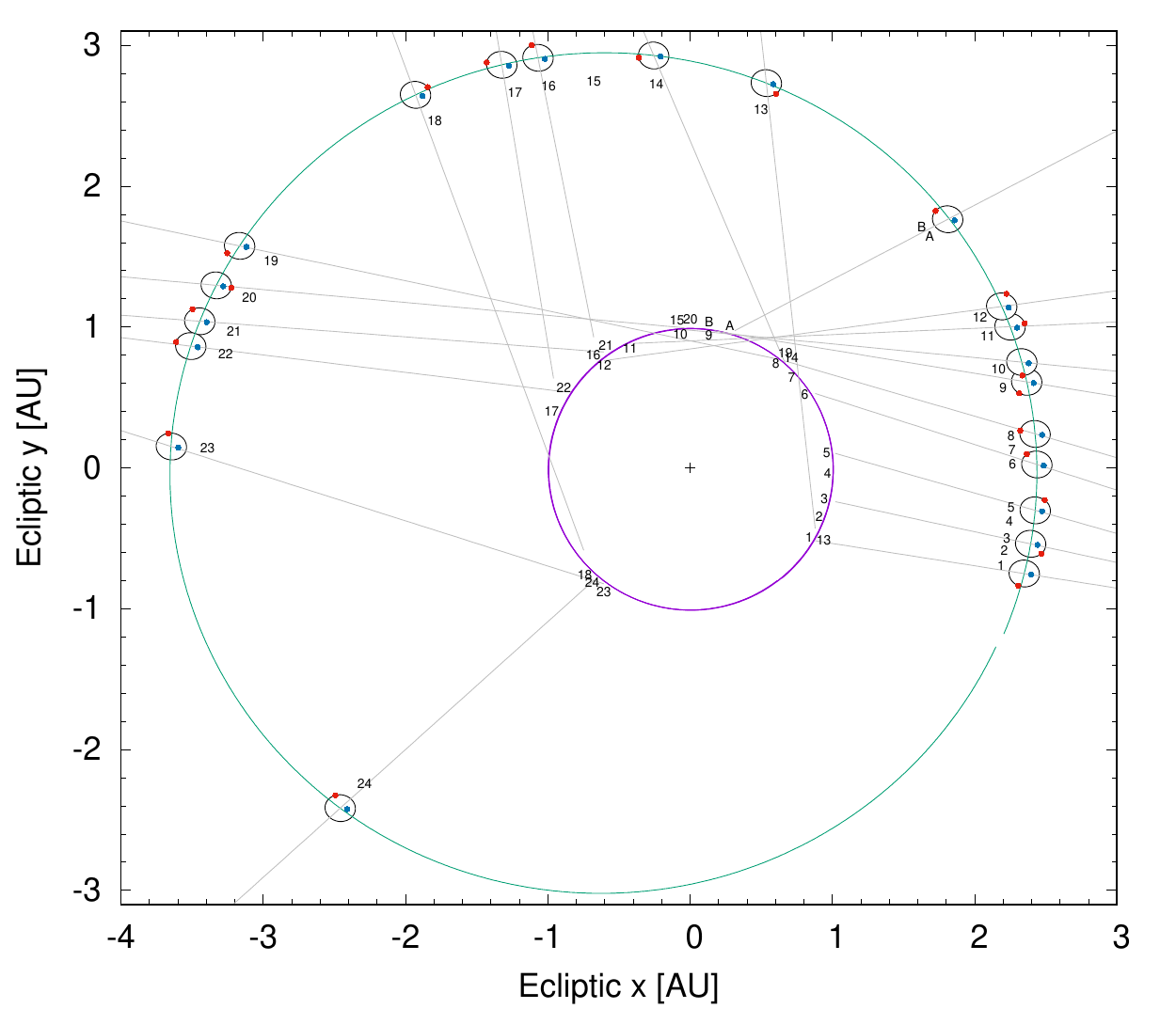}
\caption{Orbital geometry of the 25 HST observations. The heliocentric orbits of 288P (green) and Earth (violet) are shown in projection onto the ecliptic plane, viewed from North. Zero ecliptic longitude is along the positive x-axis. The Sun is located in the origin of the coordinate system. Numbers show the positions of 288P and Earth during the concerned visit, and grey lines indicate the corresponding line of sight. The blue point indicates one focal point of the binary orbit, while the red point indicates the relative position of the other component. The binary orbit is not to scale, and orbit drawings for visits 2, 4, 7, and B have been omitted for clarity.  
}
\label{fig:plane_view}
\end{figure}